\documentclass[envcountsame,envcountchap]{svmono}

\usepackage{makeidx}         
\usepackage{graphicx}        
\usepackage{multicol}        
\usepackage[bottom]{footmisc}

\makeindex             

\begin{document}

\author{Lenore R. Mullin and James E. Raynolds}
\title{
{Conformal Computing: Algebraically connecting the hardware/software
boundary using a uniform approach to high-performance computation 
for software and hardware applications}
\\
{\small SPIN Springer's internal project number, if known}}
\subtitle{-- Monograph --}
\maketitle

\frontmatter


\tableofcontents

\mainmatter
%
%
%

\chapter{Introduction}
\label{intro} 
\section{General Overview}

We present a systematic, algebraically based, design methodology for efficient
implementation of computer programs optimized over multiple levels of the 
processor/memory and network hierarchy.  Using a common formalism to 
describe the problem and the partitioning of data over processors and memory 
levels allows
one to mathematically prove the efficiency and correctness of a given algorithm
as measured in terms of a set of metrics (such as processor/network speeds, 
etc.).  The approach allows the average programmer to achieve high-level
optimizations similar to those used by compiler writers (e.g. the notion of 
{\em tiling}).

The approach is similar in spirit to other efforts using libraries of 
algorithm building blocks based on C++ template classes. In POOMA for example,
expression templates using the Portable Expression Template Engine (PETE) 
\break (http://www.acl.lanl.gov.pete)
were used to achieve efficient distribution of array indexing over scalar
operations~\cite{Humphrey97,Karmesin98,Vlot:1991:PA,Wester:1991:POS,Pooma96,musser,austern,weiss,reynders,veldhuizen}. 

As another example, The Matrix Template 
Library (MTL)~\cite{lumsdaine97,lumsdaine98} is a system that handles dense 
and sparse matrices, and uses template meta-programs to generate tiled 
algorithms for dense matrices.

For example, the addition of two $2$-dimensional arrays $A$ and
$B$

\begin{equation}
(A + B)_{ij} = A_{ij} + B_{ij},
\end{equation}
can be generalized to the situation in which multi-dimensional 
arrays are selected using a vector of indices $\vec v$.

In POOMA $A$ and $B$ were represented as classes, denoting arrays of 
any dimension, and expression templates
were used as re-write rules to efficiently carry out the translation 
to scalar operations implied by:

\begin{equation}
(A + B)_{\vec v} = A_{\vec v} + B_{\vec v}
\label{traditional2d}
\end{equation}

The approach presented in this monograph makes use of A Mathematics of Arrays 
(MoA)~\cite{mul88} and an indexing calculus (i.e. the $\psi$ calculus) to enable the 
programmer
to develop algorithms using high-level compiler-like optimizations through
the ability to algebraically compose and reduce sequences of array operations.

As such, the translation from the left hand side of Eq.~\ref{traditional2d}
to the right side is just one of a wide variety of operations that can 
be carried out using this algebra.  In the MoA formalism, the array
expression in Eq.~\ref{traditional2d} would be written:

\begin{equation}
{\vec v} \psi (A + B) = {\vec v} \psi A + {\vec v} \psi B
\end{equation}
where we have introduced the {\em psi-operator} $\psi$ to denote the 
operation of extracting an element from the multi-dimensional array using
the index vector (${\vec v}$).

In this work we demonstrate, in detail,  our approach as applied to the 
creation of efficient implementations of the Fast Fourier Transform (FFT) 
optimized
over multi-processor, multi-memory/network, environments.  Multi-dimensional
data arrays are {\em reshaped} through the process of {\em dimension lifting}
to explicitly add dimensions to enable indexing blocks of data over
the various levels of the hierarchy.  A sequence of array operations 
(represented by the various operators of the algebra acting on arrays)
is algebraically composed to achieve the {\em Denotational Normal Form} (DNF).
The DNF is a semantic normal form expressed in terms of cartesian coordinates. 
Then the DNF is transformed into the ONF ({\em Operational
Normal Form}) which explicitly expresses the algorithm in terms of loops
and operations on indices (i.e. {\em starts, stops, and strides} are explicitly
indicated).  This reduction is based on an index calculus we call the 
$\psi$-calculus.
The ONF can thus be directly translated into
efficient code in the language of the programmer's choice be it for hardware
or software application.

The algebra we use is a {\em Universal Algebra} based on Sylvester's 
work~\cite{sylvester} which was later embodied in the programming language
APL.  The algebra of MoA is a functional, enhanced subset of APL.  We
emphasize, however, that {\em \bf MoA is not a programming language} but
rather is a mathematical theory.  Also it is important to note that 
{\em \bf APL does not have an index calculus} similar to the $\psi$-calculus.
The notion of an index calculus was suggested by Abrams in 
$1970$~\cite{abrams70} and it was extended and closed by Mullin~\cite{mul88}.

The application we use as a first demonstration vehicle -- the Fast Fourier 
Transform -- is of significant interest in its own right. 
The Fast Fourier Transform (FFT) is one of the most
important computational algorithms and its use is pervasive in science and
engineering. This work traces the development and refinement of efficient
implementations of the FFT including one
in which the FFT was optimized in terms of in-cache
operations leading to factors of {\em two} to {\em four} speedup in comparison
with our previous records.~\cite{cpc} 
Further background material including comparisons with library routines can 
be found in Refs.~\cite{lenss3,lenss4,mullin.small:,mul03} and~\cite{cpc},
and will be reviewed in a following section. 

Our approach can be seen to be a generalization of similar work aimed at
out-of-core optimizations~\cite{cormen}. Similarly, tensor decompositions
of matrices (in general) are special cases of our {\em reshape-transpose} design.
Most importantly, our designs are general for any partition size,
i.e. not necessary blocked in squares, and any number of dimensions.
Furthermore, our designs use linear transformations from an algebraic
specification and thus they are {\bf verified}. Thus, by specifying
designs (such as Cormen's and others~\cite{cormen}) using these 
techniques, these designs too could be verified.

In the context of the cache-optimized algorithm we {\bf DO NOT} attempt any serious analysis of the
number of cache misses incurred by the algorithm in the spirit of
of Hong and Kung and others~\cite{hongkung,savage,vitter93algorithms}.
Rather, we  present an {\bf algebraic} method
that achieves (or is competitive) with such  optimizations
{\bf mechanically}. Through linear transformations we
produce a {\bf normal form}, the ONF, that is directly implementable in any
hardware or software language and is realized in any of the processor/memory
levels~\cite{fftharry}.
Most importantly, our designs are completely general in that through {\bf
dimension lifting} we can produce any number of levels in the processor/memory
hierarchy.

One objection to our approach is that one might incur an unacceptable
performance cost due to the periodic rearrangement of the data that is often 
needed.  This will not, however, be the case if we  pre-fetch data before
it is needed.  The necessity to pre-fetch data also exists in other
similar cache-optimized schemes.  Our algorithm does what the
compiler community calls {\em tiling}. Since we have analyzed the loop
structures,  access patterns, and speeds of the processor/memory levels,
pre-fetching becomes a deterministic cost function that can easily
be combined with  {\em reshape-transpose} or {\em tiling} operations.

Again we make no attempt to optimize the algorithm for any particular
architecture.  We provide a general algorithm in the form of an Operational
Normal Form that allows the user to specify the blocking size at run time.
This ONF therefore enables the individual user to choose the blocking
size that gives the best performance for any individual machine, 
{\em \bf assuming
this intentional information can be processed by a compiler}\footnote{Processing intentional information will be the topic of a future
paper}.

It is also important to note the importance of running {\em reproducible
 and deterministic} experiments. Such experiments are only possible
when dedicated resources exist {\em AND} no interrupts or randomness effects
memory/cache/communications behavior. This means that multiprocessing
and time sharing must be turned off for both OS's and Networks.

Conformal Computing\footnote{The name Conformal Computing \copyright $\;$ is 
protected.  Copyright 2003,
The Research Foundation of State University of New York, University at Albany.}
 is the name given by the authors to this algebraic approach 
to the construction of computer programs for array-based computations in science
and engineering. The reader should not be misled by the name.  Conformal in
this sense is not related to {\em Conformal Mapping} or similar constructs
from mathematics although it was inspired by these concepts in which
certain properties are preserved under transformations.  In particular,
by {\em Conformal Computing} we mean a mathematical system that {\it conforms}
as closely as possible to the underlying structure of the hardware.
Details of the theory including discussion of MoA and $\psi$-calculus
are provided in the following chapter.

The bulk of the material in this monograph is devoted to the FFT.  In the 
fourth chapter a generalized approach is presented based on the notion of 
a hyper-cube data structure.  In the fifth chapter we extend this approach to
consider and streamline certain key computational steps in a density-matrix
based algorithm for simulation of quantum computers.

\section{Summary of Results: Comparing with Established Library Routines}
\label{summary}

In the following sections we review previous success of the approach in which 
our routines were found to be competitive with, or outperformed 
well-established library routines.  Further background can be found in 
Refs.~\cite{lenss3,lenss4,mullin.small:,mul03} and~\cite{cpc}.

We focus on the performance of the code fragment listed in 
Fig.\ref{fftpsirad2}.

\begin{figure}[h]
\vspace*{.15in}
\begin{tt}
\hspace*{1em}  do q = 1,t \\
\hspace*{2em}    L = 2**q  \\
\hspace*{2em}    do row = 0,L/2-1 \\
\hspace*{3em}      weight(row) = EXP((2*pi*i*row)/L) \\
\hspace*{2em}    end do  \\
\hspace*{2em}    do col$'$ = 0,n-1,L \\
\hspace*{3em}       do row = 0,L/2-1 \\
\hspace*{4em}   c = weight(row)*x(col$'$+row+L/2)  \\
\hspace*{4em}   d = x(col$'$+row) \\
\hspace*{4em}   x(col$'$+row) = d + c  \\
\hspace*{4em}   x(col$'$+row+L/2) = d - c  \\
\hspace*{3em}       end do \\
\hspace*{2em}    end do  \\
\hspace*{1em}    end do
\end{tt}
\begin{center}
\end{center}
\caption{Radix 2 FFT with in--place butterfly computation from 
Ref.~\cite{mullin.small:}. We refer to this code fragment as
{\em fftpsirad2} in comparisons with other routines.}
\label{fftpsirad2}
\end{figure}

This piece of code is the radix-2 version of a general-radix algorithm
developed and tested in Ref.~\cite{mullin.small:}. In the following 
comparisons, we refer to this routine as {\em fftpsirad2}.

\section{ Performance of the Radix-2 FFT}

Our radix 2 experiments were run on three dedicated systems:

\begin{enumerate}
\item
a SGI/Cray Origin2000 at  NCSA\footnote{This work was partially supported by National Computational Science Alliance, and utilized the NCSA SGI/CRAY Origin2000} in Illinois, with 48, 195Mhz R10000 processors, and 14GB of memory. The L1 cache size is 64KB (32KB Icache and 32 KB Dcache).  The Origin2000 has a 4MB L2 cache. The OS is IRIX 6.5.
\item
an IBM SP2 at the MAUI High Performance Computing Center\footnote{We would like to thank the Maui High Performance Computing Center for access to their IBM SP2.}, with 32, P2SC 160Mhz processors, and 1 GB of memory.  The L1 cache size is 160KB (32KB Icache and 128KB Dcache), there is no L2 cache.  The OS is AIX 4.3.  
\item
a SUN SPARCserver1000E,  with 2, 60Mhz processors, and 128MB of memory.  Its L1 cache size is 36KB (20KB Icache and 16KB Dcache) and it is one-way set associative.  The OS is Solaris 2.7.   
\end{enumerate}

We tested against the FFTW on all three machines and against the  math libraries supported on the IBM SP2 and the Origin 2000 machines; on the Origin 2000:
IMSL Fortran Numerical Libraries version 3.01,
NAG version Mark 19, and SGI's SCSL library, and on the SP2: IBM's ESSL 
library.  

\subsection{Experiments}

Experiments on  the Origin 2000 were run using {\tt bsub}, SGI's batch processing environment. Similarly, on the SP2 experiments were run using {\tt loadleveler}. In both cases we used 
{\em \bf dedicated networks and processors with ALL multiprocessing
and time sharing DISABLED}.  
The SPARCserver 1000E was a machine dedicated to running our experiments.
 Experiments were repeated a minimum of three times and averaged for each vector size ($2^3$ to $2^{24}$). 
Vendor compilers were used with the {\tt -O3} and {\tt-Inline} flags, for improved 
optimizations (i.e. no special flags, noting the combinatorial explosion
of combinations of flags one might use in an ad hoc approach).   We used Perl 
scripts to automatically compile, run, time all experiments, and to plot our 
results for various problem sizes.

We believe that: 
\begin{quote}
``the only consistent and reliable measure of performance is the execution time of real programs, and that all proposed alternatives to time as the 
metric or to real programs as the items measured have eventually led to misleading claims or even mistakes in computer design.''~\cite{HennPatt}
\end{quote}

Therefore, we time the execution of 
the entire executable, which includes the creation of the FFTW plan.  Although FFTW claims a plan can be reused,
the plan is a data structure and not easily saved.  Therefore, they have 
developed a utility called {\em wisdom}, which still requires partial plan 
regeneration.

\begin{quote}``FFTW implements a method for saving plans to disk and restoring them.  The mechanism is called {\em wisdom}. There are pitfalls to using {\em wisdom}, in that it can negate FFTW's ability to adapt to changing hardware and other conditions.  Actually, the optimal plan may change even between runs of the same binary on identical hardware, due to differences in the virtual memory environment.  It is therefore wise to recreate {\em wisdom} every time an application is recompiled.''~\cite{fftwdoc}
\end{quote}

This is further confirmed by the fact that
when we ran FFTW's test program on our dedicated machines we found surprising differences in time for the same vector size.
Our dedicated machines ran a default OS, e.g. no special quantum, queue settings, or kernel tuning.  Due to this, a plan should be as current as possible.

IMSL, NAG, SCSL, and ESSL required no plan generation.  The entire execution of code created to use each library's FFT was timed.

\subsection{Evaluation of Results}
\begin{figure}
\begin{center}
\rotatebox{-90}{\includegraphics[height=8cm]{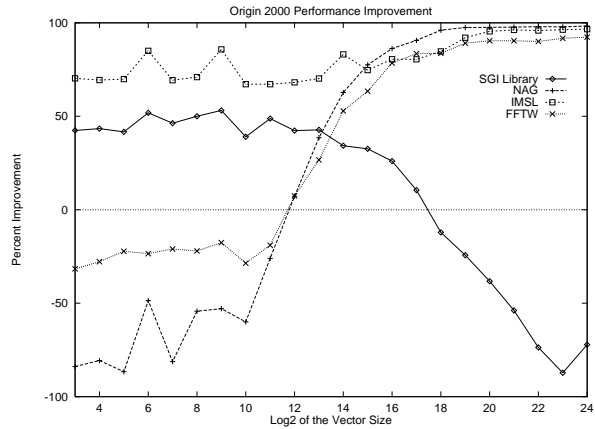}}
\end{center}
\caption{Percent improvement of fftpsirad2 over comparative libraries 
and FFTW on NCSA's SGI/CRAY Origin 2000. 
}
\label{percentncsa}
\end{figure}

\begin{figure}
\begin{center}
\rotatebox{-90}{\includegraphics[height=8cm]{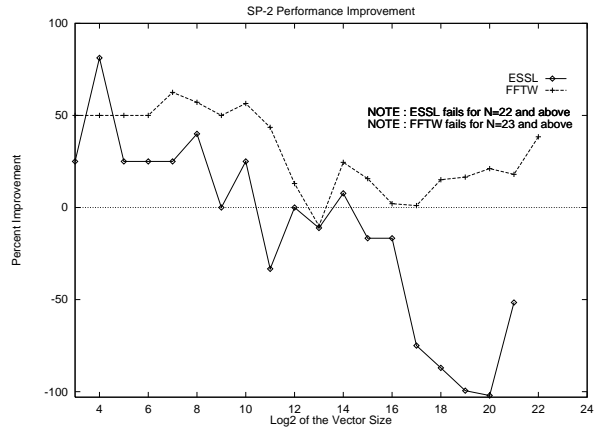}}
\end{center}
\caption{Percent improvement of fftpsirad2 over ESSL and FFTW on Maui HPCC 
IBM SP2.}
\label{percentmaui}
\end{figure}

Our results on the Origin 2000 and the SP2, shown in Figs.~\ref{percentncsa} 
and~\ref{percentmaui}, and Tables~\ref{ncsa} and~\ref{maui}, indicate a performance improvement over our competition in a majority of cases.

\subsection{Origin 2000 Results}
Performance results for our monolithic FFT code, {\em fftpsirad2}, indicate a
doubling of time when the vector size is doubled for all vector
sizes.
IMSL doubled its performance up to $2^{19}$. At $2^{19}$ there is an apparent memory
change causing a $400\%$ degradation in performance. For NAG this
degeneration begins at $2^{18}$. The SGI library, SCSL,  is apparently doing
very machine specific optimizations, perhaps out of core techniques
similar to Cormen~\cite{cormen} as evidenced by nearly identical
performance times for $2^{17}$ and $2^{18}$.

Against the FFTW, we achieved a performance improvement for vector sizes greater than $2^{11}$.  We ran slightly slower for the vectors $2^2$ through $2^{11}$, with a maximum difference in speed of 0.013 seconds.  

Both programs were able to run for the vector size $2^{24}$ (inputs of $2^{25}$ and greater, result in {\em stack frame larger than system limit} error on compile).
Therefore, we achieved a large performance increase for the upper vector sizes, while our performance remained competitive for the lower sizes.

\begin{table}
\begin{center}
\vspace*{.15in}
\begin{tabular}{|c|c|c|c|c|c|} \hline
\multicolumn{6}{|c|} {\bf Origin 2000} \\ \hline
Size & fftpsirad2 & IMSL & NAG & SCSL & FFTW\\ \hline
$2^{3}$     & 0.190	& 0.064		& 0.010		& 0.065 & 0.013\\
$2^{4}$     & 0.018	& 0.061		& 0.010 	& 0.047 & 0.013\\
$2^{5}$     & 0.018	& 0.062		& 0.010 	& 0.065 & 0.014\\
$2^{6}$     & 0.017	& 0.116		& 0.011 	& 0.073 & 0.013\\
$2^{7}$     & 0.019	& 0.063		& 0.010 	& 0.068 & 0.015\\
$2^{8}$     & 0.018	& 0.062		& 0.011 	& 0.105 & 0.014\\
$2^{9}$     & 0.017	& 0.122		& 0.011  	& 0.069 & 0.014\\
$2^{10}$     & 0.021	& 0.065		& 0.013 	& 0.056 & 0.015\\
$2^{11}$     & 0.021	& 0.064		& 0.016 	& 0.058 & 0.017\\
$2^{12}$     & 0.021	& 0.067		& 0.023 	& 0.067 & 0.023\\
$2^{13}$     & 0.022	& 0.075		& 0.036 	& 0.065 & 0.030\\
$2^{14}$     & 0.024	& 0.144		& 0.065 	& 0.066 & 0.051\\
$2^{15}$     & 0.030	& 0.120		& 0.135 	& 0.110 & 0.082\\
$2^{16}$     & 0.040	& 0.209		& 0.296 	& 0.080 & 0.189\\
$2^{17}$     & 0.065	& 0.335		& 0.696 	& 0.072 & 0.395\\
$2^{18}$     & 0.126	& 0.829		& 3.205 	& 0.075 & 0.774\\
$2^{19}$     & 0.238	& 3.007		& 9.538 	& 0.096 & 2.186\\
$2^{20}$     & 0.442	& 9.673		& 18.40 	& 0.143 & 4.611\\
$2^{21}$     & 0.884	& 23.36		& 38.93 	& 0.260 & 9.191\\
$2^{22}$     & 1.910	& 46.70		& 92.75 	& 0.396 & 19.19\\
$2^{23}$     & 4.014	& 109.4		& 187.7 	& 0.671 & 48.69\\
$2^{24}$     & 7.550   	& 221.1		& 442.7	        & 1.396 & 99.10 \\ \hline
\end{tabular}
\\
\vspace*{.15in}
\end{center}
\caption{Real Execution Times (seconds) of fftpsirad2, comparative 
libraries and FFTW on NCSA's SGI/CRAY Origin 2000.}
\label{ncsa}
\end{table}

\subsection{SP2 Results}
Our routine {\em fftpsirad2} outperforms ESSL for vector sizes up to $2^{15}$, except for two cases.  For $2^{15}$ and $2^{16}$ our performance is nearly identical.  Our times continue to double from $2^{17}$ to $2^{23}$.
Comparatively, for $2^{16}$ through $2^{17}$ ESSL appears to be doing some machine specific optimizations, which subsequently double in performance through $2^{21}$, and fail for $2^{22}$ and higher.  Consequently, we perform competitively for all but five vector sizes, and we are able to process vectors up to 6,291,456 elements larger.

Against the FFTW, we achieved
a performance improvement for every vector size from $2^2$ through $2^{22}$, 
except for the single vector size $2^{13}$.  Due to our optimizations, our code, {\em fftpsirad2}, ran successfully
for the vector size $2^{23}$ (inputs of $2^{25}$ and greater, result in a {\em not enough memory available to run} error at runtime).  The largest size vector the FFTW was able to 
run on this system was $2^{22}$.  
Therefore we achieved a performance increase for every size except one, and we were able to run successfully for a  vector 4,194,304 elements larger.

\begin{table}
\vspace*{.2in}
\renewcommand{\baselinestretch}{1}
\begin{center}
\begin{tabular}{|r|r|r|r|} \hline
\multicolumn{4}{|c|} {\bf IBM's SP2 }\\ \hline 
Size & fftpsirad2 & ESSL & FFTW \\ \hline
$2^{3}$     & 0.010 &0.013 & 0.020 \\
$2^{4}$     & 0.010 &0.053 & 0.020 \\
$2^{5}$     & 0.010 &0.013 & 0.020 \\
$2^{6}$     & 0.010 &0.013 & 0.020 \\
$2^{7}$     & 0.010 &0.013 & 0.027 \\
$2^{8}$     & 0.010 &0.016 & 0.023 \\
$2^{9}$     & 0.010 &0.010 & 0.020 \\
$2^{10}$    & 0.010 &0.013 & 0.023 \\
$2^{11}$    & 0.013 &0.010 & 0.023 \\
$2^{12}$    & 0.020 &0.020 & 0.023 \\  
$2^{13}$    & 0.033 &0.030 & 0.030 \\
$2^{14}$    & 0.040 & 0.043 & 0.053\\
$2^{15}$    & 0.070 & 0.060 & 0.083\\
$2^{16}$    & 0.140 & 0.120 & 0.143\\
$2^{17}$    & 0.280 & 0.160 & 0.283\\
$2^{18}$    & 0.580 & 0.310 & 0.683\\
$2^{19}$  & 1.216 & 0.610 & 1.456\\
$2^{20}$  & 2.580 & 1.276 & 3.273\\
$2^{21}$  & 5.553 & 3.663 & 6.770\\
$2^{22}$  & 12.12 & Failed & 15.553\\
$2^{23}$  & 25.25 & Failed & Failed\\ \hline
\end{tabular}

\vspace*{.15in}
\end{center}
\caption{Real Execution Times (seconds) of fftpsirad2 and ESSL on Maui 
HPCC IBM SP2.}
\label{maui}
\end{table}

\subsection{SUN SPARCserver 1000E results}

\begin{figure}
\begin{center}
\rotatebox{-90}{\includegraphics[height=8cm]{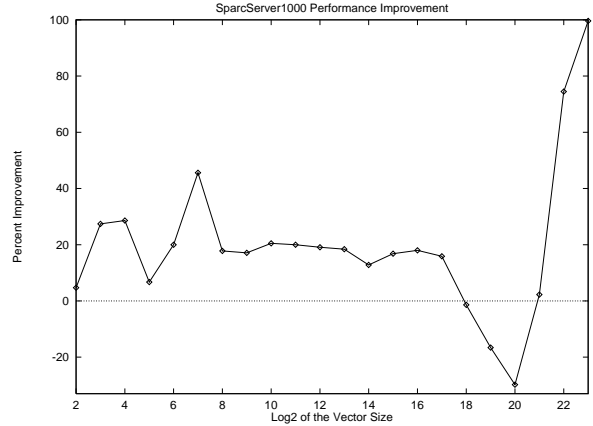}}
\end{center}
\caption{Percent improvement of fftpsirad2 over FFTW on SPARCserver 1000E.
}
\label{einpercent}
\end{figure}

Our results on the SPARCserver 1000E, shown in Fig.~\ref{einpercent}, 
achieved performance improvement for every vector size except three.  Additionally, we were able to run for the vector size $2^{24}$ (inputs of $2^{25}$ and greater, result in an {\em integer overflow} error on compile).  The FFTW failed for $2^{24}$, and for $2^{23}$ it ran for over 34 hours.

\section{Performance of the General-Radix FFT}

In this and following sections we review the performance of our general
radix algorithm, developed and tested in Ref.~\cite{mullin.small:}.

We've discussed how to design and build a faster radix 2 FFT, and the results 
we've achieved.  Using the same theoretical framework we can extend this design 
and subsequent implementations to handle any radix.  Given a particular 
architecture and its memory hierarchy, a radix other than 2 may be more 
appropriate. For example, if a machine has an $n$-way associative cache, then
radix-$(na)$ should be used (with $a > 1$), thus decreasing 
control time.

By generalizing the algorithm, we can investigate how a change in radix affects performance without changing the algorithm and subsequent executable.
In what follows, we refer to our radix-n FFT as {\em fftpsiradn}.  
Fig.~\ref{rad2butterfly} illustrates our F90 software realization for the body of the radix 2 butterfly.  Observe that the variable {\tt c} is a scalar and 
{\tt ebase} is implicitly calculated based on radix 2, i.e. 1 and -1.  Also, observe that the butterfly is explicitly done using F90 syntax.  In the generalized code, Fig.~\ref{radnbutterfly}, {\tt c} becomes a vector whose length is equal to the 
({\tt radix - 1}).  Hence for radix 2, {\tt c} is a one component vector, the 
variable {\tt base} (see Fig.~\ref{radnbutterfly}), is used to designate the 
radix desired.

\begin{figure}
\begin{verbatim}

                c = ww(j_)*z(i_+j_+L)               
                z((/ i_+j_ , i_+j_+L /)) = 
                   (/ z(i_+j_)+c , z(i_+j_)-c /) 

\end{verbatim}
\begin{center}
\end{center}
\caption{ Fragment from Radix 2 butterfly F90 code}
\label{rad2butterfly}
\end{figure}

\begin{figure}
\scriptsize
\begin{verbatim}
          c(start) = z(i_+j_)
          ebase(start) = 1
          do k_=start+1,base-1
             c(k_) = ww(j_*k_)*z(i_+j_+(k_*L))
             z(i_+j_) = z(i_+j_)+c(k_)
             ebase(k_)=EXP((-2*pi*i)/base)**k_
          end do

          do k_=start+1,base-1
             temp = c(start)
             do l_=start+1,base-1
                temp = temp+c(l_)*
                    ebase(modulo((k_*l_),base))
             end do
             z(i_+j_+(k_*L))=temp
          end do

\end{verbatim}
\footnotesize
\begin{center}
\end{center}
\caption{Fragment from Radix n F90 code.}
\label{radnbutterfly}
\end{figure}

\subsection{General Radix Issues}

When a vector has length $2^n$, a radix 2 FFT may be used.  Whenever the same vector is a power of 4, 8, 16, etc. that radix FFT may also be used: i.e.: when $2^n$ = $(2^{m})^{l}$ where $n = m+l$.  For example, $2^5$ = 64 = $(2^{3})^{2}$ = $8^2$ = $(2^{2})^{3}$ = $4^3$.  Hence, for a vector length 64, radix 2, 4, 8 or 64 may be used.  Depending on the size of the cache, the number of cache lines, associativity of the cache, etc.  one radix may perform better on one architecture over another.  The reason for this is when radix 2 is used on an input vector of length n, the butterfly takes two inputs and requires $log_2 n$ iterations.  

For radix 4, there are four inputs $log_4 n$ iterations, etc. 
Consequently, the butterfly width and the number of elements becomes the deciding factor in performance.  Using one executable increases flexibility across machines.

Incidentally, the implications of an n-way n-radix FFT is that
when a quantum machine is built, i.e. when all bits can interact with
all bits at the same time, this algorithm will scale to that machine.

\subsection{Experimental Environment}
Our general radix experiments were run on two dedicated systems:
\footnotesize
\begin{enumerate}
\item
a SUN SPARCserver1000E.  This machine has two 60Mhz processors, and 128MB of memory.  Its L1 cache size is 36KB (20KB Icache and 16KB Dcache) and it is one-way set associative.  The OS is Solaris 2.7.   
\item
a SUN 20.  This machine has a single 50 Mhz processor, and 64 MB of memory. The OS is Solaris 2.7.
\end{enumerate}

\subsection{General Radix Experiments}
\footnotesize
The SPARCserver 1000E and the SUN 20 were both machines dedicated to running our experiments.
We ran our experiments for a variety of different radices,  
given the following constraints on the input vector length: $2^n$ = $(2^{m})^{l}$, $n = m+l$, 
$n,m,l \; \epsilon \; Z^+$, $1 \leq m \leq 8$.  
  In determining what results to discuss and represent in our graphs, we removed the following radices: 128, and 256. Their non-competitive performance results is plausibly a consequence of no machines having a cache with 128 or 256 way associativity.  If this changes in the future, our executable could adapt.  Table 3 illustrates  performance times given a vector size 16,777,216,  using an FFT radix of 2, 4, 8, 16 and 64.

\footnotesize
\subsection{Evaluation of Results}
\footnotesize

\begin{figure}
\begin{center}
\rotatebox{-90}{\includegraphics[height=8cm]{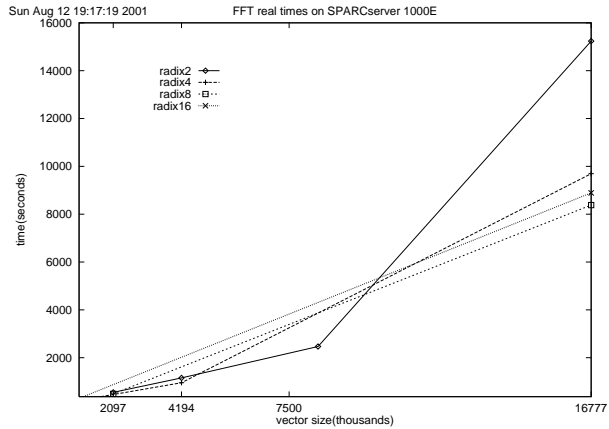}}
\end{center}
\caption{Comparison of radices performance on the SUN SPARCserver 1000E: One 
Processor}
\label{foob}
\end{figure}

\begin{figure}
\begin{center}
\rotatebox{-90}{\includegraphics[height=8cm]{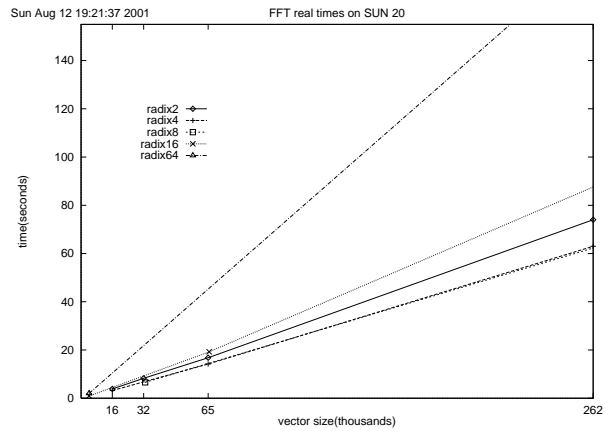}}
\end{center}
\caption{Enlarged image of Fig~\ref{foof} lower left corner.}
\label{foof}
\end{figure}

We now turn to a comparison of the general radix algorithm of various platforms
as illustrated in Figs.~\ref{foob}, through~\ref{foof}.

Radix 8 performed best on the SUN 20 with a maximum vector size of 16,777,216.  The performance of radix 2 and 4 experiments were impacted greatly by the number of page faults.  This is, due to the fact that radix 4 performs twice as many iterations as radix 8, and radix 2 performs four times as many as radix 8.  Each iteration will have about the same number of page faults as the previous, due to the fact that every sample is accessed on every iteration.  The number of page faults recorded for these experiments is increased by almost these same factors.

Radix 8 has twice as many iterations as radix 16, and as expected the number of page faults is higher, but radix 8 outperforms radix 16.  We conjecture that this is due to the large number of cache misses occurring at radix 16 and higher, as can be seen by an increase in user time in Table 4.
The cache misses are occurring at this higher vector size because 16 samples are being processed at the same time rather than 8.  Either all 16 components don't fit in cache or many are mapping to the same cache block, hence they must wait to be loaded.  Therefore, although there are fewer page faults, the higher number of cache misses are causing radices greater than or equal to 16 to incur performance degradation.  Notice when comparing the SPARCserver 1000E one processor performance to the SUN 20 for vector sizes $\leq 2^{18}$, the graphs are nearly identical.  Performance also appears similar from $2^{18}$ through $2^{22}$.  Since the SUN 20's memory was half the size of the SPARCserver 1000E's we observe virtual behavior sooner.
More memory is available on the SPARCserver 1000E even though we are using one processor.

\subsection{SPARCserver 1000E's Results (One 60 MHz Processor)}

\begin{figure}
\begin{center}
\rotatebox{-90}{\includegraphics[height=8cm]{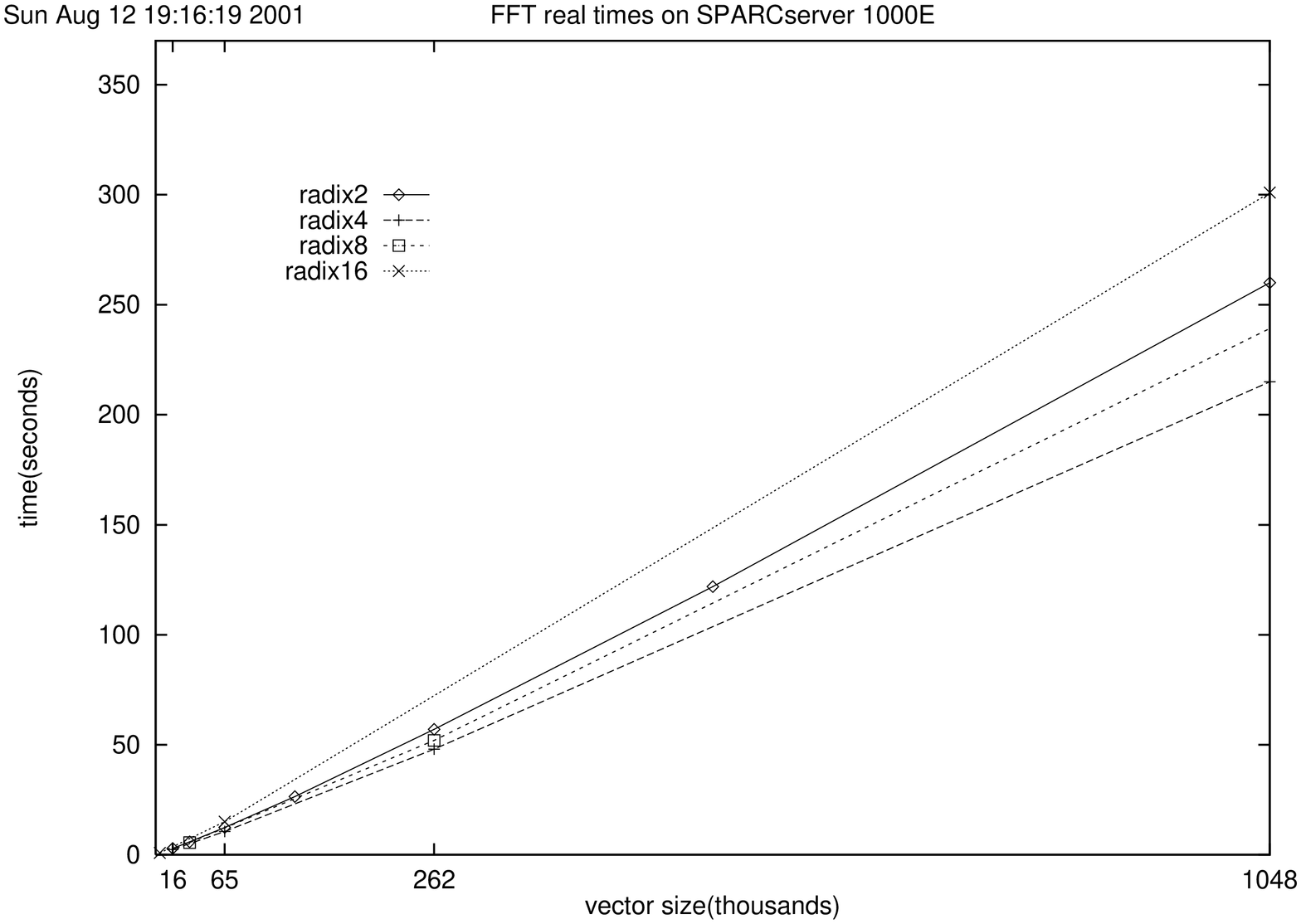}}
\end{center}
\caption{Enlarged image of Fig.~\ref{foob}, lower left corner.}
\label{fooa}
\end{figure}

\begin{tabular}{|r|r|r|r|r|r|} \hline
      &      &      &         &        & \\
      & Data    &      & \%      & \% & Total\\
      & Fault     &Page  & Data    & User   & Run \\ 
Radix & Time   &Faults& Time    & Time   &Time \\ \hline
2     & 9318		& 779628	& 50  & 28 & 15235\\
4     & 5198		& 411978	& 46  & 35 & 9693\\
8     & 3649		& 289973	& 35  & 43 & 8381\\
16    & 2903		& 237271	& 26  & 52 & 8890\\
64    & 2582		& 195693 	& 13  & 66 & 16040\\ \hline
\end{tabular}
\begin{center}
Table 4: Paging Statistics for vector size 16,777,216
as Reported by SPARCworks, times are in Seconds
\end{center}                                  
\footnotesize

Notice the steady decrease in the number of page faults as the radix increases, this is caused by the number of iterations decreasing by a factor of two as each radix increases by a factor of two.
As expected the amount of time the application spends in page faulting is directly related to the radix.  Equally important is the steady increase in the percentage of user time as the radix increases, thus implying an increase in cache misses.

\begin{figure}
\begin{center}
\rotatebox{-90}{\includegraphics[height=8cm]{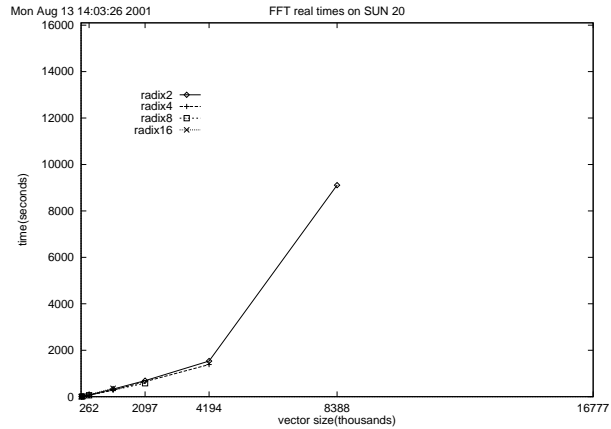}}
\end{center}
\caption{Comparison of Radices performance on SUN 20.}
\label{fooe}
\end{figure}

\subsection{ Conclusions for the General Radix Approach}
We have succeeded in simplifying a solution to a complex problem: faster and bigger FFTs.  Reproducible experiments indicate that our designs outperform 
all tested packages in either all, or a majority of the cases, while remaining competitive in the others. 

Our {\em single, portable, scalable} executable of approximately 2,600 bytes 
(which can be easily built in software, hardware, or {\bf both}) 
also must be compared with the large suite of machine--specific software required by NAG, IMSL, SCSL, ESSL, and FFTW.
For example, FFTW's necessary libraries, codelets, etc. total approximately 7,120,000 bytes.

The FFTW install is potentially complicated.  It requires knowledge
 of makefiles, compiler options, and hand-tuning to work around variants of OSs.  For example, the FFTW install fails with the default Solaris cc compiler, requiring a path change to Solaris's 
SUNWspro cc compiler.  The user must know that a new plan is needed for every vector size and they need to decide whether to attempt to use {\em wisdom} to save a plan.

The time and investment it takes to create a scientific library also needs to be factored into an analysis of the software.  New machine designs require the reprogramming of these large libraries in order to maintain their performance.  This will result in an increase of the library size as new machines are added.  Also, the skill level of the scientific programmer who supports these libraries, and their knowledge of each machine is very high.
The user of these machine-specific libraries must have a deep understanding of the package itself and the system on which it runs.  Many of these packages require numerous parameters;  fifteen for ESSL, and eight for SCSL.  An incorrect decision for these parameters can result in poor performance and even possibly incorrect results.  Consequently, the learning period to adapt to new software 
usage can be enormous from a user perspective.

In comparison, our monolithic algebraic approach to design is based on {\em mechanizable} linear transformations, which begin with a high-level specification, e.g. MATLAB, Mathematica, Maple, etc.  We have found that scientists prefer these high-level languages to rapidly prototype their designs~\cite{lincoln}, as opposed to interfacing with programmers.  Consequently, it is imperative that languages such as MATLAB compile to efficient code.  This would
necessitate such languages identifying a functional subset that
embodied MoA and Psi Calculus.

Our monolithic approach addresses these issues.  Our approach has no learning curve, a single executable can be used independent of vector size or machine.  Additionally, a naive user of {\em fftpsirad2} and {\em fftpsiradn} need only know the vector size on any platform.
Furthermore, our monolithic algebraic approach to design is extensible to cache and other levels of the memory hierarchy,  as well as parallel and distributed computing~\cite{fftharry}.

We have discovered that radix 2 is not always the best radix to choose for vectors whose length is a power of two. 
  This opens other interesting questions:  is it better to pad to radix 2 or use another radix?\footnote{Padding is the usual way of handling vector lengths not equal to $2^{n}$}
 We believe our research into a general radix formulation may yield further optimizations for FFTs.

Our research shows a systematic way of analyzing memory access patterns and subsequent performance of the FFT on one dimensional\footnote{ Multi-dimensional FFTs may be built from the one dimensional FFT.} arrays whose length is $2^n$.  

  Besides determining the constituent components of the sequential FFT, e.g. bit reversal or butterfly, the radix used is of paramount importance.  By developing algorithms for software, e.g., {\em fftpsiradn}, we are guaranteed that all instructions, loops and variables remain constant during the execution of the program.  Hence, {\bf through monolithic analysis we can in concert, analyze the algorithm, the program, and the environment}.

\section{Effects Due to Specilized Hardware}

In this section we review published work in which the effects due to the 
use of specialized hardware are illustrated~\cite{cpc}. Specifically we 
compare our routine with one that was specifically designed to exploit
hardware with the capability to carry out the operation of {\em multiply}
and {\em add} in one step.  We find that our generic radix-2 FFT 
(see Fig.~\ref{fftpsirad2}) is competitive on such hardware and is superior 
on a machine lacking the specialized hardware.

The study presented in Ref.~\cite{cpc} is a benchmark of our 
radix-2 FFT {\em fftpsirad2}~\cite{mullin.small:} in the context
of the plane-wave based electronic structure code Abinit~\cite{**key*}.  
Such codes rely heavily on the use of the FFT and considerable work has been 
expended to optimize performance~\cite{goedecker:}. Therefore this context 
serves as a further stringent benchmark test of our routine.  
The purpose of this work was NOT to claim that we have the best FFT.  Rather 
we emphasize that our approach leads to efficient code based on general 
principles without any hardware-specific optimization (which naturally can be 
included as a further refinement).  

We now turn our attention to a study of the one-dimensional FFT.  In 
comparisons between our routine, which we denote as the ``CC" routine, we will 
refer to the Abinit routine as the ``Ab"~\cite{**key*,goedecker:}.

We have carried out benchmark performance tests of our 
$1$-dimensional FFT in comparison with the routine taken from the Abinit 
code~\cite{**key*,goedecker:} using the same platforms as were used in 
previous tests: the Origin 2000 at NCSA~\cite{**key*1}
and the IBM SP2 at Maui~\cite{mullin.small:,**key*2}.
The benefit of our focus on the $1$-dimensional transform, lies in the fact 
that it serves as a control test for later developments (multi-dimensional 
arrays and multi-level, multi-processor memories).  

The following tests were carried out.  We ran both routines as single 
processor jobs with the -O3 compiler option (f90 compiler on NCSA, and xlf on
Maui) in a dedicated environment.  
For comparison, therefore, the $3$-dimensional transform of Ref.~\cite{goedecker:} 
was run as a $1$-dimensional transform by considering 
arrays of size $(N,2,2)$, $(2,N,2)$, and $(2,2,N)$ where $N=2^n$ with
$n = 1,2,3,\cdots$. Slightly better performance was obtained for the $(N,2,2)$
case so these are the results that we will use for comparisons with the CC
routine. Perl scripts were used to compile the jobs for each size, time the
execution, and collect the results.  The timings from the runs involving the
$3$-dimensional routine were divided by a factor of $4$ to account for the
fact that two of the $3$ dimensions were held fixed with the 
value $2$ (dimensions of length $1$ are not permitted in the Ab routine). 

\begin{figure}
\small

\begin{center}
\includegraphics[height=8cm]{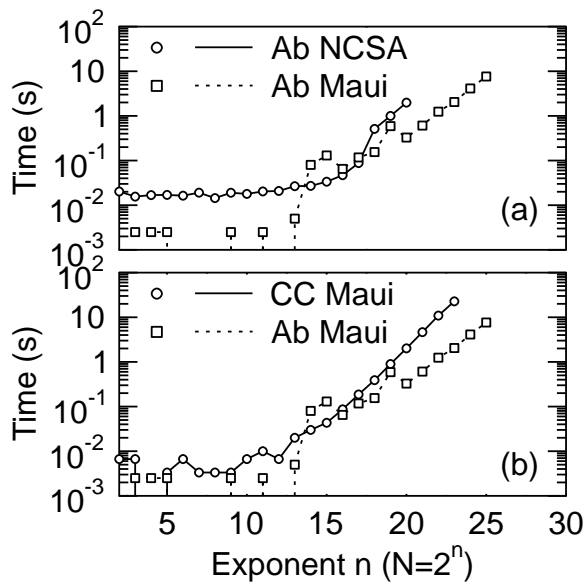}
\end{center}

\caption{\label{SG_n_m} Comparison of the Ab routine (Abinit) and the CC 
routine on the Origin (NCSA) vs. the SP2 (Maui).  (a) The Ab routine has a 
performance advantage on the SP2 at Maui in comparison to the Origin 2000 at 
NCSA (b) The Ab routine out-performs the CC routine on the SP2 for large 
systems but for small systems, the CC routine is quite competitive.}
\normalsize
\end{figure}

Qualitatively different results were obtained for the same tests carried
out on the two different platforms emphasizing the important role
played by  compilers and hardware. The Abinit routine was specifically designed
to utilize performance improvements through the use of the (specialized
hardware) {\it Multiply-Add Instruction}~\cite{goedecker:}. \rm As mentioned
in Ref.~\cite{goedecker:} this gives the Abinit routine an advantage on Maui
which is not available on NCSA.  This behavior is illustrated in
Fig.~\ref{SG_n_m} (a) in which the Abinit routine clearly is faster and can 
handle larger sizes on the SP2 (Maui) in comparison to the Origin (NCSA).  We also find the Abinit routine outperforms the CC routine (which makes no use of the specialized 
hardware) for large systems on the SP2 at Maui as illustrated in Fig.~\ref{SG_n_m} (b). For small systems, the CC routine is quite competitive.

\begin{figure}
\begin{center}
\includegraphics[height=8cm]{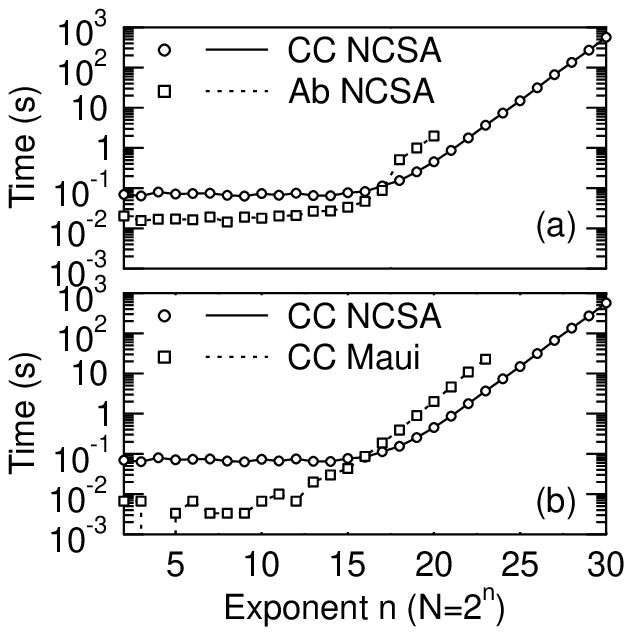}
\end{center}
\caption{\label{psi_SG_n} Performance of the Conformal Computing routine
(CC) on both platforms and with the Abinit (Ab) routine on the Origin 2000 at 
NCSA.  
(a) The CC routine outperforms the Ab routine for very large systems and is 
competitive for small systems. (b) Similar results are found for the CC routine on both platforms
for large systems.  Apparently the SP2 is faster but has less memory than the
Origin.
}
\end{figure}

In contrast, however, the simple and efficient design of the CC 
routine (see Fig.~\ref{fftpsirad2})
allows it to greatly outperform the Ab routine on the Origin (NCSA) for 
which the
Ab routine cannot take advantage of the specialized hardware.
Fig.~\ref{psi_SG_n} (a) illustrates this behavior.  For small systems, both
routines have similar slopes with a nearly constant offset most likely due to
startup costs.  For larger systems, however, the CC routine clearly wins.

Lastly by comparing the behavior of the CC routine (which is expected to
make use of similar hardware on both machines) we can conjecture relationships
between hardware and performance. Figure \ref{psi_SG_n} (b) illustrates the
performance of the CC routine on the SP2 at Maui vs. the Origin 2000 at NCSA.
Apparently the SP2 is faster but has less memory than the Origin.  This
conclusion follows from the fact that the slopes are similar for large systems
but the turn-over point (where the slope begins to increase) occurs earlier
for the SP2.  This is an important point: changes in performance (i.e. in slope)
correspond to various cache, memory, and virtual memory boundaries.

The tests presented in this section are consistent with previous
findings~\cite{mullin.small:,lenss3,lenss4}
in that our routine, constructed based on Conformal Computing
techniques, is competitive with other well tuned code despite the absence of
any special optimizations such as cache loops, or the reliance on
a specific piece of hardware, etc.  By observing the behavior of timing vs.
size (changes in slope etc.) we are able to identify various details of the
hardware (boundaries on cache, memory, virtual memory, etc.).  Such information
can be used for further refinement of the design using the general principles
of Conformal Computing.

\section{Organization of the Monograph}

In this chapter we have introduced considerable background material on our
approach and given ample material demonstrating competative to superior 
performance of resulting implementations.  The remainder of this monograph
is organized as follows.

In the following two chapters, the reader is introduced to the algebraic
formalism and important similarities and differences with standard linear
algebra are emphasized.  The following three chapters presents 
never-before-published work devoted to the development of a cache-optimized
FFT. The first of these presents our approach in a manner that bridges the
gap between the language of standard linear algebra and the methods of 
Conformal Computing.  The second of the three, presents the same problem 
in full detail using the machinery of Conformal Computing.  The third of 
these expands and illustrates the algorithm emphasizing the natural role played
by the hyper-cube data structure.  The following chapter builds on this 
hyper-cube view with an application to simulation of quantum computers.  The 
final chapter concludes this review with a discussion of important related 
developments in this newly-emerging field along with a discussion of next 
steps and the proposal of a number of {\em Grand Challenges}.


%
%
%

\newcommand{\cat}{+\!\!\!\!+}
\newcommand{\Take}{\mbox{Take}}
\newcommand{\take}{\,\bigtriangleup\,}
\newcommand{\Drop}{\mbox{Drop}}
\newcommand{\drop}{\,\bigtriangledown\,}
\newcommand{\rshp}{\,\widehat{\Rho}\;}
\newcommand{\Rho}{\,\rho\,}
\newcommand{\Tau}{\,\tau\,}
\newcommand{\Dim}{\,\delta\,}
\newcommand{\transpose}{\bigcirc\!\!\!\!\!\backslash\;}
\newcommand{\Ravel}{\,\mbox{\tt rav}\,}
\newcommand{\Gradeup}{\,\mbox{\tt gu}\,}

\chapter{Conformal Computing Techniques: A Mathematics of Arrays (MoA)
and the $\psi$-calculus}
\label{chap2}

In this chapter we introduce the two cornerstones of the Conformal Computing
approach: (1) A Mathematics of Arrays (MoA) and (2) the $\psi$-calculus. MoA
is an algebra of multi-dimensional monolithic arrays that subsumes the notions
of {\em matrix}, {\em array}, {\em tensor}, etc., of traditional mathematical
and computational approaches.  Array expressions are manipulated through the 
use of a collection of operators which are defined in this chapter.  
Expressions are manipulated by linear transformations that compose indices
using structural information of arrays (i.e. shapes) and operations on 
arrays (e.g. inner product, outer product, etc.). The consequence of these 
linear transformations is an optimized normal form.  This is the essence of 
the $\psi$-calculus.  

The first step is a cartesian normal form: the {\em Denotational Normal Form} 
(DNF) which gives the semantics of what to do but not how optimally 
{\em build the code}.  From the DNF, the process we call $\psi$-reduction
leads to the {\em Operational Normal Form} (ONF) which is an explicit recipe 
for how to build the code.
As such it describes all loops and iteration structures
in terms of {\em starts}, {\em stops} and {\em strides}.  As such, the 
ONF contains specific information regarding the algorithm and data layout
on a particular architecture in terms of memory, processor, and network layout.
It can thus be directly translated into computer code in any hardware or
software language of choice.  That is, the ONF is a {\em \bf specification} 
given:
Iteration, Sequence, and Control for each level of processor/memory
hierarchy desired. These three FUNDAMENTAL issues are the basis of ALL 
architecures and thus, this is the most abstract way to define them.

This chapter introduces the necessary techniques with some examples.  The 
rest of this monograph extends and illustrates the use of these techniques 
in the context of previously unpublished work devoted to the development
of a cache-optimized FFT and for an application to the simulation of a 
quantum computer.

\section{ Elements of the Theory}
\subsection{Indexing and Shapes}
\label{indshp} 

The $\psi$ operator is central to MoA and is used as follows.
We write:
\begin{equation}
{\vec p} \, \psi A, 
\end{equation}
to denote the operation in which a vector of $n$ integers, $\vec p$, is used to 
select an item of the $n$-dimensional array $A$.
The operation is generalized to select a partition of $A$, 
so that if $\vec q$ is a vector having only $k < n $ components
then, $\vec q \, \psi A$,
is an array of dimensionality $n-k$ and $\vec q$ selects among the possible 
choices for the first $k$ axes. In MoA zero origin indexing is assumed.
For example, if $A$ is the $3$ by $5$ by $4$ array\footnote{In all examples, 
as above, we use consecutive integers as array elements although in 
practice, array elements can be arbitrary integer, real, or complex numbers.}

\[ \left [
\begin{array}{lrrrr}
0 & 1 &  2 & 3  \\
4 & 5 & 6 & 7 \\
8 & 9 & 10 & 11  \\
12 & 13 & 14 & 15 \\
16 & 17 & 18 & 19 
\end{array} \right ] 
\left [ \begin{array}{rrrr}
20 & 21 & 22 & 23 \\
24 & 25 & 26 & 27 \\
28 & 29 & 30 & 31 \\
32 & 33 & 34 & 35 \\
36 & 37 & 38 & 39 
\end{array} \right ]
\left [ \begin{array}{rrrr}
40 & 41 & 42 & 43 \\
44 & 45 & 46 & 47 \\
48 & 49 & 50 & 51 \\
52 & 53 & 54 & 55 \\
56 & 57 & 58 & 59
\end{array} \right ] \]
then
\[
     <1> \psi A = \left [  \begin{array}{rrrr}
                   20 & 21 & 22 & 23 \\
                   24 & 25 & 26 & 27 \\
                   28 & 29 & 30 & 31 \\
                   32 & 33 & 34 & 35 \\
                   36 & 37 & 38 & 39 
                    \end{array} \right ] \]

\[     <2 \; 1> \psi A \; \; = \; \; < \; 44 \;\; 45 \;\; 46 \;\; 47 \; > \]

\[      <2\; 1\; 3> \psi A \; \; = \; \; 47 \]

Most of the common array manipulation operations found in languages like 
Fortran 90, Matlab, ZPL, etc., can be defined from $\psi$ and a few elementary 
vector operations.

	We now introduce notation to permit us to define $\psi$ formally and to develop the
{\em Psi Correspondence Theorem}~\cite{muljen94}, which is central to the 
transformation of the DNF into the ONF.
We will use $A, B, ...$ to denote an array of numbers of any type
(integer, real, complex, boolean, etc.).
An array's dimensionality will be denoted by $d_A$ and will be assumed to 
be $n$ if not specified.

	The shape of an array $A$, denoted by $\vec s_A$, is a vector of integers of length $d_A$,
each item giving the length of the corresponding axis. The total number of items in an
array, denoted by $t_A$, is equal to the product of the items of the shape. The subscripts
will be omitted in contexts where the meaning is obvious.

A full index is a vector of $n$ integers that describes one position 
in an $n$-dimensional
array. Each item of a full index for $A$ is less than the corresponding item of
$\vec s_A$ (due to a zero index origin). There are precisely $t_A$ indices for 
an array. A partial index of $A$ is a vector of $0 \leq  k < n $
integers with each item less than the corresponding item of $\vec s_A$.

	We will use a tuple notation (omitting commas) to describe vectors of a fixed length.
For example,
\[                  <i \; j \; k> \]
denotes a vector of length three. $<>$ will denote the empty vector which is 
also sometimes written as $\Theta$.

	For every $n$-dimensional array $A$, there is a vector of the items of $A$,
which we denote by the corresponding lower case letter, here $\vec a$. The length of the vector of
items is $t_A$. A {\em vector} is itself a one-dimensional array, whose shape is the one-item
vector holding the length. Thus, for $\vec a$, the vector of items of $A$, the
shape of $\vec a$ is
\[            \vec s_{\vec a} =  \; < t_A> \]
and the number of items or total number of components\footnote{We also use $\tau \vec a , \delta \vec a,$ and $\rho \vec a$ to 
denote total number of components, dimensionality and shape of a. } 
of $\vec a$
is
\[                    t_{\vec a} =  t_A . \]

	The precise mapping of $A$ to $\vec a$ is determined by a one-to-one 
ordering function: $\gamma$ ({\em gamma}).  Although the choice of ordering is arbitrary, 
it is 
essential in the following that a specific one be assumed. By convention 
we assume the items of $A$ are placed in $\vec a$ according to the lexicographic 
ordering of the indices of $A$. This is often referred to 
as {\em row major ordering}.
Many programming languages lay out the items of multidimensional
arrays in memory in a contiguous segment using this ordering. 
Fortran uses the ordering corresponding to a transposed array in which the 
axes are reversed, that is, {\em column major}. Scalars are introduced as arrays with an {\em empty shape vector}.  This way of viewing scalars (i.e. as 
{\em empty arrays}) is crucial to the consistency of the theory and will be 
discussed more fully in a later section.

There are two equivalent ways of describing an array $A$:
\begin{description}
\item[(1)] by its shape and the vector of items, i.e.  $A = \{\vec s_A , \vec a\}$, or
\item[(2)] by its shape and a function that defines the value at every index 
$\vec p$.
\end{description}
These two forms have been shown to be formally equivalent~\cite{jenk94}.
We wish to use the second form in defining functions on multidimensional
arrays using their Cartesian coordinates (indices). 
The first form is used in describing
address manipulations to achieve effective computation.

To complete our notational conventions, we assume that 
$\vec p$, $\vec q$, $\dots $, will 
be used to denote
indices or partial indices and that $\vec u$, $\vec v$, $\dots$, will be 
used to denote 
arbitrary vectors of integers.
In order to describe the $i_{th}$ item of a vector $\vec a$, either 
$\vec a_i$ or 
$\vec a[i]$ will be used. If $\vec u$ is a
vector of $k$ integers all less than $t_A$, then $\vec a[\vec u]$
will denote the vector of length $k$, whose items are the items of $\vec a$ at 
positions $\vec u_j$, $j = 0,...,k-1.$

Before presenting the formal definition of the $\psi$ indexing function we 
define a few functions on vectors:
\begin{tabbing}

\hspace{2cm}\= $\vec u \;\cat\; \vec v$ \hspace{1cm} \=		\=catentation of vectors $\vec u$ and $\vec v$ \\
\>$\vec u$ + $\vec v$\>		\>itemwise vector addition assuming 
$t_{\vec u} = t_{\vec v}$ \\
\>$\vec u \times \vec v$ \>	\>	itemwise vector multiplication \\
\>$n$ + $\vec u$, $\vec u$ + $n$ \>	addition of a scalar to each item of a vector \\
\>$n \times \vec u$, $\vec u \times n$ \>   multiplication of each item of a vector by a scalar \\
\>$\iota \; n$ \>\>  the vector of the first n integers starting from $0$ \\
\>$\pi \; \vec v$ \> \>a scalar which is the product of the components of $\vec v $ \\
\>$k \;\take \;\vec u$ \>\>	when $k \geq 0$ the vector of the first $k$ items of $\vec u$, (called {\em take}) \\
\> \> \>and when $k<0$ the vector of the last $k$ items of $\vec u$ \\
\>$k \;\drop \;\vec u$ \>  \> when $k \geq 0$ the vector of $t_{\vec u} - k$ last items of $\vec u$, (called{ \em drop}) \\
\> \> \>and when $k<0$ the vector of the first $t_{\vec u} - |k|$ items of $\vec u$ \\
\>$k \;\theta \;\vec u$ \>  \>  when $k \geq 0$  the vector of $(k  \drop  \vec u) \cat ( k \take \vec u ) $\\
\> \> \>and when $k<0$ the vector or $(k \take \vec u ) \cat ( k \drop \vec u ) $
\end{tabbing}

\begin{definition}
\label{gammadef}
Let A be an n-dimensional array and \vec p a vector of integers.
If \vec p is an index of A, 
\[             \vec p \, \psi A = \vec a[\gamma(\vec s_A,\vec p)], \]
   where 
\begin{eqnarray}
      \gamma(\vec s_A,\vec p) & = & \vec x_{n-1}  \;\;\;\;\; \mbox{defined by 
the recurrence}    \nonumber \\ 
                 \vec x_0 & = & \vec p_0, \nonumber \\ 
              \vec x_j & = & \vec x_{j-1} * \vec s_j + \vec p_j, \; \; \; j=1,...,n-1. \nonumber
\end{eqnarray}

If $\vec p$ is partial index of length $k < n,$
\[           \vec p \, \psi A = B \]
where the shape of $B$ is
\[         \vec s_B = k \drop \vec s_A,    \] 
and for every index $\vec q$ of $B$,
\[         \vec q \, \psi B = (\vec p \cat \vec q) \, \psi A \]
\end{definition}
The definition uses the second form of specifying an array to define the result
of a partial index. For the index case, the function $\gamma(\vec s,\vec p)$ 
is used to convert an index $\vec p$ to an
integer giving the location of the corresponding item 
in the row major order list of items of an array of shape $\vec s$. The
recurrence computation for $\gamma$ is the one used in most compilers 
for converting an
index to a memory address~\cite{djr}.

\begin{corollary}
$<> \psi$ A = A.
\end{corollary}

The following theorem shows that a $\psi$ selection with a partial index can
be expressed as a composition of $\psi$ selections.

\begin{theorem}
Let A be an n-dimensional array and {\vec p} a partial index so 
that $\vec p = \vec q \cat \vec r$. Then
\[                 \vec p \, \psi A = \vec r \, \psi (\vec q \, \psi A). \]

Proof: The proof is a consequence of the fact that for 
vectors $\vec u$, $\vec v$, $\vec w$
\[           (\vec u \cat \vec v) \cat \vec w = \vec u \cat (\vec v \cat \vec w). \]
If we extend $\vec p$ to a full index by $\vec p \cat \vec p',$ then

\begin{eqnarray}
             \vec p' \, \psi (\vec p \, \psi A) & = & (\vec p \cat \vec p') \, \psi A \nonumber \\
                             & = & ((\vec q \cat \vec r) \cat \vec p') \, \psi A \nonumber \\
                             &  =  & (\vec q \cat (\vec r \cat \vec p')) \, \psi A \nonumber \\
                            &  =  & (\vec r \cat \vec p') \, \psi \, (\vec q \, \psi A) \nonumber \\
                            &   = & \vec p' \, \psi ( \vec r \, \psi (\vec q \, \psi A)) \nonumber \\
                          \vec p \, \psi A & = & \vec r \, \psi \, (\vec q \, \psi A) \nonumber                   
\end{eqnarray}

\end{theorem}
{\em which completes the proof.}

We can now use $\psi$ to define other operations on arrays. 
For example, consider definitions of {\em take} ($\take$) and {\em drop} 
($\drop$) for multidimensional arrays.
\begin{definition}[take: $\take$]
Let A be an n-dimensional array, and k a non-negative integer
such that $0 \leq k < s_0$. Then
\[            k \take A = B \]
where
\[          \vec  s_B = <k> \cat (1 \drop \vec s_A) \]
and for every index $\vec p$ of B,
\[            \vec p \, \psi B = \vec p \, \psi A. \]

\end{definition}
(In MoA $\take$ is also defined for negative integers and is generalized to any vector $\vec u$ with its
absolute value vector a partial index of A. The details are omitted here.)
\begin{definition}[reverse: $\Phi$]
 Let A be an n-dimensional array. Then
\[             \vec s_{(\Phi A)} = \vec s_A \]
and for every integer i, $0 \leq i < s_0, $
\[             <i> \psi \, (\Phi A) = <s_0 - i - 1> \psi A. \]
\end{definition}
This definition of $\Phi$ does a reversal of the $0th$ axis of A. 

Note
also that all operations are over the $0th$ axis. The operator $\Omega$~\cite{mul88} extends operations over all other dimensions.
\subsection{Example}
Consider the evaluation of the following expression using the
3 by 5 by 4 array, $A$, introduced in Section~\ref{indshp}.
\begin{eqnarray}
             <1\;\; 2> \, \psi \, (2 \take (\Phi A)) 
\end{eqnarray}
where A is the array given in the previous section.
The shape of the result is: 
\begin{eqnarray}
                & &  2 \drop \vec s_{(2 \take (\Phi A))} \nonumber \\
                & = & 2 \drop (<2> \cat (1 \drop \vec s_{(\Phi A)} )), \nonumber \\
               & = & 2 \drop (<2> \cat (1 \drop \vec s_A)), \nonumber \\
               & = & 2 \drop (<2> \cat <5\; 4>), \nonumber \\
               & = & 2 \drop <2\; 5 \;4>, \nonumber \\
              &  = &<4> . \nonumber
\end{eqnarray}
The expression can be simplified using the definitions:
\begin{eqnarray}
          &&   <1\; 2> \, \psi \, (2 \take (\Phi A)) \nonumber \\
           & = &  <1 \; 2> \, \psi \, (\Phi A), \nonumber \\
         &   = & <2> \, \psi \, (<1> \psi \, (\Phi A)), \nonumber \\
         &  = & <2> \psi \, (<3 - 1 - 1 > \psi A), \nonumber \\
        &   = & <1 \; 2> \psi A.  \nonumber \\
\end{eqnarray}
This process of simplifying the expression for the item in terms of
its Cartesian coordinates is called {\em Psi Reduction}. The operations of MoA
have been designed so that all expressions can be reduced to a minimal
normal form~\cite{mul88}.

Some MoA operations defined by $\psi$ are found in Fig.~\ref{moa}.

\break
\begin{figure}[h]
\scriptsize
\begin{tabular}{|l|l|l|}
\hline\hline
{\bf Symbol} & {\bf Name}& {\bf Description} \\ \hline
$\delta$ & Dimensionality & Returns the number of dimensions of an array.\\ \hline
$\rho$ & Shape & Returns a vector of the upper bounds \\
& &or sizes of each dimension in an array. \\ \hline
$\iota \xi^n$  & Iota & When $n=0$ (scalar),
returns a vector containing elements\\
& & $0$, to $\xi^0 - 1$. When $n = 1$ (vector), returns an \\
& & array of indices  defined by the shape vector $\xi^1$ \\ \hline
$\psi$ & Psi & The main indexing function of the Psi Calculus \\
& & which defines all operations in MoA.   Returns a scalar \\
& & if a full index is provided, a sub-array otherwise. \\ \hline
\mbox{rav} & Ravel & vectorizes a multi-dimensional array based   \\
&&on an  array's layout ($\gamma_{row},\gamma_{col},\gamma_{sparse}, ...$)\\ \hline
$\gamma$ & Gamma & Translates indices into offsets given a shape. \\ \hline
$\gamma^{'}$ & Gamma Inverse & Translates offsets into indices given a shape. \\ \hline
$\vec s \rshp \xi $ & Reshape& Changes the shape vector of an array,
 possibly affecting  \\
& &its dimensionality. Reshape depends on
layout ($\gamma$). \\ \hline
$\pi \vec x$ & Pi & Returns a scalar and is equivalent to $\prod_{i=0}^{(\tau x)-1} \vec x[i]$ \\ \hline
$\tau$& Tau & Returns the number of components in an array, ($\tau \xi \equiv
\pi (\rho \xi) $) \\ \hline
$\xi_l \cat \xi_r $ & Catenate& Concatenates two arrays over their
primary axis. \\ \hline
$\xi_l f \xi_r $ & Point-wise  & A data parallel application of $f$ is performed \\
&Extension &  between all elements of the arrays.\\ \hline
$\sigma f \xi_r $ & Scalar Extension & $\sigma$ is used with
every component of $\xi_r$ in the data parallel \\
$\xi_l f \sigma$ & & application of $f$. \\ \hline
$\take$&Take& Returns a sub-array from the beginning or end of an array\\
& & based on its argument being positive or negative. \\ \hline
$\drop$ & Drop & The inverse of Take \\ \hline
$_{op} \mbox{red}$&Reduce& Reduce an array's dimension by one by applying \\
& & op over the primary axis of an array. \\ \hline
$\Phi$& Reverse& Reverses the components of an array. \\ \hline
$\Theta $& Rotate & Rotates, or shifts cyclically, components of an
array. \\ \hline
$\transpose$ & Transpose & Transposes the elements of an array based on \\
&&a given permutation vector \\ \hline
$\Omega$& Omega& Applies a unary or binary function to array argument(s) \\
&&given partitioning information. $\Omega$ is used to perform all operations \\&& (defined over the primary axis only) over all dimensions. \\ \hline
\end{tabular}
\caption{\label{moa}Summary of MoA Operations}
\end{figure}

\subsection{Higher Order Operations}

Thus far operation on arrays, such as catenation, rotation, etc., have been 
performed over their $0th$ dimensions.  We introduce the higher order binary 
operation $\Omega$, which is defined when its left argument is a unary or binary
operation and its right argument is a vector describing the dimension upon 
which operations are to be performed, or which sub-arrays are used in 
operations.  The dimension upon which operations are to be performed is often 
called the {\it axis} of operation.  The result of $\Omega$ is a unary or 
binary operation.

\subsection{Definition of $\Omega$}
\label{omegadef}

$\Omega$ is defined whenever its left argument is a unary or binary operation,
$f$ or $g$ respectively ($f$ and $g$ include the outcome of higher order
operation).  $\Omega$'s right argument is a vector, $\vec d$, such that 
$\rho {\vec d} \equiv <1>$ or $\rho {\vec d} \equiv <2>$ depending on whether 
the operation is unary or binary. Commonly, $f$ (or $g$) will be an operation
which determines the shape of its result based on the shapes of its arguments, 
not on the values of their entries, i.e. for all appropriate arguments 
$\rho (f \xi)$ is determined by $\rho \xi$ and $\rho (\xi_l g \xi_r)$ is 
determined
by $\rho \xi_l$ and $\rho \xi_r$.

\begin{definition}
${}_f\Omega_{\vec d}$ is defined when $f$ is a one 
argument function, ${\vec d} \equiv <\sigma>$, with $\sigma \ge 0$. 

For any non-empty array $\xi$, 
\begin{equation}
{}_f\Omega_{\vec d} \xi 
\end{equation}
is defined provided
({\it i}) $(\delta \xi) \ge \sigma$, and provided certain other conditions,
stated below, are met.  Let
\begin{equation}
{\vec u} \equiv (-\sigma) \drop \rho \xi.
\end{equation}
We can write
\begin{equation}
\rho \xi \equiv {\vec u} \cat {\vec z}
\end{equation}
where ${\vec z} \equiv (-\sigma) \take \rho \xi$.

We further require ({\it ii}) there exists ${\vec w}$ such that for 
$0\le^\star{\vec i} <^\star {\vec u}$,
\begin{equation}
f({\vec i}\, \psi \xi)
\end{equation}
is defined and has shape $\vec w$.  The notation $0\le^\star{\vec i} 
<^\star {\vec u}$, is a shorthand which implies that we are comparing
two vectors $\vec i$ and $\vec u$ component by component.
With this
\begin{equation}
\rho ({}_f\Omega_{\vec d})\xi \equiv {\vec u}\cat {\vec w}
\end{equation}
and for $0\le^\star {\vec i} <^\star {\vec u}$,
\begin{equation}
{\vec i}\, \psi ({}_f\Omega_{\vec d}\xi) \equiv f({\vec i}\, \psi \xi)
\end{equation}
\end{definition}

Note that condition ({\it ii}) is easily satisfied for common $f$'s.

\begin{definition}
We similarly define $\Omega$ when its function argument is a
binary operation $g$. ${}_g\Omega_{\vec d}$ is defined when $g$ is a two
argument function, ${\vec d} \equiv <\sigma_l \;\sigma_r>$, with $\sigma_l 
\ge 0$, and $\sigma_r \ge 0$.

For any non-empty arrays, $\xi_l$, and $\xi_r$,
\begin{equation}
\xi_l({}_g\Omega_{\vec d})\xi_r
\end{equation}
is defined provided ({\it i}) $(\delta \xi_l) \ge \sigma_l$ and 
$(\delta \xi_r) \ge \sigma_r$, and provided certain other conditions, stated
below, are met.

We let $\lfloor$ denote the binary operation minimum and let
\begin{equation}
m\equiv ((\delta\xi_l)-\sigma_l) \lfloor ((\delta \xi_r)-\sigma_r).
\end{equation}
We require that ({\it ii}) $((-m) \take (-\sigma_l) \drop \rho \xi_l) \equiv
((-m) \take (-\sigma_r) \drop \rho \xi_r)$.

Let
\begin{equation}
{\vec x} \equiv ((-m)\take (-\sigma_l)\drop \rho \xi_l) \equiv ((-m) \take
(-\sigma_r) \drop \rho \xi_r),
\end{equation}
\begin{equation}
{\vec u} \equiv (-m)\drop (-\sigma_l) \drop \rho \xi_l,
\end{equation}
\begin{equation}
{\vec v} \equiv (-m)\drop (-\sigma_r) \drop \rho \xi_r.
\end{equation}

Note that ${\vec u} \equiv <>$ or ${\vec v} \equiv <>$ (both could be empty).
We can now write
\begin{equation}
\rho \xi_l \equiv {\vec u} \cat {\vec x} \cat {\vec y},
\end{equation}
and,
\begin{equation}
\rho \xi_r \equiv {\vec v} \cat {\vec x} \cat {\vec z}
\end{equation}
where ${\vec y} \equiv (-\sigma_l) \take \rho \xi_l$ and
${\vec z} \equiv (-\sigma_r)\take \rho \xi_r$. Any of the vectors above
could be empty.

We also require ({\it iii}) there exists a fixed vector $\vec w$ such that for
$0\le^\star {\vec i} <^\star {\vec u}$, $0\le^\star {\vec j} <^\star {\vec v}$,
$0\le^\star {\vec k} <^\star {\vec x}$,
\begin{equation}
(({\vec i} \cat {\vec k}) \, \psi \xi_l) \, g \, (({\vec j} \cat {\vec k})\, \psi \xi_r)
\end{equation}
is defined and has shape $\vec w$.

With all this

\begin{equation}
\rho (\xi_l ({}_g\Omega_{\vec d}) \xi_r) \equiv {\vec u}\cat {\vec v} \cat 
{\vec x} \cat {\vec w}
\end{equation}

and for $0\le^\star {\vec i} <^\star {\vec u}$, 
$0\le^\star {\vec j} <^\star {\vec v}$, $0\le^\star {\vec k} <^\star {\vec x}$,
\begin{equation}
({\vec i}\cat {\vec j} \cat {\vec k}) \, \psi \, (\xi_l 
({}_g\Omega_{\vec d}) \xi_r) \equiv (({\vec i}\cat{\vec k})\, \psi \xi_l) \, g \,
(({\vec j}\cat {\vec k})\, \psi \xi_r)
\end{equation}

\end{definition}

Since at least one of $\vec u$, $\vec v$ is empty, the corresponding one of 
$\vec i$, $\vec j$ must also be empty.  We note the condition ({\it iii}) is
easily satisfied for common $g$'s.

Consider the following example.  The operator $\theta$ is defined for scalar 
and vector left arguments and n-dimensional array right arguments.  Thus for
$\theta$, valid $\vec d$'s are $<0\, n>$ and $<1\, n>$.  
Let 
\begin{equation}
\xi^2 \equiv
\left [
\begin{array} {cccc}
1 & 2 & 3 & 4 \\
5 & 6 & 7 & 8 \\
\end{array}
\right ].
\end{equation}
then
\begin{equation}
<2 1> {}_\theta \Omega_{<0 1>} \xi^2 \equiv
\left [
\begin{array} {cccc}
3 & 4 & 1 & 2 \\
6 & 7 & 8 & 5 \\
\end{array}
\right ].
\end{equation}

\section{Contrasting MoA with Linear Algebra}

As stated previously, one can think of MoA as a generalization and 
extension of standard Linear Algebra.  In this section we draw the 
readers attention to a few important differences between MoA and Linear Algebra.

\subsection{Scalars as Arrays}

{\bf \em In MoA every object is an array including a scalar}.  Scalars are 
considered to be zero-dimensional arrays.  Often we use the greek letter 
$\sigma$ to denote a scalar. The shape of a scalar is the {\em empty vector} 
$<\;>$ as:
\begin{equation}
\rho \sigma = <\;>.
\end{equation}
In general there is an infinite collection of empty arrays.  Any 
multi-dimensional array with one or more empty dimensions (i.e. the shape
vector contains at least one zero element) is called an {\em empty array}.  More
formally, we say that an empty array is one for which the product of the 
elements of the shape vector is zero.  That is:
\begin{equation}
\pi (\rho \xi) = 0.
\end{equation}
Thus for scalars we have:
\begin{equation}
\rho (\rho \sigma) =<\!0\!>, 
\end{equation}
that is, a scalar is a zero-dimensional array.  In general the shape of the 
shape gives the number of dimensions.

The notion of a scalar as an array will undoubtedly seem strange to most 
readers and may seem to be an arcane construct, however, the  distinction is
essential for consistency of the theory just as the number $0$ is essential
to the system of integers under the operation of addition or the number $1$
is under the operation of multiplication.  It is the analog of the empty set
in set theory and the identity operation in group theory. 
Note: there is a difference between a zero dimensional array (a scalar) and a
one-dimensional array with one element!  In the first case the shape is
empty and in the second the shape is a one-element vector containing the 
single element $1$.  

To illustrate this concept consider the following arrays: $\sigma$, 
$<\!\sigma \!>$ and $[\sigma]$. The first $\sigma$, is a scalar or 
zero-dimensional array, the second $<\!\sigma \!>$ is a one element vector,
and the third $[\sigma]$ is a $1\times 1$ array (square brackets denote
two-dimensional arrays as is common in Linear Algebra).
In traditional Linear Algebra, there is no
distinction between these three examples.  In MoA however, the three arrays 
are distinguished
by their shapes:
\begin{equation}
\rho \sigma = <\;>,
\end{equation}
\begin{equation}
\rho \! <\!\sigma\!> = <\!1\!>,
\end{equation}
and,
\begin{equation}
\rho  [\;\sigma\;] = <\!1\;1\!>.
\end{equation}

Thus $\sigma \ne \, <\!\sigma\!> \, \ne [\sigma ]$, because the corresponding 
dimensionalities, respectively given by: 
\begin{equation}
\rho (\rho \sigma) = <\!0\!>,
\end{equation}

\begin{equation}
\rho (\rho \! <\!\sigma\!>) = <\!1\!>,
\end{equation}
and,
\begin{equation}
\rho (\rho  [\;\sigma\;]) = <\!2\!>,
\end{equation}
are not equal.

\subsection{Need for Empty Arrays}
In the previous section we introduced the notion of scalars as arrays and of
empty arrays. These subtle distinctions are essential in that our theory is
based on shapes.  In general, the dimensionality of an array changes as 
operators act on them.  As a simple example, think of the operator corresponding
to the standard {\em inner product} ({\em dot product}).  This operator takes
two  vectors and produces a scalar.  The operator corresponding to the 
standard {\em outer product} ({\em direct product} or {\em cartesian product}) 
takes two vectors and produces a matrix.
Another example exists in the concept of a {\it functional}.  A functional
takes a {\it function} (which can be thought of as a vector in an infinite
dimensional space) and returns a {\it scalar}.

In MoA this concept is completely general.  One can imagine a sequence of 
operations that convert an $n$-dimensional array into a $m$ dimensional 
array (for $m$ {\em smaller or larger than $n$}).  If such a sequence of
operations acts to reduce the dimensionality of the result with each step, the
natural stopping point (i.e. the boundary condition) is the {\em scalar} (i.e. a zero-dimensional array).

\subsection{Graphical Representation}

Any multi-dimensional array can be represented graphically using the 
vector angle brackets ($<$ and $>$) and the square brackets ($[$ and $]$).
In section~\ref{indshp} we represented a three-dimensional array as a series of 
two-dimensional arrays next on one another.  We often find it convenient to
represent arrays by nesting two-dimensional arrays.  We illustrate this for
the hyper-cube, below. An $n$-dimensional {\em hyper-cube} is an 
$n$-dimensional array in which the length of each dimension is $2$.
For a two-by-two array we write:
\begin{equation}
\xi^{(22)} = \left [
\begin{array}{cc}
0 & 1\\
2 & 3\\
\end{array}
\right ].
\end{equation}
This is an example of a two-dimensional hyper-cube,
and a four-dimensional hyper-cube would be written as:
\begin{equation}
\xi^{(2222)} = \left [
\begin{array}{cc}
   \left [
       \begin{array}{ccc}
        &  0 \;  & 1\; \\
        &  2 \;  & 3\; \\
       \end{array}
   \right ]
& 
   \left [
      \begin{array}{ccc}
         & \; 4 \; & 5\;  \\
         & \; 6 \; & 7\;  \\
      \end{array}
   \right ]
\\
   \left [
      \begin{array}{ccc}
        &   8  &   9\\
        & 10 & 11\\
      \end{array}
   \right ]
& 
   \left [
      \begin{array}{cc}
          12  & 13\\
          14  & 15\\
      \end{array}
   \right ]
\\
\end{array}
\right ].
\end{equation}

\subsection{ Notational Subtleties}

It is essential to be always aware of the shape of the array in order to avoid 
notational confusion.  For example, the array 
\begin{equation}
{\vec v} = <\!a\;b\!>,
\end{equation}
is a one-dimensional array (i.e. a vector) with two elements, while the array
\begin{equation}
\xi = [\; a\;b\;],
\end{equation}
is a two-dimensional array (i.e. it is a $1\times 2$ array) with two elements.
The difference is determined by their shapes.  Explicitly we have:
\begin{equation}
\rho {\vec v} = <\!2\!>,
\end{equation}
and,
\begin{equation}
\rho \xi = <\!1\;2\!>.
\end{equation}
As discussed in previous sections,, we use an index vector and the $\psi$ operator
in order to select elements of the arrays.  Thus:
\begin{equation}
<\!0\!> \psi {\vec v} = a,
\end{equation}
and
\begin{equation}
<\!1\!> \psi {\vec v} = b,
\end{equation}
for the one-dimensional representation, and for the two-dimensional 
representation we have:
\begin{equation}
<\!0\;0\!> \psi \xi = a,
\end{equation}
and,
\begin{equation}
<\!0\;1\!> \psi \xi = b.
\end{equation}

Consider now the important difference between MoA and standard Linear Algebra.
The concept of a {\em row vector} exists in MoA as in standard Linear Algebra:
\begin{equation}
\left [
   \begin{array}{cc}
      a  & b\\
      c  & d\\
   \end{array}
\right ]
\equiv
\left [
   \begin{array}{c}
      <\!a\;b\!>\\
      <\!c\;d\!>\\
   \end{array}
\right ].
\label{row-vec1}
\end{equation}
In contrast to Linear Algebra, however, there is no concept of a 
{\em column vector}.  To access the elements of what would normally
be called a column vector we use the higher-order $\Omega$ operator (see the
appendices for the definition of this operator).

Note also the following inequality, 
\begin{equation}
\left [
   \begin{array}{c}
      <\!a\;b\!>\\
      <\!c\;d\!>\\
   \end{array}
\right ]
\ne
\left [
   \begin{array}{c}
      \left [ \;a\;b\; \right ] \\
      \left [ \;c\;d\; \right ] \\
   \end{array}
\right ].
\label{row-vec2}
\end{equation}

\subsection{ Addition and Multiplication of Arrays: Comparing and Contrasting
with Linear Algebra}

The following operation on two arrays of shape $<\!2\; 2\!>$ (i.e. $2\times 2$
matrices) is identical in MoA and standard Linear Algebra:

\begin{equation}
   \left [
       \begin{array}{ccc}
        &  a \;  & b\; \\
        &  c \;  & d\; \\
       \end{array}
   \right ]
+
   \left [
       \begin{array}{ccc}
        &  e \;  & f\; \\
        &  g \;  & h\; \\
       \end{array}
   \right ]
=
   \left [
       \begin{array}{ccc}
        &  (a + e) \;  & (b + f)\; \\
        &  (c + g) \;  & (d + h)\; \\
       \end{array}
   \right ].
\end{equation}
Subtraction of two arrays is defined in a similar way.  In both cases we find
elements of the two arrays are combined in a point-wise fashion.

With matrix multiplication, however, we find an important distinction.  In 
MoA, multiplication, like addition and subtraction, occurs also in point-wise
fashion:

\begin{equation}
   \left [
       \begin{array}{ccc}
        &  a \;  & b\; \\
        &  c \;  & d\; \\
       \end{array}
   \right ]
\times
   \left [
       \begin{array}{ccc}
        &  e \;  & f\; \\
        &  g \;  & h\; \\
       \end{array}
   \right ]
=
   \left [
       \begin{array}{ccc}
        &  (a \times e) \;  & (b \times f)\; \\
        &  (c \times g) \;  & (d \times h)\; \\
       \end{array}
   \right ].
\end{equation}
Similar definitions exist for all scalar operations (e.g. $+$, $-$, $\times$,
$/$).

The operation corresponding to standard matrix multiplication:
\begin{equation}
   \left [
       \begin{array}{ccc}
        &  a \;  & b\; \\
        &  c \;  & d\; \\
       \end{array}
   \right ]
\;
   \left [
       \begin{array}{ccc}
        &  e \;  & f\; \\
        &  g \;  & h\; \\
       \end{array}
   \right ]
=
   \left [
       \begin{array}{ccc}
        &  (a \times e + b \times g) \;  & (a \times f + b \times h)\; \\
        &  (c \times e + d \times g) \;  & (c \times f + d \times h)\; \\
       \end{array}
   \right ],
\label{standard}
\end{equation}
in MoA corresponds to the following sequence of operations.  First we form
the following two matrices:
\begin{equation}
A = 
   \left [
       \begin{array}{ccc}
        &  (a \times e) \;  & (a \times f)\; \\
        &  (c \times e) \;  & (c \times f)\; \\
       \end{array}
   \right ],
\end{equation} 
and,
\begin{equation}
B = 
   \left [
       \begin{array}{ccc}
        &  (b \times g) \;  & (b \times h)\; \\
        &  (d \times g) \;  & (d \times h)\; \\
       \end{array}
   \right ].
\end{equation}
The matrices $A$ and $B$ are then added to produce the result of Eq.~\ref{standard}.

The matrices $A$ and $B$ can be seen to be constructed as {\em outer products}
of the vectors $<\!a\;c\!>$ and $<\!e\;f\!>$ to form $A$ and 
$<\!b\;d\!>$ and $<\!g\;h\!>$ to form $B$.  Thus by considering the notions
of {\em matrix addition} and {\em matrix multiplication} in standard linear 
algebra we are naturally led to the MoA operations: (1) {\em point-wise
extension of scalar operation} and (2) {\em outer product}.  These
constructs are made precise in the following two definitions.

\begin{definition}[Point-wise extension of scalar operations]
Point-wise extension of a binary operation ``op" between two 
non-empty arrays $\xi_1$ and $\xi_2$, such that $\rho \xi_1 = \rho \xi_2$,
has shape:
\begin{equation}
\rho (\xi_1 op \, \xi_2) = \rho \xi_1,
\end{equation}
and for valid indices $0 \le^\star \vec i <^\star \rho \xi_1$, is given by:
\begin{equation}
\vec i \, \psi \, (\xi_1 op \, \xi_2) \equiv (\vec i \, \psi \, \xi_1) \, op \, (\vec i \, \psi \, \xi_2).
\end{equation}
Examples of valid binary operations, $op$, include include $+$, $-$, $\times$,
$/$, etc.
\label{scalar_extend}
\end{definition}
\begin{definition}[Outer product]
The outer product, $\bullet_{op}$ of two arrays $\xi_l$ and $\xi_r$ has 
shape:
\begin{equation}
\rho \, (\xi_l \bullet_{op} \xi_r) \equiv (\rho \xi_l) \cat (\rho \xi_r)
\end{equation}
and for valid indices $0 \le^\star \vec i <^\star \rho \xi_l$, and 
$0 \le^\star \vec j <^\star \rho \xi_r$, is given by:
\begin{equation}
(\vec i \cat \vec j) \, \psi \, (\xi_l \bullet_{op} \xi_r) \equiv (\vec i \, \psi \xi_l) \, op \, (\vec j \, \psi \xi_r).
\end{equation}
\label{outer}
\end{definition}

Thus we see that {\em matrix multiplication} of standard Linear Algebra is a 
special case of Def.~\ref{scalar_extend} with $op = +$, and the standard
{\em tensor product} is a special case of Def.~\ref{outer} with $op = \times$.


%
%
%

\newcommand{\directsum}{\bigcirc\!\!\!\!\!\!\!+\;}
\newcommand{\directsuma}{\bigcirc\!\!\!\!\!+\;}

\chapter{A Cache-Optimized Fast Fourier Transform: Part I}
\label{part1}

\section{Chapter Summary}

The material in this chapter is taken from a (larger) paper that is to 
appear in the Journal of Computational Physics.

Our subject in this and the following two chapters is the design and 
a Fast Fourier Transform algorithm designed to optimize data
locality in the cache.  The algorithm is presented and discussed using
traditional concepts familiar to scientists and engineers.  In this chapter
new concepts based on Conformal Computing
techniques are introduced gradually and illustrated in context.  The following
chapter, serves as a stand-alone tutorial on
Conformal Computing techniques that are developed and illustrated in the
context of the new FFT algorithm.  We find favorable performance of the new
algorithm without any machine-specific optimizations. In particular we find
the new routine to be a factor of $2$ to $4$ times faster than our previous
design that often outperformed well-tested library routines such as ESSL, IMSL,
FFTW, or NAG (see Chap.~\ref{intro} and references therein).

The results presented in these chapters are promising for further developments
in terms of optimizations over processor/memory hierarchies because the
algorithm can be generalized to arbitrary partitioning over any number of
levels of the processor/memory hierarchy.
More importantly, this research illustrates the power of a uniform, mechanical,
mathematically based design strategy that leads to
portable, scalable, and {\em verifiable} software {\em or} hardware.

\section{Introduction}

Our new Fast Fourier Transform algorithm represents a significant application 
of a mathematically rigorous, systematic, design protocol that the authors have
named Conformal Computing.  The vision of Conformal Computing is to 
algebraically connect the hardware and software through linear 
transformations from high-level specifications of the problem to the 
low-level instruction sets of the underlying hardware.  In the early days of 
computing, in which programs were written directly in assembler, this vision 
was more easily realized on single-processor, single-memory systems. Today, 
however, the situation is considerably more complex in that there are many 
levels of software, and processor-memory separating the high-level problem 
specification and the hardware.

The Fast Fourier Transform (FFT) is one of the most important computational
algorithms and its use is pervasive in science and engineering.
The research presented in this chapter represents a significant improvement
to the FFT algorithm resulting in a factor of four speed-up for some of the 
largest systems tested in comparison with our previous 
records. Our previously published work indicates that our
FFT is competitive with or outperforms standard library routines without
any machine-specific optimizations (see Chap.~\ref{intro} 
and Ref.~\cite{mullin.small:}). 
This success is achieved through optimizing in-cache operations.  In the 
traditional FFT, data access becomes progressively 
remote (leading to cache misses and page faults) as the algorithm proceeds.  
In our new approach data is periodically rearranged so as to maximize data 
locality.

Our algorithm can be seen to be a generalization of similar work aimed at
out-of-core optimizations~\cite{cormen}. Similarly, block decompositions
of matrices (in general) are special cases of our {\em reshape-transpose} design.
Most importantly, our designs are general for any partition size, 
i.e. not necessary blocked in squares, and any number of dimensions. 
Furthermore, our designs use linear transformations from an algebraic 
specification and thus they are {\bf verified}. Thus, by specifying
designs (such as Cormen's and others) using Conformal Computing techniques,
these designs too could be verified.

A general algebraic framework for Fourier and related transforms, including 
their discrete versions, is discussed in~\cite{ElliotRao}.  As discussed in 
\cite{tolimieri2,tolimieri1} and using this framework, many algorithms for 
the FFT can be viewed in terms of computing tensor product decompositions
of the matrix $B_L$, discussed below (see Fig.~\ref{vanloan_fft}).
Subsequently, a number of additional algorithms for the FFT and related 
problems have been developed centered around the use of tensor product 
decompositions~\cite{DaiGKL94,GranataCT92,GuptaHSJ92,GuptaHSJ96,JohnsonJPX,XiongJJP01}.  The work done under the acronym FFTW is based on a compiler that 
generates efficient sequential FFT code that is adapted to a target 
architecture and specified problem size~\cite{Frigo97,FrigoJo98,Frigo99,GatlinCa00,MirkovicMahJoh00}.  A variety of techniques have been used to construct 
efficient parallel algorithms for the FFT~\cite{AgarwalGZ94,CullerEtAL93,GuptaK93,GuptaHSJ94,Miles93,ThuTKG00}. Other important FFT implementations are 
discussed in~\cite{goedecker:} and~\cite{goedecker3d:}.

The purpose of this paper {\bf IS NOT} to attempt any serious analysis of the 
number of cache misses incurred by the algorithm in the spirit of 
of Hong and Kung and others~\cite{hongkung,savage,vitter93algorithms}.
Rather, we  present an {\bf algebraic} method
that achieves (or is competitive) with such  optimizations 
{\bf mechanically}. Through linear transformations we
produce a  {\bf normal form}, the ONF, that is directly implementable in any
hardware or software language and is realized in any of the processor/memory 
levels~\cite{fftharry}. 
Most importantly, our designs are completely general in that through {\bf 
dimension lifting} we can produce any number of levels in the processor/memory
hierarchy. 

One objection to our approach is that one might incur an unacceptable 
performance cost due to the periodic rearrangement of the data.  This will
not, however, be the case if we  pre-fetch data before
it is needed.  The necessity to pre-fetch data also exists in other
similar cache-optimized schemes.  Our algorithm does what the 
compiler community calls {\em tiling}. Since we have analyzed the loop
structures,  access patterns, and speeds of the processor/memory levels,
pre-fetching becomes a deterministic cost function that can easily
be combined with  {\em reshape-transpose} or {\em tiling} operations.

Again we make no attempt to optimize the algorithm for any particular 
architecture.  We provide a general algorithm in the form of an Operational
Normal Form that allows the user to specify the blocking size at run time.  
This ONF therefore enables the individual user to choose the blocking
size that gives the best performance for any individual machine.

We now begin our discussion of the new algorithm with a discussion of previous
efforts applied to index optimizations of the traditional FFT.

\section{Index Optimizations for the Traditional FFT}

\subsection{\label{traditional}Traditional FFT Algorithm}

We begin by reviewing the traditional FFT and its recent refinements
using the $\psi$-calculus.  Some of the following discussion is excerpted 
from Ref.~\cite{mullin.small:}.

We began with Van Loan's \cite{vanloan}
high--level MATLAB\footnote{MATLAB is commonly used in the
scientific community as a high-level prototyping language} program
for the radix 2 FFT, shown in Fig.~\ref{vanloan_fft}.  This program denotes 
a single loop program, with high level array/vector operations
and reshaping.

\renewcommand{\baselinestretch}{1}
\begin{figure}[hbt]
\scriptsize
\begin{quotation}
\noindent
{\bf Input:} $\; x$ in $C^n$ and  $n=2^t$,
where $ t \geq 0 $ is an integer. \\
\noindent
{\bf Output:} The FFT of $x$.

\medskip
\noindent
\hspace*{2em} $x \leftarrow P_n \; x$ \hfill (1) \\
\hspace*{2em} for $q = 1 \mbox{ to } t$ \hfill (2) \\
\hspace*{3em} begin \hfill (3) \\
\hspace*{4em} $L \leftarrow 2^q$ \hfill (4) \\
\hspace*{4em} $r \leftarrow n / L$ \hfill (5) \\
\hspace*{4em} $x_{L \times r} \leftarrow B_L \;  x_{L \times r}$ \hfill (6) \\
\hspace*{3em} end \hfill (7) \\
Here, $P_n$  is a $n \times n$ permutation matrix,
$B_L =  \left [
\begin{array}{cc}
I_{L_*} & \Omega_{L_*} \\
I_{L_*} & - \Omega_{L_*}
\end{array}
\right ],
L_* = L/2$,
and $\Omega_{L_*}$ is a diagonal matrix with values
$1, \omega_L, \ldots, \omega_L^{L_*-1}$ along the diagonal,
where $\omega_L$ is the $L$'th root of unity.
\end{quotation}
\caption{\label{vanloan_fft} High-level program for the radix $2$ FFT.}
\end{figure}
In Line 1 of Fig.~\ref{vanloan_fft}, $P_n$ is a permutation matrix that performs
the bit--reversal permutation on the $n$ elements of vector $x$.
In Line 6, the $n$ element array $x$ is regarded as being
{\em reshaped} to be a $L \times r$ matrix
consisting of $r$ columns,
each of which is a vector of $L$ elements.
Line 6 can be viewed as treating each column of this matrix as a pair
of vectors, each with $L/2$ elements,
and doing a {\em butterfly} computation that combines the two vectors
in each pair to produce a vector with $L$ elements.

The reshaping of the data matrix $x$ in Line 6 is column--wise,
so that each time Line 6 is executed, each pair of adjacent columns of the
preceding matrix are concatenated to produce a column of the
new matrix.
The butterfly computation, corresponding to multiplication of the data matrix
$x$ by the weight matrix $B_L$, combines each pair of $L/2$ element
column vectors from the old matrix
into a new $L$ element vector of values for each new column.

Let us now interpret the algorithm of Fig.~\ref{vanloan_fft} using traditional
concepts of matrix algebra. The algorithm of Fig.~\ref{vanloan_fft}
is equivalent to an iterative sequence of operations in which an 
$n\times n$ matrix is multiplied by an $n$-dimensional data vector. 
The number of such operations is given by $t = \log_2(n)$ and the matrix 
is different for each iteration as is discussed below. 

The first vector $x^{(1)}$ is given by 
\begin{equation}
x^{(1)} = M^{(1)} x^{(0)},
\label{step1}
\end{equation} 
where $x^{(0)}$ is the initial bit-reversed data vector $x^{(0)} = P_n x$.
The first-iterate matrix $M^{(1)}$ is an $n\times n$ block-diagonal matrix
consisting of $n/2$, $2\times 2$ matrices $B_2$ where the matrix $B_L$ is 
defined in Fig.~\ref{vanloan_fft}.  
The next step of the algorithm produces the second iterate vector $x^{(2)}$
\begin{equation}
x^{(2)} = M^{(2)} x^{(1)},
\label{step2}
\end{equation}
from the first-iterate vector $x^{(1)}$ through multiplication by the 
second-iterate matrix $M^{(2)}$.  The matrix $M^{(2)}$ is a $n\times n$
block-diagonal matrix consisting of $n/4$, $4\times 4$ matrices $B_4$ along
the diagonal.  This process continues in a straightforward fashion.
The final step of the process is given by 
\begin{equation}
x^{(t)} = M^{(t)} x^{(t-1)},
\end{equation}
where the $t$-th iterate vector $x^{(t)}$ is the Fourier Transform of the 
original data vector:
\begin{equation}
x^{(t)} = FFT(x),
\end{equation}
and the matrix $M^{(t)} = B_n$.

\subsection{Index Optimization Using the $\psi$-Calculus}
The discussion of the previous subsection is useful for the purposes of 
illustrating the basics of the FFT but is inefficient.  The essence of the 
developments of Ref.~\cite{mullin.small:} is the removal of all temporary 
arrays through the use of Conformal Computing techniques.  The concise 
algorithm illustrated in Fig.~\ref{vanloan_fft} is the starting point for
the developments of Ref.~\cite{mullin.small:}.  Essential to this analysis is
the notion of partitioning and {\it reshaping} the data to maximize efficiency
by taking advantage of the sparseness of the matrices $M^{(q)}$ 
(for $1 \le q \le t$).

The final result for the radix-$2$ FFT is presented in Fig.~\ref{rad2_fft}
(reproduced from Fig.~\ref{fftpsirad2} and Ref.~\cite{mullin.small:}).  As was
demonstrated in Ref.~\cite{mullin.small:} and reproduced in 
Chap.~\ref{intro}, this implementation is competitive
with or outperforms a variety of standard library routines.  Such high 
performance is a consequence of the fact that, through the optimal use of
matrix/vector indexing, no temporary arrays are used. 

\renewcommand{\baselinestretch}{1}
\begin{figure}[ht]
\footnotesize
\begin{quotation}
\begin{tt}
\noindent
\hspace*{1em}  do q = 1,t \\
\hspace*{2em}    L = 2**q  \\
\hspace*{2em}    do row = 0,L/2-1 \\
\hspace*{3em}      weight(row) = EXP((2*pi*i*row)/L) \\
\hspace*{2em}    end do  \\
\hspace*{2em}    do col$'$ = 0,n-1,L \\
\hspace*{3em}       do row = 0,L/2-1 \\
\hspace*{4em}   c = weight(row)*x(col$'$+row+L/2)  \\
\hspace*{4em}   d = x(col$'$+row) \\
\hspace*{4em}   x(col$'$+row) = d + c  \\
\hspace*{4em}   x(col$'$+row+L/2) = d - c  \\
\hspace*{3em}       end do \\
\hspace*{2em}    end do  \\
\hspace*{1em}    end do
\end{tt}
\end{quotation}
\caption{\label{rad2_fft}Final index-optimized radix-$2$ FFT with 
in-place butterfly computation.  This implementation eliminates the need
for temporary arrays through the optimized use of array indexing.}
\end{figure}

\subsection{Optimizing Array Access Patterns}
The key result of the present paper is a generalization of the algorithm of
Fig.~\ref{rad2_fft} in which performance is increased (factor of two to four
speedup for moderate to large FFT's) through repeated restructuring of the
data so as to minimize cache misses and page faults.

Note the array access patterns implied by the code fragment of 
Fig.~\ref{rad2_fft}. In particular, as the outer loop variable increases
(i.e. $q = 1$,~$2$,~$\cdots$) the stride of the data access (i.e. the difference
between the index of {\tt x(col$'$+row)} and that of {\tt x(col$'$+row+L/2)}) 
is doubling with each increment of $q$. For sufficiently small values of 
$q$, both {\tt x(col$'$+row)} and {\tt x(col$'$+row+L/2)} reside in cache
and access is fast.  However, as $q$ increases, at some point $L/2$ is larger
than the cache size (first L1 cache and subsequently L2 cache, etc) and 
accessing both {\tt x(col$'$+row)} and {\tt x(col$'$+row+L/2)} results in
a cache miss.  For large FFT's (i.e. those that don't fit into cache, main
memory, paged memory, etc.) the performance continues to deteriorate with 
increasing $q$ as the increasing separation of {\tt x(col$'$+row)} and 
{\tt x(col$'$+row+L/2)} requires accessing higher levels of the memory 
hierarchy (main memory, paged memory, etc) leading to page faults etc.
This problem, resulting from data non-locality is common to all previous 
implementations of the FFT.

The essence of the new cache-optimized algorithm, set forth in this paper, is 
the notion that periodic restructuring of the data array $x$ (i.e. actually 
moving the data around to achieve locality) is less costly than the penalty 
which is otherwise incurred as a result of cache misses and page faults.

\section{Cache-Optimized FFT: Key Elements of the New Approach}
\subsection{ Restructuring the Data: the Reshape-Transpose Operation}

The key data-restructuring operation used in the new algorithm is called
the {\it reshape-transpose} operation.  In order to understand this operation 
we must
first consider the data vector $\vec x$ as a one-dimensional array of 
length $n$.
Next we consider {\it reshaping} the array into a collection of 
vectors of length $c$ so as to form a two-dimensional array of dimension
$r\times c$ where the number of rows is given by $r = n/c$.  Using the 
$\psi$-calculus notation we write $x^{(a)} = <r\;c> {\hat \rho} \; {\vec x}$.
The length of each
row $c$ is chosen arbitrarily, and is specified as a parameter to the algorithm.
In practice, however, we find that optimal performance is generally obtained
for $c$ less than or equal to the cache size. 

At this point an example would be helpful.  Consider, for simplicity, the
initial data vector to be a vector of length $n=32$ consisting of sequential
integers starting from zero.  In $\psi$-calculus notation we write 
${\vec x} = \iota (32)$. Next we choose $c = 4$ and {\it reshape} the one-dimensional
array $\vec x$ to be an $8\times 4$ array $x^{(a)}$ by filling in the entries
in lexical order (i.e. in {\it row-major} order as is done for arrays in C++).
Using the $\psi$-calculus notation: $x^{(a)} = <8\; 4> {\hat \rho}\; 
\iota (32)$. Explicitly we have:

\begin{equation}
x^{(a)} = <8\;4> {\hat \rho} (\iota (32)) \equiv
\left [
\begin{array}{cccc}
0&1&2&3 \\
4&5&6&7 \\
8&9&10&11\\
12&13&14&15\\
16&17&18&19\\
20&21&22&23\\
24&25&26&27\\
28&29&30&31\\
\end{array}
\right ]
\label{reshape1}
\end{equation}

Now consider the FFT access patterns of the array $x^{(a)}$ in light of 
the algorithm of Fig.~\ref{rad2_fft}. In Fig.~\ref{rad2_fft}, elements 
{\tt x(col$'$+row)} and {\tt x(col$'$+row+L/2)} are accessed and combined
pairwise in the following order: (1) first ${\vec x}(0)$ and ${\vec x}(1)$ 
are combined,
followed by ${\vec x}(2)$ and ${\vec x}(3)$, followed by ${\vec x}(4)$ and 
${\vec x}(5)$, etc.  In the
next step the stride doubles leading to the following combinations: (2) 
elements ${\vec x}(0)$ and ${\vec x}(2)$ are combined, followed by 
${\vec x}(1)$ and ${\vec x}(3)$,
next ${\vec x}(4)$ and ${\vec x}(6)$, next ${\vec x}(5)$ and ${\vec x}(7)$ etc.

Now consider the next step of algorithm of Fig.~\ref{rad2_fft} in the context
of the {\it re-shaped} array of Eq.~\ref{reshape1}.  The next step of the 
algorithm, (3) combines elements ${\vec x}(0)$ and ${\vec x}(4)$, ${\vec x}(1)$ and ${\vec x}(5)$, ${\vec x}(2)$
and ${\vec x}(6)$, etc.  If, in Eq.~\ref{reshape1}, the row length $c$ corresponds 
to the cache size, the process of combining ${\vec x}(0)$ and ${\vec x}(4)$ leads to a
cache miss.  Likewise combining ${\vec x}(1)$ and ${\vec x}(5)$ also leads to a cache miss
etc.  In fact, all further operations lead to cache misses, page faults etc.

At this point, in order to avoid cache misses, we re-structure the data array
using the {\it reshape-transpose} as discussed in the following.  In effect
we wish to re-order the array by sequentially taking the elements of the 
{\it columns} and placing them into the rows in lexical order.  The first
row of the {\it reshape-transposed} array would therefore consist of the 
elements $0$, $4$, $8$, $12$.  The next row would consist of the elements
$16$, $20$, $24$, and $28$.  Proceeding to the next column of $x^{(a)}$ leads to the 
following elements in the third row: $1$, $5$, $9$, $13$, etc.

The operation is called {\it reshape-transpose} because we can think of
the process of occurring in two stages.  In the first stage we transpose the 
array.  In standard matrix language we would write 
$x^{(b)} = T x^{(a)}$. The corresponding operation (transpose)
in $\psi$-calculus notation is expressed by the symbol $\transpose$.
Thus we write:
\begin{equation}
x^{(b)} = \transpose (<8\;4> \rshp\; (\iota 32)) \equiv
\left [
\begin{array}{cccccccc}
0&4&8&12&16&20&24&28 \\
1&5&9&13&17&21&25&29\\
2&6&10&14&18&22&26&30\\
3&7&11&15&19&23&27&31\\
\end{array}
\right ] 
\label{reshape2}
\end{equation}
In the second stage of the {\it reshape-transpose} operation we {\it reshape}
the result of the {\it transpose} operation to give:
\begin{equation}
x^{(c)} \equiv < 8\;4> \rshp \;
(\transpose (<8\;4> \rshp\; (\iota 32))) \equiv
\left [ \begin{array}{cccc}
0&4&8&12\\
16&20&24&28\\
1&5&9&13\\
17&21&25&29\\
2&6&10&14\\
18&22&26&30\\
3&7&11&15\\
19&23&27&31\\
\end{array}
\right].
\label{reshape3}
\end{equation}
Note, in practice, the {\it reshape-transpose} operation which transforms
the array of Eq.~\ref{reshape1} into that of Eq.~\ref{reshape3} is carried
out in a {\bf single step} using $\psi$-calculus indexing techniques.  The
two-step process indicated by Eq.~\ref{reshape2} (for the transpose) and 
Eq.~\ref{reshape3} (for the subsequent reshaping) was given merely for the purpose
of illustration.

Now, by working with the data as restructured according to Eq.~\ref{reshape3}
we find that the data needed for step (3) of the algorithm of 
Fig.~\ref{rad2_fft} (i.e. elements ${\vec x}(0)$ and ${\vec x}(4)$, elements  
${\vec x}(8)$ and 
${\vec x}(12)$ etc.), are now close to one another thus reducing cache misses.
We thus continue to process the data by combining only elements that are 
within a given row.  The combinations for the first row, are therefore,
${\vec x}(0)$ and ${\vec x}(4)$, ${\vec x}(8)$ and ${\vec x}(12)$, ${\vec x}(0)$ and ${\vec x}(8)$, and lastly
${\vec x}(4)$ and ${\vec x}(12)$.  The next step would be to carry out similar operation
for the remaining rows. 

The last step of the FFT (for this example) requires the combination of 
elements ${\vec x}(0)$ and ${\vec x}(16)$, ${\vec x}(1)$ and ${\vec x}(17)$, etc., which would give rise
to cache misses using the data as structured in Fig.~\ref{reshape3}.  Therefore
we, once again, restructure the data by carrying out another 
{\it reshape-transpose} operation to yield:
\begin{equation}
x^{(d)} = \; < 8\;4> \rshp \;  (\transpose \; x^{(c)}) \equiv
\left [ \begin{array}{cccc}
0&16&1&17\\
2&18&3&19 \\
4&20&5&21\\
6&22&7&23\\
8&24&9&25\\
10&26&11&27\\
12&28&13&29\\
14&30&15&31\\
\end{array}
\right ]
\label{xdeqn}
\end{equation}
Now the last operations of the FFT (for this example) involve only elements within a given row.  These
operations combine ${\vec x}(0)$ and ${\vec x}(16)$, ${\vec x}(1)$ and ${\vec x}(17)$, ${\vec x}(2)$ and 
${\vec x}(18)$, ${\vec x}(3)$ and ${\vec x}(19)$ etc.

Once the process of computing the FFT is completed, there remains the final
step of putting all of the data back in the correct order.  One way to 
accomplish this would be to undo the multiple {\it reshape-transpose} 
operations. 
{\bf Using $\psi$-calculus indexing techniques, it is possible to put all of the data back in the proper order in one step} as is discussed in
section \ref{final_trans}.

\subsection{Re-Ordering of ``Butterfly" Operations in the Cache-Optimized FFT}
\label{butter-order}
\begin{figure}[ht]
\small
\begin{center}
\includegraphics{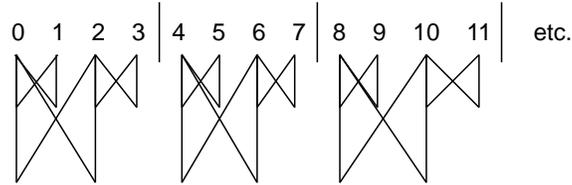}
\end{center}
\caption{\label{butterfly1} Schematic illustration of some of the access 
patterns for first two cycles of the FFT (some patterns omitted for clarity).}
\end{figure}

In addition to changing the access patterns so as to achieve data
locality, we are also changing the order in which the various
operations of the FFT are carried out. Figure~\ref{butterfly1} illustrates
pictorially the first few operations of the FFT before any data restructuring
has taken place.  The operation of, say, combining elements $x(0)$ and 
$x(1)$ to give the new values for $x(0)$ and $x(1)$, {\it in place} 
(i.e. without the need for temporary arrays) is called the ``butterfly" 
because of the way it is traditionally drawn (with vertical and 
diagonal lines) as in Fig.~\ref{butterfly1}. Even at this stage (before
any data restructuring occurs) we carry out the butterfly operations in 
non-standard order.

In the traditional FFT, all butterfly operations involving the smallest 
strides are carried out first before moving to larger strides (e.g. all 
butterfly operations involving nearest-neighbors such as $x(0)$ and $x(1)$
are carried out before moving to next-nearest-neighbors such as $x(0)$
and $x(2)$ etc.). In the new algorithm, however, we want to maximize in-cache
operations so we carry out all operations with the first set of data that
fits in cache before moving on.  This is illustrated in Fig.~\ref{butterfly1}
as follows. 

In keeping with the example of the previous subsection we are
setting $c = 4$ (nominally the cache size for this example).  The vertical
bars separating groups of four numbers (e.g. $|0\; 1\; 2\; 3|$; $|4\; 5\; 6\; 7|$, 
etc) schematically indicate cache boundaries.  Thus we perform all operations
of the FFT (with sequentially increasing strides) within a given data vector
of length $= c$ until the point at which
the stride equals $c$.  At that 
point we move the next
group and repeat the process until all groups of data have been exhausted.
To proceed to the next step (where the stride $ = c$) would lead to cache
misses for the data structured as in Fig.~\ref{butterfly1}.  At this point,
therefore, we carry out the {\it reshape-transpose} operation to restructure
the data.

\begin{figure}[ht]
\small
\begin{center}
\includegraphics{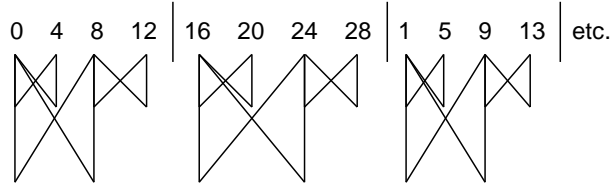}
\end{center}
\caption{\label{butterfly2}Schematic illustration of some of the access
patterns for
next two cycles of the FFT after the transpose/reshape (some patterns omitted
for clarity).}
\end{figure}

In the next two cycles of the FFT we work with the re-structured data so
as to achieve data locality as illustrated in Fig.~\ref{butterfly2}. Note that
the access patterns are identical to those of Fig.~\ref{butterfly1} and that
all operations which can be carried out with a given set of data elements 
(i.e. those which fit within a vector of length $c$) are performed before
moving on to the next group.  

At this point one can see the general pattern emerging. (1) The index-optimized
radix-$2$ FFT of Fig.~\ref{rad2_fft} is carried out within a given data
block of length $c$ albeit with modified weights (to be discussed in the 
next subsection).  (2) Then an outer loop cycles over the data blocks.  Following
that, (3) the data is re-arranged by carrying out a {\it reshape-transpose}
until the last re-arrangement is achieved.  (4) At this point, one must 
decide how many cycles of the FFT need to be done after the last 
{\it reshape-transpose} as, in
general, the number of cycles after the last {\it reshape-transpose} might be
less than $\log_2(c)$.  (5) Lastly the original ordering of the data is restored (as discussed in
section~\ref{final_trans}). 

\begin{figure}[ht]
\small
\begin{center}
\includegraphics{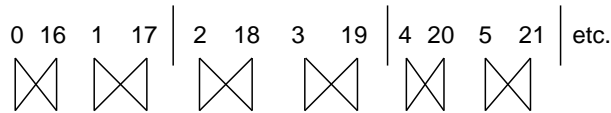}
\end{center}
\caption{\label{butterfly3} Schematic illustration of some of the access
patterns for the final stage of the FFT in which only cycles of length $2$ are
required.
}
\end{figure}

In the last paragraph, in step (4), we mentioned that after the final 
{\it reshape-transpose} the number of FFT cycles within a given data block of
length $= c$ might be fewer than for all the other stages of the calculation.
This can be clearly seen in the present example.  As illustrated in 
Fig.~\ref{butterfly3}, in the last stage of the present example, only
cycles of length $2$ are needed (as opposed to cycles of $2$ and $4$ for
previous orderings).

\subsection{The Number of Reshape-Transpose Operations}
The total number of {\it reshape-transpose} operations and the number
of FFT steps between {\it reshape-transpose} operations is determined
by the two parameters $c$ and total length of the input data array $n$.
The illustrations of the butterfly operations presented in 
Figs.~\ref{butterfly1},~\ref{butterfly2}, and~\ref{butterfly3} illustrate
the notion of the {\it stride} of the data elements being combined.  The stride
is simply the number of memory locations separating a given two elements. From 
the definition of the block matrices $B_L$ (see Fig.~\ref{vanloan_fft})
we easily identify the stride to be equal to $L/2$.  Since the stride
doubles with each iteration of the FFT, we find that a stride of length $c/2$
(the maximum stride which stays within the vector of length $c$) is reached
after $log_2(c)$ steps of the calculation. Thus there are $log_2(c)$ FFT
steps between subsequent {\it reshape-transpose} operations.

The total number of iterations of the FFT is $log_2(n)$, therefore the total
number {\it reshape-transpose} operations is the integer part of the ratio
$log_2(n)/log_2(c)$.

\subsection{Re-Ordering of the Weights}

The book-keeping for the weights required for the new algorithm is presented
in this subsection.  We wish to implement the new algorithm as 
a generalization of those illustrated in Figs.~\ref{vanloan_fft} 
and~\ref{rad2_fft}. In order to do that we must keep track of which weights
go with which elements of the data array as it is re-arranged via the
sequence of {\it reshape-transposes}.

At this point it is convenient to return to the notation of
subsection~\ref{traditional} in which we illustrated a given iteration of the
FFT as the multiplication of a block-diagonal $n\times n$ matrix by 
an $n$-dimensional vector.  As discussed in the previous subsection, we carry
out $log_2(c)$ iterations of the FFT (the last of which having a stride 
$= c/2$) and then re-arrange the data with a {\it reshape-transpose} 
operation.  Then we carry out another $log_2(c)$ steps 
before re-arranging, etc.  Thus the last iteration before the first 
{\it reshape-transpose} is written as:
\begin{equation}
x^{(log_2(c))} = M^{(log_2(c))} x^{(log_2(c/2))}.
\end{equation}
Thus the first operation requiring the use of the restructured data 
is written as:
\begin{equation}
x^{(log_2(2c))} = M^{(log_2(2c))} x^{(log_2(c))}.
\label{fft_eq_trans}
\end{equation}
Let us write the {\it reshape-transpose} operation as a linear transformation
defined by the matrix $S$ acting on the data vector $x^{(log_2(c))}$ as
\begin{equation}
y^{(0)} = S x^{(log_2(c))},
\end{equation}
where we have introduced the notation $y^{(0)}$ to indicate the data as 
re-ordered by the {\it reshape-transpose} operation.  Next we transform 
Eq.~\ref{fft_eq_trans} by multiplying both side of the equation, on the left, by 
the matrix $S$ and introducing the identity $I = S^{-1}S$ between the matrix 
$M^{(log_2(2c))}$ and the vector $x^{(log_2(c))}$ to give:
\begin{equation}
S x^{(log_2(2c))} = S M^{(log_2(2c))} S^{-1} S x^{(log_2(c))}.
\end{equation}
which, in turn we write as:
\begin{equation}
y^{(1)} =  N^{(1)} y^{(0)}.
\label{ydef}
\end{equation}
In equation~\ref{ydef} we have introduced the re-ordered weight matrix:
\begin{equation}
N^{(1)} \equiv S M^{(log_2(2c))} S^{-1},
\label{none}
\end{equation}
which defines the first iterate of 
a sequence of $log_2(c)$ transformations involving the re-structured data.  We
have also introduced the first-iterate data vector $y^{(1)}$.

This iterative procedure continues in an analogous fashion to that
outlined in subsection~\ref{traditional} using the matrices 
$N^{(j)} \equiv S M^{(log_2(c)+j)} S^{-1}$ for $1\le j \le log_2(c)$.  The last step in
this sequence of operations (i.e. before the next {\it reshape-transpose} 
is carried out) is given by:
\begin{equation}
y^{(log_2(c))} = N^{(log_2(c))} y^{(log_2(c/2))},
\end{equation}
where $N^{(log_2(c))} \equiv S M^{(log_2(c^2))} S^{-1}$.

At this point we introduce the next restructuring of the data and 
transform the next step of the FFT as:
\begin{equation}
S y^{(log_2(2c))} = S N^{(log_2(2c))} S^{-1} S y^{(log_2(c))}.
\end{equation}
and the procedure continues in an obvious fashion.

\subsection{$\psi$-Reduction}
As emphasized in subsection~\ref{traditional} we {\bf never actually materialize}
(instantiate) any $n\times n$ matrices.  The discussion of 
subsection~\ref{traditional} and of the previous subsection is merely for the purpose
of constructing the derivation.  Ultimately we use the techniques of the 
$\psi$-calculus, in particular the technique called ``$\psi$-reduction" to
arrive at an implementation analogous to that of Fig.~\ref{rad2_fft}.  The
process of $\psi$-reduction is a procedure which reduces an algebraic
expression to its simplest form by eliminating temporary arrays.
In effect, all array access operations are effected through direct indexing.
The application of this procedure to the {\bf general-radix FFT} is discussed
in Ref.~\cite{mullin.small:} and the result for the special case of the
radix-$2$ FFT is presented in Fig.~\ref{rad2_fft}

In order to effect the $\psi$-reduction for the present problem we must 
consider the structure of the transformed weight matrices, such as 
$N^{(j)} \equiv S M^{(log_2(c)+j)} S^{-1}$ (for $1\le j \le log_2(c)$) in greater detail 
as presented in the next subsection.

\subsection{Structure of the Weight Matrices}
In this subsection, we consider the structure of the weight matrices $M^{(i)}$
and the corresponding transformed weight matrices $N^{(j)}$ in some detail.
As discussed in subsection~\ref{traditional}, the matrix $M^{(i)}$ is a 
block-diagonal $n\times n$ matrix, consisting of $n/2^i$, $2^i\times 
2^i$-dimensional blocks $B_{2^i}$.  The structure of the block matrices $B_L$
is given in Fig.~\ref{vanloan_fft}.  Note, each of the block matrices $B_L$
is sparse, having only $2L$ non-zero elements. The remaining $L^2 - 2L$ 
elements are zero.  {\bf Note, by using $\psi$-calculus techniques, we only 
carry out multiplication operations involving non-zero elements}.

The effect of the {\it reshape-transpose} operation on the weight matrix
$M^{(i)}$ is to rearrange each of its sparse diagonal blocks into a series of
{\bf smaller} diagonal blocks.  {\bf Thus through the repeated use of the} 
{\it \bf  reshape-transpose} {\bf operation the weight matrix remains banded 
with a fixed maximum width dictated by the parameter $c$}.

\subsection{Transforming the Weights}
\label{trans-weights}

The $2L$ non-zero elements in a given block $B_L$ map a specific set of $L$
elements of the data array into the corresponding $L$ elements of the updated
array.  There is thus a unique $2\rightarrow 1$ homomorphism of the $2L$ 
nonzero elements of the array $B_L$ and the $L$ elements of the data vector.
The $2L$ non-zero elements of a block $B_L$ are contained within four 
$L/2$-dimensional sub-blocks (see Fig.~\ref{vanloan_fft}).  Two of these 
blocks are the $L/2$-dimensional identity matrix $I_{L_\star}$ 
(in Fig.~\ref{vanloan_fft} $L_\star = L/2$) .  The other two blocks consist of 
the $L/2$-dimensional blocks $\Omega_{L_\star}$, and  $-\Omega_{L_\star}$. 

The transformation of the weight matrix $N^{(j)} = SM^{(log_2(c)+j)}S^{-1}$
can be determined by constructing a unique $1\rightarrow 1$ isomorphism
between the non-zero elements of $M^{(loc_2(c) +j)}$ and the elements of the
data vector $x^{(log_2(2c)+j-1)}$.  Since we know how to transform the 
data vector $y^{(j)} = S x^{(log_2(2c)+j-1)}$ using the {\it reshape-transpose},
the isomorphism allows us to determine the transformation of the weight matrix.

The isomorphism is very simple.  We begin by constructing a vector $\xi$ of 
length $n$ as follows.  The first $L/2$ elements are unity, taken from the
$I_{L_\star}$ identity matrix of the first block $B_L$.  The next $L/2$
elements are the non-zero elements (taken in lexical order) of the matrix
$\Omega_{L_\star}$ also from the {\it first} block $B_L$.  The next $L/2$ 
elements are unity taken from the $I_{L_\star}$
{\it second} block $B_L$, and the next $L/2$ elements are the non-zero elements
of $\Omega_{L_\star}$ also from the {\it second} block $B_L$.  This procedure
continues until $L$ elements have been extracted from each block $B_L$ to
yield the $n$-dimensional weight vector $\xi$.  This procedure is completely
general and applies to all transformations of the weight vector.  Before the
first {\it reshape-transpose} all of the blocks $B_L$ are identical.  However,
as we will see, for all other weight matrices generated by the 
{\it reshape-transpose} operation, the sub-blocks $B_L$ will in general be
different.

The vector $\xi$ transforms in 
the same way as the data vector $\xi^\prime = S\xi$ to give a transformed weight
vector.  Thus we only need to keep track of the changes to the weight vector
$\xi$ under the influence of the {\it reshape-transpose} operation.  We follow
the procedure to transform $\xi$ exactly as was done to transform $x$.  That is
we re-shape it into a two-dimensional vector $<r\;c> {\hat \rho} \xi$ and
observe the re-ordering which occurs due to the {\it reshape-transform} 
operation.

That the isomorphism between $\xi$ and $x$ is correct can easily be seen by
the following argument.  In carrying out one butterfly operation, we update the 
data vector $x \rightarrow x^\prime$ two elements at a time.  For the 
$log_2(L)$-th FFT step, the stride is $L/2$ and therefore elements $x(i)$ and
$x(i+L/2)$ are combined with the corresponding weights in the $\xi$ array as 
follows
\begin{equation}
x^\prime(i) = \xi(i)*x(i) + \xi(i+L/2)*x(i+L/2)
\label{xi1}
\end{equation}
and
\begin{equation}
x^\prime(i+L/2) = \xi(i)*x(i) - \xi(i+L/2)*x(i+L/2)
\label{xi2}
\end{equation}
Since each element of $\xi$ gets multiplied by the corresponding element of $x$
the two vectors must transform together.  Again we emphasize that the 
multiplications by unity implied in the first terms of Eqs.~\ref{xi1} 
and~\ref{xi2} is only an intermediate step in the derivation.  In the final
implementation there is no multiplication by unity or zero.

\section{Cache-Optimized FFT Illustrated}

\subsection{Specific Examples and Patterns}

Consider now the transformed equation $N^{(1)}$ as defined by Eq.~\ref{none}.
The matrix to be transformed, $M^{(log_2(2c))}$ is a block-diagonal $n\times n$
matrix with $n/(2c)$, blocks $B_{(2c)}$. Each matrix $B_{(2c)}$, is thus
comprised of two $c$-dimensional identity matrices $I_c$ and two 
$c$-dimensional matrices $\Omega_c$ and $-\Omega_c$.  As indicated in 
Fig.~\ref{vanloan_fft}, $\Omega_c$ is a diagonal matrix with elements
$(\Omega_c)_{ii} = \omega_{2c}^i$ (i.e. the (2c)-th root of unity raised
to the $i$-th power) for $0\le i \le (c-1)$.  Thus the elements of the 
weight vector $\xi$ consist of: (1) $c$ entries equal to unity, 
(2) $c$, $(2c)$-th roots of unity: $\omega_{2c}^0$, 
$\omega_{2c}^1\cdots$~$\omega_{2c}^{(c-1)}$, 
(3) $c$ entries equal to unity, etc. Now upon reshaping the weight vector
we obtain:
\begin{equation}
<r\; c> {\hat \rho} \xi \equiv
\left[
\begin{array}{ccccc}
1 & 1 & 1 & \cdots & 1 \\
\omega_{2c}^0 & \omega_{2c}^1 & \omega_{2c}^2 & \cdots & \omega_{2c}^{(c-1)}  \\
1 & 1 & 1 & \cdots & 1 \\
\omega_{2c}^0 & \omega_{2c}^1 & \omega_{2c}^2 & \cdots & \omega_{2c}^{(c-1)}  \\
\vdots & \vdots & \vdots & \vdots & \vdots \\
1 & 1 & 1 & \cdots & 1 \\
\omega_{2c}^0 & \omega_{2c}^1 & \omega_{2c}^2 & \cdots & \omega_{2c}^{(c-1)}  \\
\end{array}
\right ]
\label{xi_one_general}
\end{equation}
Returning to our example of the $n = 32$ array, explicitly, we have:
\begin{equation}
<8\; 4> {\hat \rho} \xi \equiv
\left [
\begin{array}{cccc}
1 & 1 & 1 & 1 \\
\omega_8^0 & \omega_8^1 & \omega_8^2 & \omega_8^3 \\
1 & 1 & 1 & 1 \\
\omega_8^0 & \omega_8^1 & \omega_8^2 & \omega_8^3 \\
1 & 1 & 1 & 1 \\
\omega_8^0 & \omega_8^1 & \omega_8^2 & \omega_8^3 \\
1 & 1 & 1 & 1 \\
\omega_8^0 & \omega_8^1 & \omega_8^2 & \omega_8^3 \\
\end{array}
\right ]
\label{xi_one_n32}
\end{equation}
Now reorder with a {\it reshape-transpose} to give
\begin{equation}
<8\; 4> {\hat \rho} (\transpose
<8\; 4> {\hat \rho} \xi) \equiv
\left [
\begin{array}{cccc}
1 & \omega_8^0 & 1 & \omega_8^0 \\
1 & \omega_8^0 & 1 & \omega_8^0 \\
1 & \omega_8^1 & 1 & \omega_8^1 \\
1 & \omega_8^1 & 1 & \omega_8^1 \\
1 & \omega_8^2 & 1 & \omega_8^2 \\
1 & \omega_8^2 & 1 & \omega_8^2 \\
1 & \omega_8^3 & 1 & \omega_8^3 \\
1 & \omega_8^3 & 1 & \omega_8^3 \\
\end{array}
\right ]
\end{equation}
Now by reversing the process that was used to construct the weight vector
$\xi$ from the elements $M^{(log_2(2c))}$ (which is $M^{(3)}$ in this example) 
we obtain from the weight vector $\xi$ the weight matrix $N^{(1)}$:
\begin{equation}
\left [
\begin{array}{ccccccccccccccccccc}
1 & \;\;\;\omega_8^0 & 0            & 0                & 0                  & 0                  & 0            & 0 & 0 & 0 & 0 & 0 & 0 & 0 & 0 & 0 & \cdots & 0 & 0\\

1 & -\omega_8^0      & 0            & 0                & 0                  & 0                  & 0            & 0 & 0 & 0 & 0 & 0 & 0 & 0 & 0 & 0 & \cdots & 0 & 0\\

0 & 0            & 1                & \;\;\;\omega_8^0 & 0                  & 0                  & 0            & 0 & 0 & 0 & 0 & 0 & 0 & 0 & 0 & 0 & \cdots & 0 & 0\\

0 & 0            & 1            & -\omega_8^0          & 0                  & 0                  & 0            & 0 & 0 & 0 & 0 & 0 & 0 & 0 & 0 & 0 & \cdots & 0 & 0\\

0 & 0                & 0            & 0                &  1 & \;\;\;\omega_8^0 & 0            & 0                 & 0 & 0 & 0 & 0 & 0 & 0 & 0 & 0 & \cdots & 0 & 0\\

0 & 0                & 0            & 0                &  1 & -\omega_8^0      & 0            & 0                 & 0 & 0 & 0 & 0 & 0 & 0 & 0 & 0 & \cdots & 0 & 0 \\

0 & 0                & 0            & 0                &  0 & 0                & 1            & \;\;\;\omega_8^0  & 0 & 0 & 0 & 0 & 0 & 0 & 0 & 0 & \cdots & 0 & 0\\

0 & 0                & 0            & 0                &  0 & 0                & 1            & -\omega_8^0       & 0 & 0 & 0 & 0 & 0 & 0 & 0 & 0 &\cdots & 0 & 0\\

0 & 0 & 0 & 0 & 0 & 0 & 0 & 0 & 1 & \;\;\;\omega_8^1 & 0            & 0                & 0                  & 0                  & 0            & 0 &\cdots & 0 & 0\\

0 & 0 & 0 & 0 & 0 & 0 & 0 & 0 & 1 & -\omega_8^1      & 0            & 0                & 0                  & 0                  & 0            & 0&\cdots & 0 & 0 \\

0 & 0 & 0 & 0 & 0 & 0 & 0 & 0 & 0 & 0            & 1                & \;\;\;\omega_8^1 & 0                  & 0                  & 0            & 0 &\cdots & 0 & 0\\

0 & 0 & 0 & 0 & 0 & 0 & 0 & 0 & 0 & 0            & 1            & -\omega_8^1          & 0                  & 0                  & 0            & 0 &\cdots & 0 & 0\\

0 & 0 & 0 & 0 & 0 & 0 & 0 & 0 & 0 & 0                & 0            & 0                &  1 & \;\;\;\omega_8^1 & 0            & 0                 &\cdots & 0 & 0\\

0 & 0 & 0 & 0 & 0 & 0 & 0 & 0 & 0 & 0                & 0            & 0                &  1 & -\omega_8^1      & 0            & 0                 &\cdots & 0 & 0\\

0 & 0 & 0 & 0 & 0 & 0 & 0 & 0 & 0 & 0                & 0            & 0                &  0 & 0                & 1            & \;\;\;\omega_8^1 & \cdots & 0 & 0 \\ 

0 & 0  & 0 & 0  & 0 & 0  & 0 & 0  & 0 & 0                & 0            & 0                &  0 & 0                & 1            & -\omega_8^1 & \cdots & 0 & 0\\

\vdots & \vdots &\vdots  &\vdots  &\vdots  &\vdots  &\vdots  &\vdots  &\vdots  &\vdots  &\vdots  &\vdots  &\vdots  &\vdots  &\vdots  & \vdots & \cdots & \vdots & \vdots \\
0 & 0 & 0 & 0 & 0 & 0 & 0 & 0 & 0 & 0 & 0 & 0 & 0 & 0 & 0 & 0 & \cdots & 1 & \omega_8^3\\
0 & 0 & 0 & 0 & 0 & 0 & 0 & 0 & 0 & 0 & 0 & 0 & 0 & 0 & 0 & 0 & \cdots & 1 & \omega_8^3\\

\end{array}
\right ]
\label{bign1}
\end{equation}

\subsection{Structure of Reshape-Transposed Weight Matrix}
Note carefully the structure of $N^{(1)}$ given in Eq.~\ref{bign1}.  The
equation is now block diagonal with $2\times 2$ blocks along the diagonal.
Thus the access patterns for the matrix multiplication 
\begin{equation}
y^{(1)} = N^{(1)} y^{(0)},
\end{equation}
using the reshape-transposed quantities are {\bf the same} as those for 
\begin{equation}
x^{(1)} = M^{(1)} x^{(0)}.
\end{equation}
Now, however, we must realize that the structure of $N^{(1)}$ is more general
than that of $M^{(1)}$

Previously, in the definition of the untransformed matrices $M^{(i)}$, a block
diagonal sub-matrix $B_L$ was completely specified by its dimension $L$.  Now
however, the definition of the sub-matrices must be generalized.

We see in Eq.~\ref{bign1} that the first four blocks on the diagonal are equal
and the weight element $\omega_8^0 = 1$ is the same as that we previously found
in the definition of the sub-matrix $\Omega_1 = \omega_2^0 = 1$ (see 
Fig.~\ref{vanloan_fft}). Now, however, there are {\bf four different 
$2\times 2$} matrices.  They are distinguished by their weight elements
$\omega_8^0$, $\omega_8^1$, $\omega_8^2$, $\omega_8^3$, respectively.

Now we see that the weight matrix, requires three labels: (1) $L$
to determine the particular root of unity $L$ ($L = 8$ in this example), 
(2) $\sigma$ giving the power of the root, ($0$, $1$, $2$, and $3$ in this 
example), and (3) $d$ ($= 2$ in this example) to determine the dimensionality 
(this in turn, is 
related to the number of {\it reshape-transpose} operations which have 
occurred). We use the symbol $\Omega_L^{(\sigma,\;d)}$, which is a 
$d/2$-dimensional matrix.  Likewise we generalize
the definition of the block sub-matrices $B_L \rightarrow \beta_L^{(\sigma,\; d)}$. 
For example:
\begin{equation}
\beta_L^{(\sigma,\;d)} \equiv
\left [
\begin{array} {cc}
I_{d/2}  &  \Omega_8^{(\sigma,\;d)} \\
I_{d/2}  &  -\Omega_8^{(\sigma,\;d)} \\
\end{array}
\right ].
\label{genblock}
\end{equation}
So far we have only encountered the $1$-dimensional ($d/2 = 1$) weight matrix
$\Omega_L^{(\sigma,\;d)}$.  The generalization to higher dimensions (which
depends on how many times the {\it reshape-transpose} operation has occurred)
will be presented shortly.
 
We now summarize the discussion of $N^{(1)}$ for the $n=32$ case. The
block diagonal matrix to be transformed, $M^{(3)}$, consists of four 
{\bf identical }$8$-dimensional blocks $B_8$.  We write:
\begin{equation}
M^{(3)} = B_8 \directsum B_8 \directsum B_8 \directsum B_8,
\label{mthree}
\end{equation}
The transformed matrix, on the other hand, transforms into four {\it groups} of 
four different $2\times 2$ matrices:
\begin{eqnarray}
N^{(1)} & = & \beta_8^{(0,2)} \directsum \beta_8^{(0,2)} \directsum \beta_8^{(0,2)} \directsum \beta_8^{(0,2)} \nonumber \\
& \directsuma & \beta_8^{(1,2)} \directsum \beta_8^{(1,2)} \directsum \beta_8^{(1,2)} \directsum \beta_8^{(1,2)} \nonumber \\
& \directsuma & \beta_8^{(2,2)} \directsum \beta_8^{(2,2)} \directsum \beta_8^{(2,2)} \directsum \beta_8^{(2,2)} \nonumber\\
& \directsuma & \beta_8^{(3,2)} \directsum \beta_8^{(3,2)} \directsum \beta_8^{(3,2)} \directsum \beta_8^{(3,2)}. \nonumber\\
\end{eqnarray} 
We now state the result for the next iterate of the weight matrix 
$N^{(2)} = SM^{(4)}S^{-1}$, and then illustrate it in detail.  The matrix 
$N^{(2)}$ is a $n\times n$ sparse matrix ($n = 32$) consisting of four
groups of two, $4\times 4$ matrices as:
\begin{eqnarray}
N^{(2)} & = & \beta_{16}^{(0,\; 4)} \directsum \beta_{16}^{(0,\; 4)} \nonumber \\
        & \directsuma & \beta_{16}^{(1,\; 4)} \directsum \beta_{16}^{(1,\; 4)} \nonumber \\
        & \directsuma & \beta_{16}^{(2,\; 4)} \directsum \beta_{16}^{(2,\; 4)} \nonumber \\
        & \directsuma & \beta_{16}^{(3,\; 4)} \directsum \beta_{16}^{(3,\; 4)}. \nonumber \nonumber \\
\label{ntwodsum}
\end{eqnarray}
where the four-dimensional weight sub-blocks are given by
\begin{equation}
\beta_{16}^{(\sigma,\; 4)} \equiv
\left [
\begin{array} {cccc}
1 & 0 & \omega_{16}^\sigma & 0 \\
0 & 1 & 0           & \omega_{16}^{\sigma+c} \\
1 & 0 & -\omega_{16}^{\sigma} & 0 \\
0 & 1 & 0                   & -\omega_{16}^{\sigma+c} \\
\end{array}
\right ].
\label{bigntwo}
\end{equation}
Note that the sequence of matrices $\beta_{16}^{(\sigma,\; 4)}$, for $0 \le \sigma \le (c-1)$ ($c = 4$ for this example) in Eq.~\ref{ntwodsum}  contain all of 
the 
weights: 
$\omega_{16}^{0}$, $\omega_{16}^{1}$, $\omega_{16}^{2}$, $\omega_{16}^{3}$,
$\omega_{16}^{4}$, $\omega_{16}^{5}$, $\omega_{16}^{6}$, $\omega_{16}^{7}$,
originally present in the untransformed matrix $M^{(4)}$

\subsection{ Derivation of $N^{(2)}$ From the Weight Vector}
We now explicitly illustrate the construction of the weight matrix $N^{(2)}$
from the corresponding weight vector $\xi$.  As was discussed for the situation
illustrated by Eqs.~\ref{xi_one_general} and~\ref{xi_one_n32}, we form the 
weight vector $\xi$ from the elements of $M^{(4)}$ and reshape to give:

\begin{equation}
<8\; 4> {\hat \rho } \xi \equiv
\left [
\begin{array} {cccc}
1 & 1 & 1 & 1 \\
1 & 1 & 1 & 1 \\
\omega_{16}^{0} & \omega_{16}^{1} & \omega_{16}^{2} & \omega_{16}^{3} \\
\omega_{16}^{4} & \omega_{16}^{5} & \omega_{16}^{6} & \omega_{16}^{7} \\
1 & 1 & 1 & 1 \\
1 & 1 & 1 & 1 \\
\omega_{16}^{0} & \omega_{16}^{1} & \omega_{16}^{2} & \omega_{16}^{3} \\
\omega_{16}^{4} & \omega_{16}^{5} & \omega_{16}^{6} & \omega_{16}^{7} \\
\end{array}
\right ].
\label{blob}
\end{equation}
Next we carry out the {\it reshape-transpose} to give:

\begin{equation}
<8\; 4> {\hat \rho} (\transpose <8\; 4> {\hat \rho} \xi) \equiv
\left [
\begin{array} {cccc}
1 & 1 & \omega_{16}^{0} & \omega_{16}^{4} \\
1 & 1 & \omega_{16}^{0} & \omega_{16}^{4} \\
1 & 1 & \omega_{16}^{1} & \omega_{16}^{5} \\
1 & 1 & \omega_{16}^{1} & \omega_{16}^{5} \\
1 & 1 & \omega_{16}^{2} & \omega_{16}^{6} \\
1 & 1 & \omega_{16}^{2} & \omega_{16}^{6} \\
1 & 1 & \omega_{16}^{3} & \omega_{16}^{7} \\
1 & 1 & \omega_{16}^{3} & \omega_{16}^{7} \\
\end{array}
\right ].
\label{expandntwo}
\end{equation}
By reversing the steps that led to the formation of $\xi$ (in Eq.~\ref{blob})
from $M^{(4)}$, Eq.~\ref{expandntwo} expands to give Eq.~\ref{ntwodsum}. Each
row of Eq.~\ref{expandntwo} translates into the corresponding $4\times 4$
sub-block matrices $\beta_{16}^{(\sigma,\; 4)}$ of Eq.~\ref{bigntwo}.

\subsection{Last Step}
The last step in this example ($n = 32$) corresponds to the original weight 
matrix $M^{(5)}$:
\begin{equation}
M^{(5)} = B_{32}.
\end{equation}
Carrying out one {\it reshape-transpose} gives:
\begin{eqnarray}
N^{(3)} & = & S M^{(5)} S^{-1}  \nonumber \\
& \equiv & \beta_{32}^{(0,8)} \directsum \beta_{32}^{(1,8)} \directsum \beta_{32}^{(2,8)} \directsum \beta_{32}^{(3,8)}, \nonumber \\
\label{nthree}
\end{eqnarray}
which is a block-diagonal matrix consisting of $8\times 8$ blocks.  One can see
that Eq.~\ref{nthree} is the analog of Eq.~\ref{mthree}.  In order to 
achieve data locality (i.e. blocks of dimension less than or equal to $c = 4$)
we must {\it reshape-transpose} once more to yield: 
\begin{eqnarray}
S N^{(3)} S^{-1} & = & \beta_{32}^{(0,2)} \directsum \beta_{32}^{(1,2)} \directsum \beta_{32}^{(2,2)} \directsum \beta_{32}^{(3,2)} \nonumber \\
& \directsuma & \beta_{32}^{(4,2)} \directsum \beta_{32}^{(5,2)} \directsum \beta_{32}^{(6,2)} \directsum \beta_{32}^{(7,2)} \nonumber \\
& \directsuma & \beta_{32}^{(8,2)} \directsum \beta_{32}^{(9,2)} \directsum \beta_{32}^{(10,2)} \directsum \beta_{32}^{(11,2)} \nonumber\\
& \directsuma & \beta_{32}^{(12,2)} \directsum \beta_{32}^{(13,2)} \directsum \beta_{32}^{(14,2)} \directsum \beta_{32}^{(15,2)}, \nonumber\\
\end{eqnarray} 
which is a block-diagonal matrix with sixteen different $2\times 2$ sub-blocks
on the diagonal.

\section{The General Algorithm}
The general pattern for the multiply-restructured weight matrices was 
discovered, by continuing the analysis of the past few sections with larger 
and more general data structures, and is presented in this section.
\subsection{Numbering}
There are $log_2(c)$ operations before the first {\it reshape-transpose} 
corresponding to strides of length $v_\star = v/2 = 2^{p-1}$, where we allow
$1 \le p \le log_2(c)$. For iteration $p$ the block-diagonal sub-matrices of 
$M^{(p)}$ have dimension $v\times v$.

After the first {\it reshape-transpose} the block-diagonal sub-matrices of
array $N^{(p)}$, for $1 \le p \le log_2(c)$ are of dimension 
$v_\star = v/2 = 2^{p-1}$ as before, however there are now different types 
of weight matrices which must be specified, in addition to their dimension,
by a parameter $\sigma$.  

At this point, the length of the FFT cycle is $L = 2^p c$ which also denotes
the root of unity (e.g. $\omega_L$, and powers thereof). Because the number of
distinct roots of unity (i.e. $L/2 = 2^{(p-1)} c$ ) is greater than can fit
in a single vector of length $l \le c/2$ there are different types of block
sub-matrices for a given dimension.  There will also be copies.

We find it convenient to specify the cycle length $L$ with two 
parameters $p$ and $l$ where we define $L = 2^p c^l$ where for $l = 0$, the 
variable cycles through the values $1 \le p \le log_2(c)$.  After the first
{\it reshape-transpose}, $l = 1$ and we write $L = 2^p c$ for $1 \le p \le 
log_2(c)$, after the second {\it reshape-transpose} we have $l = 2$, and
$L = 2^p c^2$, etc.  Thus the variable $l$ counts the number of 
{\it reshape-transpose} operations.  

\subsection{General Weight Sub-Matrices}

After $l$ {\it reshape-transpose} operations, the weight matrix is 
most-conveniently specified by including $l$ as an additional parameter.  
We designate the weight matrix with {\it four} parameters: $\sigma$ 
(the type of matrix), $d$ its dimensionality, $L$ the root of unity, and $l$
the number of {\it reshape-transpose} operations which have been carried out.
Explicitly we have, 
\begin{equation}
\beta_{(L,\; l)}^{(\sigma, d)} \equiv
\left [
\begin{array} {cc}
I_{d/2} & \Omega_{(L,\; l)}^{(\sigma,\;d)} \\
I_{d/2} & -\Omega_{(L,\; l)}^{(\sigma,\;d)} \\
\end{array}
\right ]
\end{equation}
where the $j$-th component of the (diagonal) matrix 
$\Omega_{L,\; l}^{\sigma,d}$ is given by:
\begin{equation}
(\Omega_{L,\; l}^{\sigma,d})_{jj} \equiv \omega_{L}^{(\sigma + jc^l)}
\end{equation}
where $0 \le j \le (d/2 - 1) $, and the number of different types of matrices
is indicated by $0 \le \sigma \le (c^l-1)$.

\subsection{Number of Different Weight Matrices}

The number of different weight matrices is determined as follows.
We specify the total data array length by $n = 2^{p_{max}} c^{l_{max}}$. 
With this factorization we have by definition $1 \le p_{max} \le \log_2(c)$
We use
two parameters to specify the cycle length $L = 2^p c^l$  where $l$ counts
the number of transposes and $1 \le p \le log_2(c)$ (indices $p$ and $l$ cycle 
as inner and outer loops respectively).  For all $0 \le l \le l_{max}-1$
the variable $p$ cycles as $1 \le p \le log_2(c)$.  For $l = l_{max}$,
we have the restriction $1 \le p \le p_{max}$

Next determine the number of different cycles:
\begin{equation}
d_v = \frac {L}{v} = \frac {2^p c^l}{2^p} = c^l.
\end{equation}
This is to be interpreted as the number of different weight matrices of 
dimension $v\times v$.  
For cycle lengths $L \le c$ we find $l = 0$ and $d_v = 1$ meaning that for
this case (corresponding to the traditional FFT without reshaping) each matrix
is uniquely determined by its dimensionality. 

Note also that the number different weight matrices jumps by a factor of $c$
for each {reshape-transpose} operation. For example, after the first 
{\it reshape-transpose} operation, we have $c$ different $2\times 2$ matrices
corresponding to factoring the cycle length $L = 2c$. For the next FFT 
iteration we have $L = 4c$ corresponding to $p = 2$, which becomes factored
as $c$ different $4\times 4$ matrices etc.  After the next {reshape-transpose}
operation there are $c^2$ different weight matrices of a given size 
$v\times v$ where as before $v = 2^p$ and cycle length $L = 2^p c^2$.

\subsection{Number of Copies of a Given Cycle Length}

The number of copies of a given cycle length $L$ 
is given as the ratio of the total data vector length $n = 2^{p_{max}} 
c^{l_{max}}$ to the cycle length $L = v d_v$:

\begin{equation}
S_v \equiv \frac {n}{v d_v} = \frac {2^{p_{max}}c^{l_{max}}}{2^p c^l}.
\end{equation}
To shed some light on this quantity, consider the situation we encounter on 
the last step of the FFT in which $L = n$.  In this case $S_v = 1$ and there
is only {\it one} copy of each weight matrix $\beta^{(\sigma,\;d)}_{(L,\;l)}$.
For example if $p_{max} =1$ then we have the total data vector length 
$n = 2 c^{l_{max}}$ and we have each of the $c^{l_{max}}$ different
$2\times 2$ matrices represented only once.  If on the other hand we have
$p_{max} = 2$ we have $n = 4c^{l_{max}}$ and each of the $c^{l_{max}}$ 
different $4\times 4$ matrices appears only once etc.

Next consider the situation in which $L = n/2$.  In this case $S_v = 2$ and
there will be {\it two} cycles and each matrix $\beta^{(\sigma,\;d)}_{(L,\;l)}$
appears twice.  

The algorithm can be thought of as a generalized FFT algorithm which
processes vectors of length $c$.  The weight matrix thus naturally partitions 
into $c\times c$ blocks.  Thus we must consider the total number of copies of
a given $c\times c$ block.  This is computed as follows.  The length of a 
given set of copies is $v S_v = 2^{p_{max}} c^{(l_{max} - l)}$ which is 
partitioned into blocks of length $c$ is obtained by dividing $v S_v$ by $c$
to give:
\begin{equation}
S_B = 2^{p_{max}} c^{(l_{max}-l - 1)}
\end{equation}

\section{ Cache-Optimized FFT Using the $\psi$-Calculus}

At this point we turn to a brief look at two key steps in the algorithm:
the {\it reshape-transpose} and the {\it final transpose}.  These two examples
serve as vehicles to illustrate the use of Conformal Computing techniques.
The discussion here presents only the key results.  The complete details of
the derivations are presented in the following chapter (part II).

\subsection{Reshape Transpose Operation}

Recall the expression to abstract our view of the cache.  We write:

\begin{equation}
  < r\;c> \rshp \; (\transpose (<r\;c> \rshp\; (\iota n)))
\label{abstract}
\end{equation}
to represent the {\it reshape-transpose} operation on our restructured data
array.  Reading from right to left we start with the array of indices
$(\iota n)$
which is a one-dimensional array (vector) of $n$ sequential integers starting
with $0$ and ending with $n-1$. In the present example we only concern
ourselves with the manipulation of this index array.  The first reshaping of
($\iota n$) is indicated in Eq.~\ref{abstract} by the expression
$<r\;c> \rshp\; (\iota n) $ which implies a $r\times c$ array consisting
of the entries of $(\iota n)$ taken in sequential (i.e. {\it row-major})
order. For this example we require the reshaped array to contain the same
number of elements as the original array: $r\times c = n$. Using the
notation of the $\psi-calculus$ we write $\pi\! <\!r\;c\!> \; \equiv n$, where 
the
operator $\pi$ acting on a vector produces a scalar equal to the product of
the elements of the vector.

In the next step in Eq.~\ref{abstract} we apply the transpose operator
$\transpose$ to produce the $c\times r$ array
$\transpose (<\!r\;c\!> \rshp (\iota n))$.  In the last step we re-partition
the array with the  $<\!r\;c\!> \rshp $ to produce the $r\times c$ array 
obtained
by taking the elements of $\transpose (<r\;c> \rshp\; (\iota n))$ sequentially
in lexical order (i.e. {\it row-major}).

\subsection{$\psi$-Reduction of the Reshape-Transpose Operation}

Using Conformal Computing techniques, an Operational Normal Form (ONF) is 
obtained from Eq.~\ref{abstract} as is derived in detail in Part II.  For
the purposes of this paper we merely indicate the final result. The ONF
is an algebraic specification, in terms of indices (i.e. {\em starts}, 
{\em stops} and 
{\em strides}), that indicates explicitly how a given expression is to be built (in 
software or hardware).  For simplicity we define:
\begin{equation}
A \equiv (<r\;c> \rshp\; (\iota n)).
\end{equation} 
Using this notation we express the ONF of Eq.~\ref{abstract} as:
\begin{equation}
\forall\;\; i,j \;\;\mbox{s.t.}\;\; 0 \leq^* \;\;<i\;j>\;\;
<^* \;\; <r\;c>   \\
\label{reversed}
\end{equation}
\begin{equation}
<i\;j> \psi   ( <r\;c>  \rshp (\transpose A))
 \equiv  (\Ravel  \transpose A )[ j+(c \times i) ].
\label{trans}
\end{equation}
\normalsize

The last step that expresses the result directly in terms of the elements of 
the array $A$ (as opposed to $\transpose A$) is discussed in detail in the 
next chapter (Part II).  The result is simple and has been directly translated 
into C++ code as indicated in Fig.~\ref{trans_rshp_code}.

\renewcommand{\baselinestretch}{1}
\begin{center}
\begin{figure}[ht]
\scriptsize
\begin{quotation}
\begin{tt}
\noindent

\begin{verbatim}
void trans_rshp(complex *datvec,complex *temp,int nmax,
                int csize)
{
        int iind,jind,kind,kmax,jmax,imax,index,c2size;
        int rows,arg;
        int max(int a,int b);

//      The routine carries out the transpose-reshape operation
        index=0;

        rows = nmax/csize;
        imax = rows-1;

        c2size = int(pow(csize,2.0));
        jmax = max(0,nmax/c2size-1);


        for(jind=0;jind<=jmax;jind++)
        {
            for(iind=0;iind<=imax;iind++)
            {

                temp[index] = datvec[jind+csize*iind];
                index += 1;
            }
        }

        for(iind=0;iind<=nmax-1;iind++)
        {
            datvec[iind] = temp[iind];
        }
}
\end{verbatim}

\end{tt}
\end{quotation}
\caption{\label{trans_rshp_code}
C++ code fragment implementing the reshape-transpose in terms of 
index manipulations.
}
\end{figure}
\end{center}

\subsection{Reordering the Data\label{final_trans}}

After the last step of the FFT the data will not be in the correct order and
we must do something to return it to its initial order (i.e. prior to
any {\it reshape-transpose} operations).  The simplest approach would be
to rearrange the data by applying a series of inverse {\it reshape-transpose}
operations.  {\bf There is, however, a far more efficient approach in which
no data needs to be moved}.  In other words, we use Conformal Computing
techniques to determine the index vector which will select the correct
components of the array.  We present only the final result in this paper.
Complete details of the derivation and an in-depth discussion (with examples)
is presented in the following chapter (Part II.).

The input data vector $\vec y$ of length $n$ is carried into the
vector ${\vec x}$ through the $d = log_2 (n)$ steps of the FFT.  In the
process, $l_{max}$ {\it reshape-transpose} operations have been carried out.
The resulting vector $\vec x$ is thus not in the correct order (as a result
of the multiple {\it reshape-transpose} operations) and must therefore be
rearranged into its final form ${\vec \xi}$.
We now obtain $\xi_{hyper}$ as
\begin{equation}
\xi_{hyper} = ((-\sigma)\; \phi\; (\iota d)) \transpose ((<\!d\!> \rshp 2) \rshp {\vec x})
\label{xihyper}
\end{equation}
where $\sigma = l_{max} \log_2 (c)$ and $d = log_2 (n)$ and $\phi$ is the 
{\it rotate} operator that induces a cyclic permutation of the vector 
$\iota (d)$ (as will be discussed in detail in Part II).  
The operators acting on the
vector ${\vec x}$, in Eq.~\ref{xihyper} represent the composite {\it inverse}
operation of the series of {\it reshape-transpose} operations that occurred
during the FFT.
In the final step, we {\it reshape} $\xi_{hyper}$ into 
a one-dimensional array:
\begin{equation}
{\vec \xi} = <\!n\!> \rshp  \xi_{hyper} \equiv FFT({\vec y}).
\end{equation}
The ONF is now expressed as follows.  Define two new indices $t$ and $s$ with limits given by:
\begin{equation}
0 \leq t < \; t^\star \equiv (\pi  (( <\!d - \sigma\!> )\rshp 2)),
\end{equation}
and,
\begin{equation}
0 \leq s < \; s^\star \equiv (\pi  (  <\!\sigma \!> \rshp 2 )),
\end{equation}
we define a new two-dimensional array $\xi^{(2)}$ by reshaping the vector 
$\vec \xi$ as:
\begin{equation}
\xi^{(2)} \equiv <\!s^\star\;t^\star\!> \rshp {\vec \xi}
\end{equation}
The final result is then written:
\begin{equation}
<\!s\;t\!> \psi \xi^{(2)} = {\vec x} [s*2^{d-\sigma} + <\!t\!>].
\end{equation}

This expression was directly translated into C++ code as illustrated in 
Fig.~\ref{final_trans_code}. 

\renewcommand{\baselinestretch}{1}
\begin{figure}[ht]
\scriptsize
\begin{quotation}
\begin{tt}
\noindent

\begin{verbatim}
void final_trans(complex *datvec,complex *temp,int nmax,
                 int lcap,int csize)
{
/*      dvar = # of 2's in the hypercube
 *      lcap = # of transpose-reshapes (T-rho) that have to be undone
 */
        logc = log(float(csize))/log(2.0);
        sig = lcap*logc;
        smax = pow(float(2),sig);

        dvar = log(float(nmax))/log(2.0);
        tmax = pow(2.0,dvar-sig);

        index = 0;
        for(tind=0;tind<=tmax-1;tind++)
        {
            for(sind=0;sind<=smax-1;sind++)
            {
                temp[index] = datvec[sind*tmax + tind];
                index += 1;
            }
        }

        for(index=0;index<=nmax-1;index++)
        {
            datvec[index] = temp[index];
        }
}

\end{verbatim}

\end{tt}
\end{quotation}
\caption{\label{final_trans_code}
C++ code fragment implementing the final transpose bringing the data back
into the correct order.
}
\end{figure}

\section{Results and Discussion}

The performance results for our new algorithm are presented in 
Figs.~\ref{time_vs_size} and~\ref{enhance}.  These experiments were 
run in a single-processor, dedicated, non-shared environment on the 
IBM SP2 machine ``squall" at the Maui High-Performance Supercomputer 
center~\cite{maui:}.
Specifications for the machine are quoted in the caption to 
Fig.~\ref{time_vs_size}.

In the first figure (Fig.~\ref{time_vs_size}) we plot the time vs. input 
data length.  There are two curves, one for our new cache-optimized FFT
and one for a similar run with no cache optimization.  Direct comparisons
are possible since both curves are produced by the same code.  For the 
non-cache-optimized run, we simply chose the blocking size $c$ (specified as
a parameter at run time) to be greater than or equal to the length of the 
data vector $n$.
We see that the curves have essentially the same shape but the cache-optimized
one is shifted to the right by one power of two compared to the non-optimized 
one.  Thus for a {\it given run time}, we can legitimately claim a factor of two
speed-up.

The results presented in Fig.~\ref{enhance} emphasize the improved performance
for a {\it fixed data size} by taking the ratio of the run time for the 
non-optimized run to the cache-optimized one.  We see that for the some of 
the largest
sizes considered, a factor on the order of $4$ speedup is achieved.  

Figure~\ref{enhance} is also enlightening in that it highlights the various
levels of the memory hierarchy.  A change in slope of run-time vs. size 
indicates the crossing of a boundary between one level of the memory hierarchy
and another.  For example from the results of Fig.~\ref{enhance} we can make
the following estimates.  For roughly $2^1 \le N \le 2^6$ the speed is most 
likely dominated by the speed of the registers.  For
$2^6 \le N \le 2^8$ the speed is dominated by L1 and L2 cache and for
$2^9 \le N \le 2^{12}$ the speed corresponds to main memory.  For $N \ge 2^{13}$
paged memory (of size $4kB$) dominates.  The large jumps in performance (i.e.
factors of $4$ for the largest sizes) correspond to the presence or absence of 
page faults.

In general, the performance of the algorithm is a tradeoff 
between the increased speed obtained by having more data in the cache (with
increasing $c$) and the cost of actually moving the data around.  One might
naively guess that the best performance would be obtained by choosing $c$ to
be equal to the cache size.  However, we find, the best performance by choosing
blocking sizes $c$ given by small powers of $2$.  In other words, it is
more economical to move data around many times within the cache than it is
to move large blocks into and out of cache due to the extreme speed of the 
cache.  That is, direct access and movement of components within the cache
has no overhead.


%
%
%
\begin{figure}[ht]
\vspace{2.5in}
\small
\begin{centering}
\begin{center}
\includegraphics{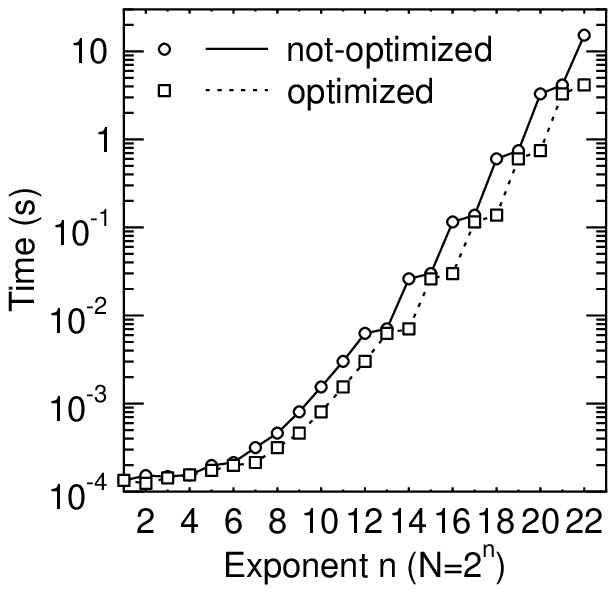}
\end{center}
\end{centering}
\caption{\label{time_vs_size} Comparison of the cache-optimized FFT with our
previous $\psi$-designed FFT~\cite{mullin.small:} indicating reproducible 
enhanced performance. The data in this figure 
represent the raw timing data obtained by running the experiments in a 
single-processor, dedicated, non-shared environment on the Maui SP2 machine 
``squall" (one of $2$, 375Mhz Nighthawk-2, IBM SP2 nodes).  
Reproducibility was demonstrated by comparison of the results of 
five separate runs which produced nearly identical results (not shown). The 
slope of the curve reveals the speed of various levels of the memory hierarchy 
as amplified in Fig.~\ref{enhance}.} 
\end{figure} 
\begin{figure}[ht]
\vspace{2.5in}
\small
\begin{centering}
\begin{center}
\includegraphics{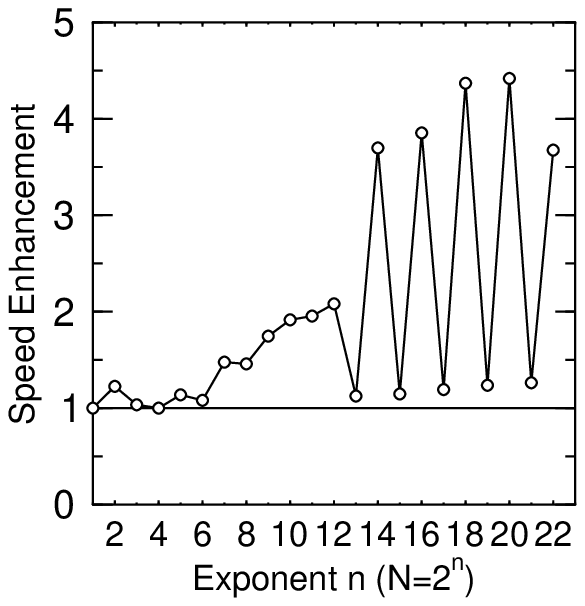}
\end{center}
\end{centering}
\caption{\label{enhance} Performance enhancement of the cache-optimized FFT 
as compared with our previous $\psi$-designed FFT showing a factor of four
enhancement for some of the largest sizes. The data plotted in this
figure is the ratio of the time for the non-optimized routine divided by that for
the optimized routine. The changing slope of the optimization curve reveals
various levels of the memory hierarchy.  For roughly $2^1 \le N \le 2^6$ 
the speed is most likely dominated by the speed of the registers.  For
$2^6 \le N \le 2^8$ the speed is dominated by L1 and L2 cache and for
$2^9 \le N \le 2^{12}$ the speed corresponds to main memory.  For $N \ge 2^{13}$
paged memory (of size $4kB$) dominates.  The jumps in performance correspond
to the presence or absence of page faults.
}
\end{figure}

\section{Conclusions}
We have presented a new algorithm for the Fast Fourier Transform that is a 
factor of $2$ to $4$ times faster than our previous records 
(that were competitive with or outperformed well-tested library routines as
shown in Chap.~\ref{intro} and Ref.~\cite{mullin.small:}).  This success was
achieved through the use of Conformal Computing techniques to devise a 
generalized partitioning scheme leading to optimized cache access.  The 
principle is very simple.  Data is periodically re-arranged so as to always
achieve locality in cache. This approach is in contrast to the traditional FFT 
in which data access becomes progressively remote (leading to cache misses 
and page faults) as the algorithm proceeds.  The key concept in the new 
algorithm is the repeated use of the {\it reshape-transpose} operation 
to move data from remote locations into the cache as needed.  A given 
one-dimensional data structure for the input vector is initially {\it reshaped} 
into a two-dimensional array of dimension $r\times c$ where $c$ is an 
arbitrary blocking size.  The blocking size $c$ is completely arbitrary and
is specified as a parameter at run time.  Based on various cost functions 
(cache speed, cost of moving data, etc.) we can predict the performance
of the algorithm vs. the blocking size $c$.  We find the best performance
for blocking sizes $c$ given by small powers of $2$.  The results presented in 
this paper are promising for further developments in terms of optimizations of 
multi-dimensional FFT's over cache in single as well as multi-processing 
environments.


%
%
%

\chapter{A Cache-Optimized Fast Fourier Transform: Part II}
\label{chap4}

\section{Chapter Summary}

This chapter explores in detail two key steps of a new cache-optimized
Fast Fourier Transform algorithm that was presented in the previous chapter
(see Chap.~\ref{part1}).
Through the use of Conformal Computing techniques, discussed herein, an
impressive performance improvement (factors of $2$ to $4$ speed-up) was
obtained.  The present chapter serves as a tutorial introduction to the
techniques of Conformal Computing: a systematic design methodology for 
hardware/software
algorithms based on a rigorous mathematical theory.  Two key aspects of the
new algorithm, the {\it reshape-transpose} and {\it final-transpose} are
explored in detail, and serve as vehicles to introduce and illustrate many
key aspects of the theory.  Although our approach is based on a rigorous
formal theory, the net result is always an efficient specification, the
{\it Operational Normal Form} (ONF) for how to build a particular algorithm in
software or hardware {\it in any programming language}.  As we explicitly
demonstrate, the ONF for each of the operations ({\it reshape-transpose} and
the {\it final-transpose}) are directly translated into computer code.

\section{Introduction}

This chapter continues the discussion of the cache-optimized FFT presented in 
the previous chapter (Part I: see Chap.~\ref{part1}) and develops the 
techniques of Conformal Computing in 
some detail.  Over the past decade, these techniques have been successfully applied
to a number of algorithms that are ubiquitous across science and
engineering disciplines, such as the Fast Fourier Transform 
(FFT)~\cite{cpc,mullin.small:,lenss3,lenss4,HuntMulRos99}, LU
decomposition~\cite{muldol94}, matrix multiplication, Time Domain convolution,
QR decomposition~\cite{mul03,MRHL03},  etc. That is to say, these and other
algorithms were first expressed algebraically using MoA then $\psi$-reduced.
These designs were realized in both hardware and
software~\cite{bms92,hardy1,hardy2,ll2,mcmahon95,manal,lee95,mul96}.

At this point we turn to a detailed look at two key steps in the new
cache-optimized FFT algorithm: the {\it reshape-transpose} and the {\em final 
transpose}.  

\section{Reshape Transpose}

\subsection{Algebraic Specification}

Recall the expression to abstract our view of the cache.  We write:

\begin{equation}
  < r\;c> \rshp \; (\transpose (<r\;c> \rshp\; (\iota n))) 
\label{abstract}
\end{equation}
to represent the {\it reshape-transpose} operation on our restructured data
array.  Reading from right to left we start with the array of indices 
$(\iota n)$
which is a one-dimensional array (vector) of $n$ sequential integers starting 
with $0$ and ending with $n-1$. In the present example we only concern 
ourselves with the manipulation of this index array.  The first reshaping of 
($\iota n$) is indicated in Eq.~\ref{abstract} by the expression 
$<r\;c> \rshp\; (\iota n) $ which implies an $r\times c$ array consisting
of the entries of $(\iota n)$ taken in sequential (i.e. {\it row-major}) 
order. For this example we require the reshaped array to contain the same
number of elements as the original array: $r\times c = n$. Using the 
notation of the $\psi$-calculus we write $\pi\! <\!r\;c\!> \; \equiv n$, where the 
operator $\pi$ acting on a vector produces a scalar equal to the product of
the elements of the vector.

In the next step in Eq.~\ref{abstract} we apply the transpose operator 
$\transpose$ to produce the $c\times r$ array 
$\transpose (<\!r\;c\!> \rshp (\iota n))$.  In the last step we re-partition
the array with the  $<\!r\;c\!> \rshp $ to produce the $r\times c$ array obtained
by taking the elements of $\transpose (<r\;c> \rshp\; (\iota n))$ sequentially
in lexical order (i.e. {\it row-major}).

\subsection{$\psi$-Reduction: Denotational Normal Form and Operational
Normal Form}
As we will see in the following, all of the operations in Eq.~\ref{abstract}
are composed to yield one final expression through the use of direct indexing.
The process of converting an expression such as that in Eq.~\ref{abstract}
into one involving only indexing operations is called $\psi$-{\it reduction}.  
The
first step is to produce the {\it Denotational Normal Form} (DNF) which 
reveals the semantic meaning of a reduced array expression such as
Eq.~\ref{abstract},  in terms of Cartesian coordinates. 

For
example, given an array A, we write: 
\begin{equation}
<i\;j> \psi A \equiv A[i,j], 
\label{psi_extract}
\end{equation}
where we've introduced the $\psi$ operator which takes an index vector
$<\!i\;j\!>$ and extracts the $i,j$-{\it th} element of the array $A$. On the right
hand side of Eq.~\ref{psi_extract} we use a common bracket notation
to denote a component of an array. For an expression involving a 
number of operations, such as the one
in Eq.~\ref{abstract}, the DNF is obtained by composing indices
using the $\psi$ operator and an index vector by applying the definitions of
the various operations ($\rshp$, $\transpose$, etc.) as will be demonstrated 
shortly.  
In essence, we view each operator ($\rshp$, $\transpose$, etc.) as
effecting a certain re-arrangement of the index vector.  Obviously such 
re-arrangements can be performed sequentially to find the re-arrangement 
corresponding to the composite operation.  The resulting expression, involving
only the starting data array (i.e. the array $(\iota n)$ in this example) and
Cartesian coordinates is, by definition, the DNF which is independent of layout.

The final step of the $\psi$-{\it reduction} process results in the 
{\it Operational Normal Form} (ONF)
which describes the underlying access patterns of the array 
operation in terms of the specific data layout\footnote{In this context,
{\em layout} means {\em row major, column major, regular sparse, etc.}} 
on a computer.
Explicitly, for the simple example of Eq.~\ref{psi_extract}, we write:
\begin{equation}
<i\;j> \psi A \equiv (\Ravel A)[\gamma_{row}(<i\;j>;(\rho A))],
\label{row_maj}
\end{equation}
if the layout is row major, and
\begin{equation}
<i\;j> \psi A \equiv (\Ravel A)[\gamma_{column}(<i\;j>;(\rho A))],
\label{col_maj}
\end{equation}
if the layout is column major.
Equations~\ref{row_maj}, and~\ref{col_maj} are examples of ONF's corresponding
to the DNF of Eq.~\ref{psi_extract}. By necessity, the ONF is dependent on 
the underlying data layout as are the operations {\it reshape} $\rshp$ 
and {\it ravel} ($\Ravel$).

In Eqs.~\ref{row_maj} and~\ref{col_maj} the operation $\Ravel$ (which we call
{\it ravel}) produces the one-dimensional vector $(\Ravel A)$ from the elements
of the array $A$.  The ordering of the elements of $(\Ravel A)$ depends on the
layout ({\it row-major} vs. {\it column-major}).  For an $r\times c$ array A in
{\it row-major} order, the first $c$ elements of $(\Ravel A)$ are the elements
of the first row, the next $c$ elements of $(\Ravel A)$ are taken from the 
elements of the next row etc.  In a similar way, if A is an $r\times c$ array, 
in {\it column-major} order, the first $r$ elements of $(\Ravel A)$ are taken from the first column of 
A, the next $r$ from the second column of $A$, etc.

In Eqs.~\ref{row_maj} and~\ref{col_maj} we have also introduced the layout 
functions, \\  $\gamma_{column}(<i\;j>;(\rho A))$, and 
$\gamma_{row}(<i\;j>;(\rho A))$ which produce scalar indices in order to
extract a single element from the one-dimensional array $(\Ravel A)$.  
In general there will be a family of layout functions for various situations.
Equations~\ref{row_maj} and~\ref{col_maj} are each, therefore, of the form
\begin{equation}
<i\;j> \psi A \equiv (\Ravel A)[index],
\label{index_rhs}
\end{equation}
where $index$ is a scalar, and we have again used the bracket notation (as is
done in $C++$) on the right hand side of Eqs.~\ref{row_maj},~\ref{col_maj} 
and~\ref{index_rhs} to denote the use of a scalar index to extract a component
of a one-dimensional array.

The indexing functions $\gamma_{row}$ and $\gamma_{column}$ take two arguments:
(1) the index vector $<i\; j>$ and (2) the {\it shape} vector $(\rho A)$.
The shape vector $(\rho A)$ is a vector consisting of the lengths of the 
various dimensions of the array.  For the present example, given an $r\times c$
array $A$, we have:
\begin{equation}
(\rho A) = <r\; c>
\end{equation}

The function $\gamma$ was given in {\bf Definition~\ref{gammadef}} and in 
all further discussions we will assume {\it row-major}
ordering and drop the $row$ subscript: $\gamma_{row}\longrightarrow \gamma$

\subsection{From the ONF to the Generic Design}
The ONF describes how to build the code independent of a particular
programming language. At this point all loop nests are revealed,
and all data flow and memory management are indicated. Each
loop nest indicates not only access patterns but levels of the processor/memory
hierarchy. We use $\forall$ ({\it for all}) as our generic loop indicator. Similarly,
bracket notation indicates { \em ``address of''} so that when indices are calculated
they are calculated relative to the address of the indicated array (i.e. the 
address of (pointer to) the first element).
All of this in conjunction with $\gamma$ gives us all essential information
to build the design in any programming language at both the hardware and
software levels.
For example, $\forall\;\; i,j \;\; s.t.\;\; 0 \leq i < 2,\;\;0 \leq j<4$
\begin{eqnarray}
( \Ravel A)[\gamma(<i\;j>\;;\;<2\;4>] \equiv  ( \Ravel A)[j+(4 \times i)]
\end{eqnarray}
denotes a generic form with two loops with
$@A+j+(4 \times i)$ in the body. At this point a mechanization to
any language is possible. Similarly, loops indicating message passing,
shared memory access can easily be instantiated by whatever libraries
support that loop level.

\subsection{$\psi$-Reduction of the Reshape-Transpose Operation}

We now return to the problem of $\psi$-reducing the {\it reshape-transpose}
operation of Eq.~\ref{abstract}.  To illustrate this example we choose
$n = 32$, $r = 8$ and  $c = 4$, as was done in earlier discussions.

\subsubsection{Step I: Determine Shape}
The first step in the $\psi$-{\it reduction} process is the determination of
the shape vector which, for this example, is given by:
\begin{equation}
(\rho A) \equiv <r\;c> = <8\; 4>.
\end{equation}
This step is crucial in order to enforce the use of valid index vectors.
In order for an index vector $<i\; j>$ to be valid, its components must not
exceed the lengths of the corresponding dimensions of the shape vector.
Specifically we say that a valid index vector satisfies:
\begin{equation}
0 \leq^* \;\;<\!i\;j\!>\;\; <^* \;\; <\!r\; c\!>,
\label{comparison}
\end{equation}
where the notation $\leq^*$ and $<^*$ implies component-wise comparison of the
two vectors $<i\; j>$ and $<r\; c>$.

\subsubsection{Step II: Perform Psi-Reduction and Reduce to Normal Form(s)}
We begin by taking Eq.~\ref{abstract} apart using the $\psi$ operator 
and an index vector $<\!i\;j\!>$. To simplify the notation we define 
the quantity within the innermost set of parentheses in Eq.~\ref{abstract}
as $A \equiv \; <\!r\;c\!> \rshp\; (\iota n)$.
By applying the definition for {\em reshape}:
$\forall\;\; i,j \;\;\mbox{s.t.}\; 0 \leq^* \;\;<\!i\;j\!>\;\; <^* \;\; <\!r\;c\!>$,
we get:
\begin{eqnarray}
<i\;j> \psi   ( <r\;c>  \rshp (\transpose A))  &\equiv&  (\Ravel  (\transpose A)) [ \gamma (<i\;j>\; ; \; <r\;c>)\; \mbox{mod}\; (\pi (<\!r\;c\!>))] 
\nonumber \\
& \equiv & (\Ravel (\transpose A))[ \gamma (<i\;j>\; ; \; <r\;c>) ] \nonumber\\
& \equiv & (\Ravel (\transpose A))[ j+(i \times c) ] 
\label{reduce}
\end{eqnarray}
In Eq.~\ref{reduce} we are applying the definition of the {\it reshape} operator
to the object $\Ravel(\transpose A)$ which is defined in terms of the 
$\gamma (<i\;j>\; ; \; <r\;c>)$ function.  In the first line of 
Eq.~\ref{reduce} the expression: $\mbox{mod}(\pi (<\!r\;c\!>))$, is part of 
the definition which handles the case in which one wants to reshape a smaller
array into a larger array from the elements of the smaller array (repeated 
appropriately).  In this case, however, the total number of components in
the reshaped array is the same, allowing us to drop the expression,
$\mbox{mod} (\pi (<\!r\;c\!>))$, in the second line of Eq.~\ref{reduce}.
In the third line of Eq.~\ref{reduce} we have inserted the explicit expression
for $\gamma$ from its definition (see {\bf Definition~\ref{gammadef}}).

Next we wish to get rid of the {\em transposed} array $\transpose A$
in favor of the original array $A$. We consider selecting an element of 
$\transpose A$ using an index $<\!i^\prime\;j^\prime\!>$ as follows:
\begin{equation}
\forall\;\; i^\prime,j^\prime \;\;\mbox{s.t.}\;\; 0 \leq^* 
\;\;<\!i^\prime\;j^\prime\!>\;\; 
<^* \;\; <\!c\;r\!>   \\
\label{reversed}
\end{equation}
\begin{equation}
<\!i^\prime\;j^\prime\!> \psi   (\transpose A)
 \equiv  (\Ravel  (\transpose A))[\gamma (<\!i^\prime\;j^\prime\!>;<\!c\;r\!>)]
= (\Ravel (\transpose A)) [j^\prime + i^\prime * r].
\label{trans}
\end{equation}
In order for this expression to agree with Eq.~\ref{reduce} we must equate
the arguments in square brackets in Eqs.~\ref{reduce} and~\ref{trans} as:
\begin{equation}
j^\prime + i^\prime * r = j + i*c.
\end{equation}
From this we find the the primed indices in terms of the unprimed indices 
(assuming $r > c$ for this example) as:
\begin{equation}
i^\prime = \mbox{int} (c*i/r),
\label{iprime}
\end{equation}
and,
\begin{equation}
j^\prime = (j + i*c)\; \mbox{mod}\;  r.
\label{jprime}
\end{equation}

Next we use the definition of {\em transpose} to simplify 
Eq.~\ref{trans}: 
\begin{equation}
<\!i^\prime\;j^\prime\!> \psi (\transpose A) \equiv
<\!j^\prime\;i^\prime\!> \psi A = (\Ravel A)[i^\prime + j^\prime * c].
\label{transsimp}
\end{equation}
Thus by using Eqs.~\ref{iprime} and~\ref{jprime} in Eq.~\ref{transsimp}
we finally express Eq.~\ref{reduce} completely in terms of $A$. Considerable
further simplification is possible, however, as is discussed in the next 
section.

\subsubsection{Step III: further Simplification}
The formulation just presented, while formally correct, is computationally 
inefficient.  In particular, we wish to avoid the 
operations {\bf int} and ${\bf mod}$ in Eqs.~\ref{iprime} 
and~\ref{jprime}.  Instead,
{\em \bf if we need to work with both the primed and un-primed indices} we 
can use the loop structure (expressed in C++ syntax assuming $r > c$ for this 
example) 
given in Fig.~\ref{loopindex}:
\begin{figure}
\begin{verbatim}

i = 0;
ratio = r/c;
for(iprime=0;iprime < c; iprime++)
{
   for(k=0; k < ratio; k++)
   {
      for(j=0; j < c; j++)
      {
         jprime = j + k*c;
      }
      i = i + 1;
   }
} 

\end{verbatim}
\caption{ Efficient loop structure to produce both sets of indices $<\!i\;j\!>$
and $<\!i^\prime\;j^\prime\!>$ assuming $r > c$. A similar loop structure
is easily constructed for the case $c > r$.}
\label{loopindex}
\end{figure}

{\em \bf In our applications, further simplification occurs}. 
Note, in Eq.~\ref{reduce}
a single element is extracted from the transposed array $\transpose A$.  In 
this case, as in many situations, we can compute the entire $\transpose A$
and select elements from 
it.\footnote{Bear in mind: in our cache-optimized FFT we {\em actually 
materialize} $\transpose A$ so as to achieve data locality.  As such it is 
most efficient to compute $\transpose A$ all at once.}  
This is most easily computed by looping over the 
values of the index $<\!i^\prime\;j^\prime\!>$ in the rightmost expression
in Eq.~\ref{transsimp}.

Thus, to construct the array
\begin{equation}
(<r\;c>  \rshp (\transpose A)),  
\end{equation}
for example,
we only need the elements of the corresponding one-dimensional array:
\begin{equation}
\Ravel (<r\;c>  \rshp (\transpose A)) \equiv \Ravel \transpose A,
\end{equation}
constructed from Eq.~\ref{transsimp}

This computation was implemented in C++ as illustrated in 
Fig.~\ref{trans_rshp_code4}.  
Note in Fig.~\ref{trans_rshp_code4} the two {\tt for}-loops correspond to the 
bounds as indicated in Eq.~\ref{reversed} upon making the substitutions:
\begin{equation}
i^\prime \rightarrow j \mbox{    (variable {\tt jind} in the code)}
\label{isub}
\end{equation}
and, 
\begin{equation}
j^\prime \rightarrow i \mbox{    (variable {\tt iind} in the code)}
\label{jsub}
\end{equation}
This substitution is made to simplify the notation and is 
allowed because: (1) the bounds on $<\!j^\prime\;i^\prime\!>$ 
(see Eq.~\ref{reversed}) coincide
with those on $<\!i\;j\!>$ (see Eq.~\ref{comparison}) and,
(2) we are constructing the entire array $\transpose A$ and we need not keep
track of the explicit relationship between the primed and unprimed indices
indicated in Eqs.~\ref{iprime} and~\ref{jprime}.

Note also that the argument of 
{\tt datvec} on the right hand side of the assignment 
{\tt temp[index] = datvec[jind+csize*iind];} in Fig.~\ref{trans_rshp_code4}  
is a direct translation of the argument that appears
in Eq.~\ref{transsimp} upon carrying out the substitutions of 
Eqs.~\ref{isub} and~\ref{jsub}.

\renewcommand{\baselinestretch}{1}
\begin{center}
\begin{figure}[ht]
\scriptsize
\begin{quotation}
\begin{tt}
\noindent

\begin{verbatim}
void trans_rshp(complex *datvec,complex *temp,int nmax,
                int csize)
{
        int iind,jind,kind,kmax,jmax,imax,index,c2size;
        int rows,arg;
        int max(int a,int b);

//      The routine carries out the transpose-reshape operation
        index=0;

        rows = nmax/csize;
        imax = rows-1;

        c2size = int(pow(csize,2.0));
        jmax = max(0,nmax/c2size-1);


        for(jind=0;jind<=jmax;jind++)
        {
            for(iind=0;iind<=imax;iind++)
            {

                temp[index] = datvec[jind+csize*iind];
                index += 1;
            }
        }

        for(iind=0;iind<=nmax-1;iind++)
        {
            datvec[iind] = temp[iind];
        }
}
\end{verbatim}

\end{tt}
\end{quotation}
\caption{\label{trans_rshp_code4}
C++ code fragment implementing the reshape-transpose in terms of 
index manipulations.
}
\end{figure}
\end{center}

\section{Reordering the Data}

After the last step of the FFT the data will not be in the correct order and
we must do something to return it to its initial order (i.e. prior to 
any {\it reshape-transpose} operations).  The simplest approach would be 
to rearrange the data by applying a series of inverse {\it reshape-transpose} 
operations.  {\bf There is, however, a far more efficient approach in which
no data needs to be moved}.  In other words, we use Conformal Computing 
techniques to determine the index vector which will select the correct
components of the array.

\subsection{Final Transpose}
We now seek to answer the following general question: after a single
{\it reshape-transpose} operation, where is the data? More precisely we 
are interested in the re-arrangement of the corresponding index vector.
The answer to this question was revealed to the authors upon viewing the
data vector as an $\log_2(n)$-dimensional {\it hyper-cube}.  
The hyper-cube is formed by reshaping the data into an 
$2\times 2\times \ldots 2$ array where the number of $2$-s is given
by $\log_2(n)$. An element of the hyper-cube is then determined by specifying
an index vector of length $\log_2(n)$ consisting of ones and zeros.  There 
is, therefore, an isomorphism between the index vector of an element and the
binary number corresponding to the index of the original one-dimensional 
array.

For example, suppose $n = 8$, then the original data vector would be:
\begin{equation}
A = \iota(8) \equiv <0\; 1\; 2\; 3\; 4\; 5\; 6\; 7>.
\end{equation}
Next we re-shape this vector into a hyper-cube by performing the following
operation
\begin{equation}
A_{hyper} \equiv (3 \rshp 2) \rshp A.
\label{a_reshape}
\end{equation}
The operation in parentheses yields 
\begin{equation}
(3 \rshp 2) \equiv <\!2\;2\;2\!>, 
\label{two_vec}
\end{equation}
which is
the argument to the second {\it reshape} operator.  Thus more explicitly,
Eq.~\ref{a_reshape} is written as:
\begin{equation}
A_{hyper} \equiv <\!2\;2\;2\!> \rshp A,
\end{equation}
which is a $2\times 2 \times 2$ array constructed by taking the elements of
$A$ in increasing order.

Incidentally, the operation of Eq.~\ref{two_vec} is an example in which a 
smaller array (in this case the scalar $2$) is {\it reshaped} into the larger
array $<\!2\;2\;2\!>$.  Such a situation was anticipated in the definition
of the {\it reshape} operator by the presence of the {\it modulo} operation
(e.g the factor $\mbox{mod} (\pi (r\times c))$ in Eq.~\ref{reduce}), as was 
discussed earlier in the context of Eq.~\ref{reduce}.

The elements of the hyper-cube $A_{hyper}$ are now selected by specifying 
an index of zeros and ones corresponding to the binary representation of 
the original array.  For example:
\begin{equation}
<\!0\;1\;0\!> \psi A_{hyper} = A[3] = 3,
\end{equation}
\begin{equation}
<\!1\;0\;1\!> \psi A_{hyper} = A[5] = 5,
\end{equation}
etc., where we have again used the bracket notation to denote selection of
elements of one-dimensional arrays.

Unfortunately, there is no truly satisfactory way to visualize the hyper-cube
construction.  However, for the purpose of discussion, we have adopted the 
following convention.  We begin by grouping four elements together at a time
and write them as $2\times 2$ matrices surrounded by square brackets.
For example suppose we have a $4$-dimensional hyper-cube $B$
defined by
\begin{equation}
B \equiv (4 \rshp 2) \rshp \iota (16)
\end{equation}
the first four elements of $B$ would be written as:
\begin{equation}
<\!0\;0\!> \psi B = \left [
\begin{array}{lr}
0&1\\
2&3\\
\end{array}
\right ],
\end{equation}
where we have used a partial index $<\!0\;0\!>$ to select the first four 
elements of $B$: 
$<\!0\;0\;0\;0\!> \psi B$, $<\!0\;0\;0\;1\!> \psi B$, $<\!0\;0\;1\;0\!> \psi B$,
$<\!0\;0\;1\;1\!> \psi B$ 
and we have written them as a two-dimensional array (assuming {\it row-major}
order).
the next four would be grouped together as:
\begin{equation}
<\!0\;1\!> \psi B = \left [
\begin{array}{lr}
4&5\\
6&7\\
\end{array}
\right ],
\end{equation}
followed by
\begin{equation}
<\!1\;0\!> \psi B = \left [
\begin{array}{lr}
8&9\\
10&11\\
\end{array}
\right ],
\end{equation}
and lastly:
\begin{equation}
<\!1\;1\!> \psi B = \left [
\begin{array}{lr}
12&13\\
14&15\\
\end{array}
\right ].
\end{equation}
The entire $2\times 2\times 2\times 2$ array $B$ can thus be visualized by
arranging the four $2\times 2$ blocks as a $2\times 2$ array of $2\times 2$
arrays and enclosing them in parentheses as:
\begin{equation}
B = \left [
\begin{array}{lr}
<\!0\;0\!> \psi B  &  <\!0\;1\!> \psi B\\
<\!1\;0\!> \psi B  &  <\!1\;1\!> \psi B\\
\end{array}
\right ].
\end{equation}

The generalization to higher dimensions is straightforward. In the present 
example we are assuming the hyper-cube $B$ to be of even dimension (i.e. $4$).
For an odd-dimensional hyper-cube, we treat the final index (on the left) as
defining a column index.

For example, consider the array $x^{(d)}$ defined 
to be the last step in the FFT for the $n = 32$ example given in 
Eq.~\ref{xdeqn} in Chap.~\ref{part1}.

\[\xi  \equiv ((\log_2 n) \rshp  2) \rshp x^{(c)} \]

\begin{equation}
\equiv
\left [
\begin{array}{lr}

\left [
\begin{array}{lr}

\left [

\begin{array}{lr}
0&16\\
1&17\\
\end{array}
\right ]
&
\left[
\begin{array}{lr}
2&18\\
3&19\\
\end{array}
\right ]      \\

\left [
\begin{array}{lr}
4&20\\
5&21\\
\end{array}
\right ] &
\left [
\begin{array}{lr}
6&22\\
7&23\\
\end{array}
\right ]

\end{array}
\right ]
&
\left [
\begin{array}{lr}

\left [
\begin{array}{lr}
8&24\\
9&25\\
\end{array}
\right ] &
\left[
\begin{array}{lr}
10&26\\
11&27\\
\end{array}
\right ]      \\
\left [
\begin{array}{lr}
12&28\\
13&29\\
\end{array}
\right ] &
\left [
\begin{array}{lr}
14&30\\
15&31\\
\end{array}
\right ]
\end{array}
\right ]
\end{array}
\right ]
\label{final_xi}
\end{equation}

Our task in the next section is to correctly identify an index vector
which rearranges Eq.~\ref{final_xi} into the correct order (i.e. ascending
integers starting with $0$) {\bf without actually moving any data}.

\subsection{\label{final_trans} General Rule}
By studying the patterns induced by the {\it reshape-transpose} operation
on arrays arranged as hyper-cubes (such as that in Eq.~\ref{final_xi}), the 
authors have discovered a rule for
returning a hyper-cube of any dimension (e.g. an input data vector for an FFT of
any length) to its correct form in one step through the use of 
indexing, without the need to actually rearrange the data.  We present and 
discuss this rule in this section.  The rule is simply stated as follows.  

{\bf Given an array $A$, that has been rearranged with the reshape-transpose
operator to give $B$, we select an element of $B$ with a
binary index $\vec p$ of the array formulated as hyper-cube. The element, so 
selected, is obtained from the array $A$ (also formulated as a hyper-cube) 
with the index $\vec q$ where $\vec q$
is simply a cyclic permutation of $\vec p$}.

At this point, we present an example to clarify the situation.  Consider
a $4\times 4$ array of integers starting with $0$ and ending with $15$ which is
written as
\begin{equation}
A \equiv <\!4\;4\!> \rshp \iota (16) = \left [
\begin{array}{cccc}
0&1&2&3\\
4&5&6&7\\
8&9&10&11\\
12&13&14&15\\
\end{array}
\right ].
\end{equation}
Now define the array $B$ obtained through a {\it reshape-transpose} operation
acting on $A$:
\begin{equation}
B \equiv <\!4\;4\!> \rshp (\transpose A) = \left [
\begin{array}{cccc}
0&4&8&12\\
1&5&9&13\\
2&6&10&14\\
3&7&11&15\\
\end{array}
\right ].
\end{equation}
Incidentally, for the special case of a square array (as we are considering 
here) the {\it reshape-transpose} operation is equivalent to the 
{\it transpose} operation.

Let us now reshape the arrays as hyper-cubes, that is:
\begin{equation}
A_{hyper} \equiv (4 \rshp 2) \rshp A = <\!2\;2\;2\;2\!> \rshp A,
\label{a_hyper}
\end{equation}
and,
\begin{equation}
B_{hyper} \equiv (4 \rshp 2) \rshp B = <\!2\;2\;2\;2\!> \rshp B,
\label{b_hyper}
\end{equation}

Now if we consider the relationship between the elements of $A$ and $B$ viewed
as hyper-cubes we find the following behavior:
\begin{equation}
<\!i\;j\;k\;l\!> \psi B_{hyper} = <\!k\;l\;i\;j\!> \psi A_{hyper}. 
\end{equation}
For example:
\begin{equation}
<\!0\;0\;1\;0\!> \psi B_{hyper} = <\!1\;0\;0\;0\!> \psi A_{hyper} = 8, 
\end{equation}
\begin{equation}
<\!0\;1\;1\;0\!> \psi B_{hyper} = <\!1\;0\;0\;1\!> \psi A_{hyper} = 6, 
\end{equation}
\begin{equation}
<\!1\;1\;0\;1\!> \psi B_{hyper} = <\!0\;1\;1\;1\!> \psi A_{hyper} = 7. 
\end{equation}

The general rule is as follows.  Suppose there are $d = \log_2 (n)$, $2$'s in 
the hyper-cube representation of an $r\times c$ array (with $n = r\times c$).
Upon transforming $A$ into $B$ via the {\it reshape-transpose} operation,
\begin{equation}
B \equiv (<\!r\;c\!>) \rshp (\transpose A),
\end{equation}
we find the following relation between the hyper-cube 
representations of $A$ and $B$: 
\begin{equation}
{\vec p} \psi B_{hyper} = {\vec q} \psi A_{hyper}, 
\label{hyper_general}
\end{equation}
where the relationship between indices $p$ and $q$ will be discussed shortly.
In Eq.~\ref{hyper_general} $A_{hyper}$ and $B_{hyper}$ are defined by:
\begin{equation}
A_{hyper} \equiv (\log_2 (n) \rshp 2) \rshp A, 
\end{equation}
and,
\begin{equation}
B_{hyper} \equiv (\log_2 (n) \rshp 2) \rshp B. 
\end{equation}

To complete the specification of Eq.~\ref{hyper_general} we need to determine
the index $q$ in terms of $p$.  {\bf In general, $q$ is a cyclic permutation
of $p$ in which $\log_2(c)$ elements of $p$ are sequentially removed from the
left and placed at the right.}  For example, if $d = \log_2(32) = 5$,  and
$\log_2(c) = 2$, for 
\begin{equation}
{\vec p} \equiv <\!i\;j\;k\;l\;m\!>,
\end{equation}
indexing the array $B_{hyper}$, the corresponding index of $A_{hyper}$
is given by:
\begin{equation}
{\vec q} \equiv <\!k\;l\;m\;i\;j\!>.
\label{q_eq}
\end{equation}
Using the formalism of the $\psi$-calculus, we rewrite Eq.~\ref{q_eq} as
\begin{equation}
{\vec q}  = <\!k\;l\;m\;i\;j\!> \equiv (2\; \phi <\!i\;j\;k\;l\;m\!> ) = {\vec p},
\end{equation}
where we have introduced the {\it rotate} operation $\phi$.  Naturally we 
are also interested in the inverse operation which is written as
\begin{equation}
{\vec p} =  <\!i\;j\;k\;l\;m\!>  \equiv  (-2\; \phi <\!k\;l\;m\;i\;j\!>)  = {\vec q}
\end{equation}

Now that we understand the relationship between $A_{hyper}$ and $B_{hyper}$
we are in a position to specify how $B$ can be re-ordered to correspond to
the ordering of $A_{hyper}$.  In order to do that we invoke the definition of 
the generalized {transpose} operation $\transpose$.  

Up to this point we have only used $\transpose$ in its traditional manner for
two-dimensional arrays.  For an $d$-dimensional array, such as $A_{hyper}$ or
$B_{hyper}$ the transpose operation invokes a permutation of the dimensions
which is specified by a permutation vector as its left argument (by convention
no left argument is required for two-dimensional arrays).
Thus, because we found the following relationship between the components of 
$A_{hyper}$ and $B_{hyper}$
\begin{equation}
 <\!i\;j\;k\;l\;m\!>  \psi B_{hyper} = <\!k\;l\;m\;i\;j\!> \psi A_{hyper},
\end{equation}
we say that $B_{hyper}$ is related to $A_{hyper}$ through the following
generalized transpose:
\begin{equation}
B_{hyper} = <\!2\;3\;4\;0\;1\!> \transpose A_{hyper},  
\label{b_vs_a}
\end{equation}
and invoking the inverse of this transpose, we can write
\begin{equation}
A_{hyper} = <\!3\;4\;0\;1\;2\!> \transpose B_{hyper}.  
\label{a_vs_b}
\end{equation}
In anticipating further developments, we use the rotate operator to write
the index vectors more abstractly in terms of the $\iota$ operation.  
Specifically, Eqs.~\ref{b_vs_a} and~\ref{a_vs_b}, respectively become:
\begin{equation}
B_{hyper} = ((2)\; \phi \; (\iota 5)) \transpose A_{hyper}
\label{b_phi}
\end{equation}
and,
\begin{equation}
A_{hyper} = ((-2)\; \phi \; (\iota 5)) \transpose B_{hyper}
\label{a_phi}
\end{equation}
The right sides of Eqs.~\ref{b_vs_a} and~\ref{a_vs_b} can be further 
abstracted by substituting the definitions of $A_{hyper}$ and $B_{hyper}$
from Eqs.~\ref{a_hyper} and~\ref{b_hyper} to yield
\begin{equation}
B_{hyper} = ((2)\; \phi \; (\iota 5)) \transpose (5 \rshp 2) \rshp A
\end{equation}
and,
\begin{equation}
A_{hyper} = ((-2)\; \phi \; (\iota 5)) \transpose (5 \rshp 2) \rshp B
\end{equation}

We now consider a general $r\times c$ array $A$ having 
$n = r\times c$ components.  Now define the array $B$ to be that which is 
obtained from $A$ through the application of $l_{max}$ {\it reshape-transpose} 
operations:
\begin{equation}
B \equiv  (<\!r\;c\!> \rshp \transpose)^{l_{max}} A
\end{equation}
we interpret the operator 
$(<\!r\;c\!> \rshp \transpose)^{l_{max}}$ to mean the operation
$(<\!r\;c\!> \rshp \transpose)$ carried out $l_{max}$ times.
Based on the principles introduced so far, the elements of $B$ are related to
the elements of $A$ through the relation:
\begin{equation}
B_{hyper} = ((\sigma)\; \phi \; (\iota d)) \transpose (d \rshp 2) 
\rshp A,
\label{lprod}
\end{equation}
where $\sigma = l_{max} \log_2(c)$ and $d = \log_2 (n)$.

For example, the hyper-cube $\xi$ considered previously in Eq.~\ref{final_xi}
corresponds to Eq.~\ref{lprod} with $c = 4$, $n = 32$, $l_{max} = 2$, $d = 5$,
$\sigma = 4$, $A_{hyper} = (5 \rshp 2) \rshp A$ where $A = 
<\!8\;4\!> \rshp (\iota 32)$

\subsection{\label{final_trans_psi} $\psi$-Reduction}

Building on the developments of the previous section, we now show how to 
effect the final rearrangement of the FFT.  In other words, we wish to 
find the inversion of an equation of the form given in Eq.~\ref{lprod}.
This is easily accomplished by simply {\it changing the sign} of the 
variable $\sigma$ as was demonstrated in Eqs.~\ref{b_phi} and~\ref{a_phi},
a fact which nicely underscores the power of the Conformal Computing approach.

The input data vector $\vec y$ of length $n$ is carried into the 
vector ${\vec x}$ through the $d = log_2 (n)$ steps of the FFT.  In the 
process, $l_{max}$ {\it reshape-transpose} operations have been carried out. 
The resulting vector $\vec x$ is thus not in the correct order (as a result
of the multiple {\it reshape-transpose} operations) and must therefore be 
rearranged into its final form ${\vec \xi}$.
We now obtain $\xi_{hyper}$ as
\begin{equation}
\xi_{hyper} = ((-\sigma)\; \phi\; (\iota d)) \transpose ((<\!d\!> \rshp 2) \rshp {\vec x})
\label{xihyper}
\end{equation}
where $\sigma = l_{max} \log_2 (c)$ and $d = log_2 (n)$.  The operators acting on 
the vector ${\vec x}$, in Eq.~\ref{xihyper} represent the composite {\it 
inverse}
operation of the series of {\it reshape-transpose} operations that occurred
during the FFT.
The final step is 
obtained by {\it reshaping} $\xi_{hyper}$ into the final one-dimensional array:
\begin{equation}
{\vec \xi} = <\!n\!> \rshp  \xi_{hyper} \equiv FFT({\vec x}).
\label{xidef}
\end{equation}

\noindent
The derivation to {\em normal  form} follows:

\subsubsection{Step I: Determine Shape}
We now wish to carry out the process of $\psi$-reduction of an expression 
of the form:
\begin{equation}
((-\sigma)\; \phi\; (\iota d)) \transpose ((<\!d\!> \rshp 2) \rshp {\vec x})
\label{xreduce}
\end{equation}
The first step is to find the {\it shape} ($\rho$) in order to enforce the 
use of valid indices.  The shape is given by:
\begin{eqnarray}
\Rho ((-\sigma) \; \phi  \; (\iota \; d)) \transpose  (( <\!d\!> \; \rshp  2) \rshp \vec x)  & \equiv & ( <\!d\!> \rshp 2)
\label{two_shape}
\end{eqnarray}
which shows that permuting the elements of the shape vector does not change the shape of a 
{\it 2-cube} (i.e. a $d$-dimensional {\it hyper-cube}. This is
true in general for a d-dimensional n-cube).  More explicitly, the 
vector of Eq.~\ref{two_shape} is a vector
consisting of $d$ entries each given by the integer $2$.  Such a shape implies
that the expression in Eq.~\ref{xreduce} is a $d$-dimensional array, the 
length of each dimension being precisely $2$.  In other words it is a 
$d$-dimensional {\it hyper-cube}.
\subsubsection{Step II: Perform Psi-Reduction and Reduce to Normal Form(s)}

For an index ${\vec i}$ to be valid it must satisfy:
\begin{equation}
0 \leq^* \vec i <^* (<\!d\!> \rshp 2),
\label{valid_index}
\end{equation}
which states that all of the components of ${\vec i}$ must be less than the
corresponding components of the shape vector (i.e. the index vector must 
consist of $0$'s and $1$'s).

We now use a valid index vector ${\vec i}$ to select an element of the 
expression given in Eq.~\ref{xreduce}.  By doing so we can apply the definition
of the {\it transpose} (which is defined in terms of the corresponding 
rearrangement of the index vector). For the moment we simplify the notation
with the definition: 
\begin{equation}
\eta \equiv (<\!d\!>\; \rshp \; 2) \rshp {\vec x}.
\label{eta_def}
\end{equation}
We thus obtain:
\begin{equation}
{\vec i} \psi (((-\sigma)\; \phi\; (\iota d)) \transpose \eta )
= ((-\sigma)\; \phi\; \vec i)) \psi \eta
\label{first_step}
\end{equation}
which shows that selecting an element of the transposed array with an index
$\vec i$ is the same thing as selecting an element from the non-transposed array
$\eta$ using an index $((-\sigma)\; \phi\; \vec i)) $ that has been permuted.
This is, in essence, the definition of the generalized {\it transpose} 
operation.

Now we further reduce the form of the index vector as:
\begin{equation}
((-\sigma)\; \phi\; \vec i))  \equiv
(((-\sigma) \take {\vec i}) \cat ((-\sigma) \drop \vec i)).
\label{index}
\end{equation}
which explicitly denotes the way in which $((-\sigma)\; \phi\; \vec i))$  
is built from two fragments of $\vec i$. We have introduced the operations
{\it take} $\take$ and {\it drop} $\drop$ which select sub-vectors from $\vec i$.
Specifically, $((-\sigma) \take {\vec i})$ forms a vector from the {\it last}
(i.e. rightmost) $\sigma$ elements of $\vec i$ and $((-\sigma) 
\drop {\vec i})$ is the
vector which remains after dropping the {\it last} $\sigma$ elements of 
$\vec i$.  In the expression on the right hand side of Eq.~\ref{index} we have 
introduced the 
operation {\it cat} $\cat$ which concatenates the two fragments together.
Thus Eq.~\ref{first_step} becomes:
\begin{equation}
{\vec i} \psi (((-\sigma) \phi (\iota d)) \transpose \eta )
= (((-\sigma) \take {\vec i}) \cat ((-\sigma) \drop \vec i)) \psi \eta.
\label{second_step}
\end{equation}

To make further progress, we now rely on the concept of a partial index.  For 
example, suppose we have a three dimensional array $\cal Z$. An individual
component is specified with a valid {\it full} index containing three elements 
such as:
\begin{equation} 
<\!k\;l\;m\!> \psi \cal Z.
\end{equation}
However, we can also extract sub-arrays of $\cal Z$ by using partial indices.
Thus:
\begin{equation} 
<\!k\!> \psi \cal Z,
\end{equation}
is a two-dimensional sub-array, while 
\begin{equation} 
<\!k\;l\!> \psi \cal Z,
\end{equation}
is a one-dimensional array.  Therefore, the result of selection with a 
{\it full}
index can be written as a composition, such as:
\begin{equation} 
<\!k\;l\;m\!> \psi {\cal Z} \equiv <\!l\;m\!> \psi (<\!k\!> \psi {\cal Z})
\equiv <\!m\!> \psi (<\!k\;l\!> \psi \cal Z).
\end{equation}
Thus the right hand side of Eq.~\ref{second_step} becomes:

\begin{equation}
(((- \sigma ) \take \vec i) \cat (( -\sigma ) \drop  \vec i)   )
\psi \eta)
\equiv (( -\sigma ) \drop  \vec i) \psi
(((- \sigma ) \take \vec i) \psi \eta)
\label{third_step}
\end{equation}
The next step is to apply the {\it Psi Correspondence Theorem}, which specifies
the manner in which to construct indexing functions (such as the $\gamma$'s 
introduced earlier) from arbitrary {\it full} or {\it partial} index vectors.

\subsection{The $\psi$ Correspondence Theorem (PCT)}
\label{psicorrespond}

We now state the $\psi$ {\it Correspondence Theorem} algebraically and then 
pause to take it apart piece by piece.

\begin{definition}
Given an array $\xi$ with shape $\rho \xi$, 
$\forall {\vec j}$ 
\hbox{ s.t. } 
$(\tau {\vec j}) < (\delta \xi)$,
and
$0 \le^* {\vec j} <^* ((\tau {\vec j}) \take (\rho \xi))$ 

\begin{equation}
\Ravel ({\vec j} \psi \xi) \equiv
(\Ravel \xi) [\gamma ({\vec j};((\tau {\vec j}) \take (\rho \xi)))*\pi ((\tau 
{\vec j}) \drop (\rho \xi)) + \iota (\pi ((\tau {\vec j}) \drop (\rho \xi)))
].
\label{psi_corr}
\end{equation}
\end{definition}
The operators {\it tau} ($\tau$), and {\it delta} ($\delta$) in 
Eq.~\ref{psi_corr} give the number of elements of a vector, and the 
dimensionality of an array, respectively. 

Consider first, the left hand side of Eq.~\ref{psi_corr}.  The expression
$({\vec j} \psi \xi)$ is a sub-array of $\xi$ obtained using the (partial or 
full) index
${\vec j}$.  For example, suppose $\xi$ is a three dimensional array with 
shape $\rho \xi = <\!4\;6\;7\!>$, and ${\vec j} = <\!2\!>$, then we have
the two dimensional array:
${\vec j} \psi \xi = <\!2\;m\;n\!> \psi \xi$ for valid indices
$0 \le  m  <  (\rho \xi) [1]$, and  
$0 \le  n  < \; \; (\rho \xi) [2]$.
In this example $(\rho \xi) [1] = 6$ and $(\rho \xi) [2] = 7$.  
For this example, we have four such arrays since $(\rho \xi)[0] = 4$.

Generally the convenient bracket notation for a one-dimensional vector $A$ 
has the 
following equivalence:
\begin{equation}
A[r] \equiv <\!r\!> \psi A.
\end{equation}
for some scalar $0 \le r < (\rho A)[0]$.  Often we use standard vector 
notation $(e.g. \; \;{\vec A})$ for emphasis.  We stress, however, that the 
consistent structure of the algebraic system is designed in such a way that
scalars and vectors are simply {\it zero-dimensional} and {\it one-dimensional}
arrays and as such don't require any special symbols.  For convenience, however
we often use such symbols as $r$ to indicate a scalar and $\Theta \equiv
<>$ to denote the empty vector.  The existence of the empty vector is necessary
in order that the scalar $r$ has the correct shape $\rho (r) = <>$.
The existence of the empty vector may seem strange to some readers.  Its use,
however, is necessary in order to have a consistent algebra. The need for the
empty vector is analogous to the need for the {\it zero} $0$ in the set of 
integers and the empty set in set theory.  

Note a related aspect of the theory is often misunderstood by newcomers: a 
vector $<\!r\!>$ is NOT equivalent to a 
$1\times r$ array (i.e. a {\it row-vector} in traditional matrix theory)
NOR is it equivalent to a $r \times 1$ array (i.e. a {\it column-vector}
in traditional matrix theory).  In the present theory, these three objects
each have different shapes, namely, $<\!r\!>$, $<\!1\;r\!>$ and 
$<\!r\; 1\!>$ respectively.  In traditional matrix theory, the distinction
plays no essential role.

Continuing our analysis of Eq.~\ref{psi_corr}, on the left hand side, the 
operation {\it ravel} ($\Ravel$) takes
its argument, the sub-array ${\vec j} \psi \xi$ and flattens it into the 
one-dimensional array $\Ravel ({\vec j} \psi \xi)$.  
On the right hand side of Eq.~\ref{psi_corr} the expression is written in the
form $(\Ravel \xi)[r + {\vec a}\;]$ where $r$ is a scalar offset
which is added component-wise to the vector $\vec a$ to produce the vector 
index ${\vec b} = r + {\vec a}$.  The components of the index vector
$\vec b$ are integers which select the components of $\xi$ in lexical order.
Explicitly the scalar $r$ and vector ${\vec a}$ in Eq.~\ref{psi_corr}
are given by:
\begin{equation}
r \equiv \gamma ({\vec j};((\tau {\vec j}) \take (\rho \xi)))*(\pi ((\tau 
{\vec j}) \drop (\rho \xi))) 
\label{sigma_def}
\end{equation}
and,
\begin{equation}
{\vec a} \equiv \iota (\pi ((\tau {\vec j}) \drop (\rho \xi)))
\label{avec_def}
\end{equation}
respectively. These expressions will be described in detail shortly.  First
we continue the example of a three dimensional array introduced above.

In the example considered above, 
\begin{equation}
\Ravel ({\vec j} \psi \xi) = \Ravel (<\!2\;m\;n\!> \psi \xi),
\end{equation}
is a one dimensional array consisting of the following elements:
$<\!2\;0\;0\!> \psi \xi\; \;$,
$<\!2\;0\;1\!> \psi \xi$,
$<\!2\;1\;0\!> \psi \xi$,
$<\!2\;1\;1\!> \psi \xi$,
$<\!2\;2\;0\!> \psi \xi$,
$<\!2\;2\;1\!> \psi \xi$, etc.

Now the significance of $r$ reveals itself.  {\bf It is the index of 
the first element of our sub-array ${\vec j} \psi \xi$}. {\bf Likewise the 
vector ${\vec a}$ is a vector of integers 
starting with $0$ which serves as a vector of regularly spaced offsets}. The 
total number of integers in the vector ${\vec a}$  is equal to the number of 
elements in the sub-array ${\vec j} \psi \xi$.

The shape vector naturally partitions into two sub-vectors given by
\begin{equation}
((\tau {\vec j})\take (\rho \xi)),
\end{equation}
and,
\begin{equation}
((\tau {\vec j})\drop (\rho \xi)).
\end{equation}
In the first expression $\tau {\vec j}$ counts the number of elements of the
partial index $\vec j$ and the {\it take} operation ($\take$) forms a 
vector of length equal to that of the index vector composed of the first
$\tau {\vec j}$ elements of the shape vector $\rho \xi$ (i.e. the $\tau 
{\vec j}$ leftmost elements of $\rho \xi$). The second expression
forms the corresponding vector of the remaining elements.
The product of the elements of 
\begin{equation}
((\tau {\vec j})\drop (\rho \xi))
\end{equation}
gives the total number of elements of the sub-array ${\vec j}\psi \xi$
and is written as:
\begin{equation}
n = \pi (((\tau {\vec j})\drop (\rho \xi)))
\label{ndef}
\end{equation}

Now we complete the description of the vector of offsets $\vec a$ by using
the {\it iota} operation $\iota$ to create the vector of integers
${\vec a} = \iota (n)$ as given in Eq.~\ref{avec_def}.

Likewise, the starting index $r$ has a simple explanation.  In 
Eq.~\ref{sigma_def}, the expression
\begin{equation}
\gamma ({\vec j};((\tau {\vec j}) \take (\rho \xi)))
\end{equation}
counts the number of sub-arrays that precede the one of interest 
${\vec j}\psi \xi$ in the array $\xi$.  Thus, in order to obtain the index
of the first element of our chosen sub-array ${\vec j}\psi \xi$ we multiply
by the total number of such elements, given in Eq.~\ref{ndef} to give
Eq.~\ref{sigma_def}

\subsection{Applying the $\psi$ Correspondence Theorem}

We now apply the PCT, developed in the previous section, to the expression
given in Eq.~\ref{third_step}.  We begin by writing the right hand side of
Eq.~\ref{third_step} in the form required by the theorem:
\begin{equation}
(( -\sigma ) \drop  \vec i) \psi (((- \sigma ) \take \vec i) \psi \eta)
\equiv {\vec j} \psi \xi
\end{equation}
where the partial index $\vec j$ of the PCT has been defined to be
\begin{equation}
{\vec j}\equiv ((-\sigma)\drop {\vec i}),
\label{j_for_psi}
\end{equation}
and the array $\xi$ of the PCT has been taken to be the subarray:
\begin{equation}
\xi \equiv ((-\sigma)\take {\vec i}) \psi \eta.
\label{xi_for_psi}
\end{equation}

We now work out the various quantities appearing in the PCT.  The index
$\vec i$ is chosen to be a {\it full index} for the array $\eta$. From 
Eqs.~\ref{two_shape},~\ref{valid_index}, and~\ref{eta_def} we 
find the 
shape of the index $\vec i$ to be:
\begin{equation}
\rho ({\vec i}) = <\!d\!>. 
\end{equation}
From Eq.~\ref{j_for_psi} we 
find the shape of $\vec j$ to be:
\begin{equation}
\rho ({\vec j}) = <\!d - \sigma\!> 
\end{equation}
which follows from the definition of {\it drop} $\drop$ as applied to 
Eq.~\ref{j_for_psi}.
Formally we say:
\begin{equation}
\tau ({\vec j}) = d - \sigma,
\end{equation}
where the {\it tau} operator counts the number of elements in the vector 
$\vec j$. Next from Eq.~\ref{xi_for_psi} we find the shape of $\xi$ to be
\begin{equation}
\rho (\xi) = \rho(((-\sigma)\take {\vec i}) \psi \eta)
 = <\!d - \sigma\!> \rshp 2,
\label{xi_shape}
\end{equation}
which shows explicitly that the index vector $\vec j$ is a valid {\it full 
index} for the sub-array $\xi$ as required.

Next we compute the following quantity appearing in the PCT:
\begin{equation}
\tau ({\vec j}) \drop (\rho \xi) = (d - \sigma) \drop 
(<\!d - \sigma\!>  \rshp 2) = < > \equiv \Theta.
\end{equation}
We obtain the empty vector $\Theta$ because the {\it drop} operation $\drop$
is dropping $d - \sigma$ elements from a vector that contains precisely
$d - \sigma$ elements. In the next step, the {\it product} operator $\pi$
acts on this quantity to give:
\begin{equation}
\pi (\tau ({\vec j}) \drop (\rho \xi)) = \pi (\Theta) \equiv 1.
\label{pi_unit}
\end{equation}
In general, the {\it product} operator $\pi$ multiplies the elements of 
a vector to obtain an scalar quantity.  The equivalence on the right hand
side of Eq.~\ref{pi_unit} defines the {\it product of the empty vector} to 
be unity.  

We have now computed all the quantities needed to evaluate the scalar
index $r$ of Eq.~\ref{sigma_def}.
With the above quantities,
the scalar index appearing in the PCT is now given by:
\begin{equation}
r = \gamma (((-\sigma) \drop {\vec i}); <\! d - \sigma\!> \rshp 2)
\label{r_def}
\end{equation}

Now we need to compute the offset vector $\vec a$ appearing in 
Eq.~\ref{avec_def}. In order to do that we form the quantity:
\begin{equation}
\iota (\pi (\tau ({\vec j}) \drop (\rho \xi))) = \iota (\pi (\Theta)) = 
\iota (1) = <\!0\!>.
\end{equation}
We see that the offset vector $\vec a$ contains only one entry $<\!0\!>$.
Thus the combination $r + {\vec a}$ selects only one element.  This is because
although $\vec j$ is a {\it partial index} of $\vec i$, it is indexing the 
subarray $\xi$.  Therefore, the index $\vec j$ is a {\it full index} of the 
subarray $\xi$.  This is consistent with the fact that, although we are dealing
with a partial index $\vec j$  and a subarray $\xi$ it is one step in the 
overall calculation of the action of $\vec i$ on the array $\eta$, the action
of which is to select a single element.

We now summarize the calculation so far.  We now have:
\begin{equation}
\Ravel ({\vec j} \psi \xi) = 
\Ravel (((- \sigma ) \take \vec i) \psi \eta)
[(\gamma(((- \sigma ) \drop \vec i) ; <\!\; d - \sigma\!> \rshp 2) + <\!0\!>]. 
\label{psi_first}
\end{equation}

Now we can simplify the scalar index $r$ of Eq.~\ref{r_def} by noting that
as the index $((- \sigma ) \drop \vec i)$ cycles through all possible 
values (in order), the scalar $r$ takes on all values from $0$ to 
$(\pi  (( <\!d - \sigma\!> )\rshp 2) - 1)$.  We thus define a new variable
$t$, and the right hand side of Eq.~\ref{psi_first} ($RHS$) simplifies as:
$\forall t, \mbox{ s.t.} \; 0 \leq t < \; (\pi  (( <\!d - \sigma\!> )\rshp 2))$

\begin{equation}
RHS = \Ravel (((- \sigma ) \take \vec i) \psi \eta) [t + <\!0\!>] 
\equiv \Ravel (((- \sigma ) \take \vec i) \psi \eta) [<\!t\!>]
\label{rhs_def}
\end{equation}

Now, we must apply the PCT again, this time to the quantity
\begin{equation}
((- \sigma ) \take \vec i) \psi \eta,
\label{sub_expression}
\end{equation}
that appears in Eq.~\ref{rhs_def}.  In order to facilitate the application
of the PCT, we redefine the variables $\vec j$ and $\xi$ appearing in
the PCT as follows.  We write:
\begin{equation}
{\vec j} \equiv ((- \sigma ) \take \vec i), 
\label{new_j}
\end{equation}
and,
\begin{equation}
\xi \equiv \eta = (<\!d\!> \rshp 2 ) \rshp \vec x,
\label{new_xi}
\end{equation}
where we have used the definition of $\eta$ given in Eq.~\ref{eta_def}.

We now proceed to evaluate, step by step, the quantities appearing in the 
PCT based on the new definitions of $\vec j$ and $\xi$ given in 
Eqs.~\ref{new_j} and~\ref{new_xi}.  The shape of the new index $\vec j$
is given by:
\begin{equation}
\rho ({\vec j}) = <\! \sigma \!>,
\end{equation}
and the total number of components of $\vec j$ is
\begin{equation}
\tau ({\vec j}) = \sigma.
\end{equation}
The shape of $\xi$ is given by
\begin{equation}
\rho \xi = <\! d\!> \rshp 2.
\end{equation}
Using these results we have
\begin{equation}
\tau ({\vec j}) \take (\rho \xi) = <\!\sigma\!> \rshp 2,
\end{equation}
followed by
\begin{equation}
\tau ({\vec j}) \drop (\rho \xi) =\; <\!d - \sigma\!> \rshp 2,
\end{equation}
and
\begin{equation}
\pi (\tau ({\vec j}) \drop (\rho \xi)) = \pi (<\!d - \sigma\!> \rshp 2)
= 2^{d - \sigma }.
\label{pi_def}
\end{equation}
We now have computed all the quantities needed to evaluate the 
scalar $r$ of Eq.~\ref{sigma_def}.  Explicitly we have
\begin{equation}
r \equiv \gamma (((-\sigma) \take {\vec i}); (<\!\sigma\!> \rshp 2))*2^{d-\sigma}
\end{equation}
We can also simplify this expression by noting that as we cycle through
all possible values of the {\it partial} index vector 
$(-\sigma) \take {\vec i}$, the function
$\gamma (((-\sigma) \take {\vec i}); (<\!\sigma\!> \rshp 2))$
takes on all values from $0$ to $(\pi (<\!\sigma\!> \rshp 2) - 1) = 2^{\sigma} - 1$.
Thus we define a new variable $s$ and we note that 
$\forall s,$ s. t.  $0 \leq s < \; (\pi  (  <\!\sigma \!> \rshp 2 )) $ the
scalar variable $r$ is written as:
\begin{equation}
r = s*2^{d-\sigma}
\end{equation}

Lastly, to complete the application of the PCT to the expression of 
Eq.~\ref{sub_expression} we compute the offset vector:
\begin{equation}
\iota (
(\pi (\tau ({\vec j}) \drop (\rho \xi)))) = \iota(2^{d - \sigma })
=<\!0\;1\;\ldots (2^{d -\sigma} - 1)\!>.
\end{equation}

Now we summarize this second application of the PCT.  By acting on the 
expression in Eq.~\ref{sub_expression} with the {\it ravel} $\Ravel$ operator 
and applying the PCT we obtain:
\begin{equation} 
\Ravel (((- \sigma ) \take \vec i) \psi \eta) 
= {\vec x} [s*2^{d-\sigma} + <\!0\;1\;\ldots (2^{d -\sigma} - 1)\!>].
\label{psi_sub_expression}
\end{equation}
We see that for each value of $s$, the result of Eq.~\ref{psi_sub_expression}
is a vector of length $2^{d -\sigma}$.  Returning to Eq.~\ref{rhs_def}, however,
we find that to obtain the final result, we must select a component of 
Eq.~\ref{psi_sub_expression} using the index $<\!t\!>$.  Thus the final expression
in Eq.~\ref{rhs_def}, using Eq.~\ref{psi_sub_expression}, is written
\begin{equation}
\Ravel (((- \sigma ) \take \vec i) \psi \eta) [t]
= {\vec x} [s*2^{d-\sigma} + <\!0\;1\;\ldots (2^{d -\sigma} - 1)\!>][<\!t\!>],
\end{equation}
which is simply equivalent to the expression:
\begin{equation}
\Ravel (((- \sigma ) \take \vec i) \psi \eta) [t] 
= {\vec x} [s*2^{d-\sigma} + <\!t\!>],
\end{equation}
which is the final result. 

\subsection{Summary of Final-Transpose Operation}

The ONF is now expressed as follows.  Define two new indices $t$ and $s$ with 
limits given by:
\begin{equation}
0 \leq t < \; t^\star \equiv (\pi  (( <\!d - \sigma\!> )\rshp 2)),
\end{equation}
and,
\begin{equation}
0 \leq s < \; s^\star \equiv (\pi  (  <\!\sigma \!> \rshp 2 )),
\end{equation}
we define a new two-dimensional array $\xi^{(2)}$ by reshaping the vector $\vec \xi$ as:
\begin{equation}
\xi^{(2)} \equiv <\!s^\star\;t^\star\!> \rshp {\vec \xi},
\end{equation}
where $\vec \xi$ is defined in Eq.~\ref{xidef}.
The final result is then written:
\begin{equation}
<\!s\;t\!> \psi \xi^{(2)} = {\vec x} [s*2^{d-\sigma} + t].
\end{equation}
This result was translated directly into code as
shown in Fig.\ref{final_trans_code}.

\renewcommand{\baselinestretch}{1}
\begin{figure}[ht]
\scriptsize
\begin{quotation}
\begin{tt}
\noindent

\begin{verbatim}
void final_trans(complex *datvec,complex *temp,int nmax,
                 int lcap,int csize)
{
/*      dvar = # of 2's in the hypercube
 *      lcap = # of transpose-reshapes (T-rho) that have to be undone
 */
        logc = log(float(csize))/log(2.0);
        sig = lcap*logc;
        smax = pow(float(2),sig);

        dvar = log(float(nmax))/log(2.0);
        tmax = pow(2.0,dvar-sig);

        index = 0;
        for(tind=0;tind<=tmax-1;tind++)
        {
            for(sind=0;sind<=smax-1;sind++)
            {
                temp[index] = datvec[sind*tmax + tind];
                index += 1;
            }
        }

        for(index=0;index<=nmax-1;index++)
        {
            datvec[index] = temp[index];
        }
}

\end{verbatim}

\end{tt}
\end{quotation}
\caption{\label{final_trans_code}
C++ code fragment implementing the final transpose bringing the data back
into the correct order.
}
\end{figure}

\section{Conclusions}

We have presented a tutorial introduction to the techniques of Conformal
Computing illustrated in the context of the new cache-optimized FFT algorithm
presented in the preceding chapter (Part I: see Chap.~\ref{part1}).  Two key aspects of the new
algorithm, the {\it reshape-transpose} operation and the {\it final-transpose}
operation were presented and discussed in detail.  These two examples
are excellent vehicles for developing and illustrating the new techniques in
that many of the most important concepts of the theory (such as the notion
of {\it shapes}, {\it re-shaping}, {\it indexing}, the {\em $\psi$ function}, the {\it 
$\psi$-correspondence theorem}, etc.) are presented.  Indeed, the 
{\it reshape-transpose} operation is an extremely important concept which is
applicable in many other situations.  In addition, the {\it 
$\psi$-correspondence theorem} is a cornerstone of the Conformal Computing
approach. It is the key link between the Mathematics of Arrays (MOA),
which provides a means for reasoning about the algebraic properties of 
array-based algorithms, and the $\psi$-calculus which allows one to reduce 
an algebraic expression to an explicit form that can be directly translated
into computer code in any computer language.  

The Conformal Computing approach
is leading to important new insights by allowing one to view multi-dimensional 
arrays, their decompositions and mappings in a unified, 
general way.  In other words, one can change the 
dimensionality of a given array, through the use of the {\it reshape} operation
to suit the needs of the application at hand (without necessarily moving the 
data).  In particular, the 
multi-dimensional {\it hyper-cube} played an important role in the development
of the insights leading to the {\it final-transpose} operation appearing in
the new FFT.  
In another ongoing investigation, the use of the {\it hyper-cube} 
representation is shedding new light on problems related to quantum computing:
an area in which the {\it hyper-cube} is a natural data structure in a context
based on two-state qubits (see Chap.~\ref{chap6}).


%
%
%

\chapter{A Cache-Optimized Fast Fourier Transform: Part III}
\label{chap5}

\section{Chapter Summary}
This chapter continues to develop the techniques of Conformal Computing as
applied to the Fast Fourier Transform (FFT) that were introduced in the two
preceding chapters.  In these previous chapters, a new cache-optimized algorithm
was presented that was two to four times faster than our previous records
(which beat or were competitive with well-tuned library routines).
This chapter presents a new hyper-cube representation that is contrasted with
that of the preceding chapters. We argue that any arbitrary partitioning of the
data over cache, processors, etc. can be efficiently handled in a hyper-cube
representation.  The rearrangements of the data (virtual or
physically-realized) are represented in the hyper-cube in terms of direct
indexing, thus avoiding most temporary arrays.  Implementation and performance
details, presented in the two preceding chapters are also reviewed.  In
addition to the presentation of this new hyper-cube view, this chapter also
serves as a continuing tutorial introduction to the methods of Conformal
Computing.

\section{Introduction}

The focus of this chapter is the formulation of the FFT in a generalized 
hyper-cube representation using the high-level techniques of the Conformal
Computing approach.  

The Fast Fourier Transform (FFT) is one of the most 
important computational algorithms and its use is pervasive in science and 
engineering. The work in this chapter builds on that of the two previous 
chapters
in which the FFT was optimized in terms of in-cache 
operations leading to factors of {\em two} to {\em four} speedup in comparison 
with our previous records.  Further background material including 
comparisons with library routines can be found in 
Refs.~\cite{lenss3,lenss4,mullin.small:,mul03} and~\cite{cpc} and in Chap.~\ref{intro}. 

{\bf It is also important to note the importance of running {\em reproducible
 and deterministic} experiments. Such experiments are only possible
when dedicated resources exist {\em AND} no interrupts or randomness affects
memory/cache/communications behavior. This means that multiprocessing
and time sharing must be turned off for both OS's and Networks.}

The MoA is a consistent mathematical system in which operators act on arrays
to carry out arbitrary rearrangements of the array elements. Arrays can be
repartitioned ({\em reshaped}) in arbitrary ways. The MoA bears similarities to
Linear Algebra and Group Theory but was 
designed specifically to allow reasoning about the mathematical problem to be
solved (i.e. the application) and layout of the underlying hardware using a 
common formalism.  By using the MoA one obtains high-level monolithic array
expressions.

The second cornerstone of the Conformal Computing approach is the 
{\em $\psi$-calculus}.  Each of the various operators in the MoA is defined in 
terms of its action on the {\em indices} of the array on which it operates,
as defined by the array {\em shape}.
The $\psi$-calculus allows one to translate the high-level MoA expressions 
into the so-called {\em Denotational Normal Form } (DNF), an expression 
involving only cartesian indices of the array, i.e. {\it the semantics}.
In this way temporary arrays are virtually eliminated. Also, due to the 
mathematical properties of the $\psi$-calculus (i.e. the Church-Rosser 
property~\cite{church}) two expressions may be proven to be equivalent by demonstrating
that they reduce to the same normal form.  This ability will become 
increasingly important as issues of performance necessarily extend to power
consumption, heat generation, etc.

To take the DNF into a form (i.e. the
{\em Operational Normal Form}) (ONF) that can be directly translated into 
efficient computer code in any hardware/software language, one then employs 
the so-called {\em Psi Correspondence Theorem} (PCT) as discussed in the 
Chap.~\ref{chap4}.
The resulting ONF can be directly translated into efficient
computer code because the ONF explicitly shows how data should be manipulated.
Loops are revealed in terms of {\em stops, starts, and strides}.  

The techniques of Conformal Computing have a long history dating back to the
work of Sylvester in the nineteenth century.  The Universal Algebra of 
Sylvester, introduced in 1894~\cite{sylvester}, and reintroduced by 
Iverson~\cite{iverson62}, formed the basis for the programming 
language APL and subsequent machine design~\cite{abrams70}.  
Abrams'~\cite{abrams70} revolutionary insight into the use of shapes to
define array operations was the inspiration for the $\psi$-calculus.
Unfortunately, APL had too many mathematical anomalies~\cite{gerhart} to be
used as a formal mathematical tool.  
{\em In addition, APL never had an associated indexing calculus like the 
$\psi$-calculus.}
Similarly, closure was not obtained for 
Abrams' indexing rules despite 10+ years of 
research~\cite{lyon,guibas,miller,perlis,budd84}.

Mullin's introduction of MoA and $\psi$-calculus removed all anomalies in 
Iverson's algebra and put closure on Abrams' indexing through the 
introduction of the indexing function, $\psi$. She also combined MoA and 
$\psi$-calculus with the $\lambda$-calculus~\cite{mul91} to achieve
full reasoning capabilities computationally building upon recommendations
from Perlis~\cite{tu2}, Berkling~\cite{berkling}, and Budd~\cite{budd91}.

The most difficult aspect of the Conformal Computing approach is the need 
for one to learn to think in the space of multi-dimensional arrays
{\bf and to envision an algorithm in which the architecture and network are 
viewed as one data structure.}  This is
the part of the approach (that is still somewhat of an art-form) that leads to
the high level formulation of the problem in monolithic MoA constructs 
The techniques of the $\psi$-calculus are more straightforward and can 
be applied mechanically since all transformations are linear.

This chapter continues to introduce higher-level concepts of the theory as
required for the application at hand: the hyper-cube formulation of the 
FFT. A tutorial style is adopted as in previous chapters as the concepts are, 
no-doubt, unfamiliar to most readers.  Thus we present all steps of every  
calculation.  As such, there is considerable mathematical detail which 
may appear formidable at first glance.  The determined reader, however, will
no-doubt be rewarded by going through each step in detail.
All fundamentals of the theory needed to approach this material were
presented in Chap.~\ref{chap2}.  Again we emphasize that Conformal Computing 
{\bf \em is not a programming language} but rather is an algebraic approach
to an efficient construction of computer programs for implementation 
{\bf \em in any programming language}. 

\section{Extreme Generality of Representation: Contrasting MoA with 
Linear Algebra}

\subsection{Linear Arrays and Hyper-Cubes}

The Mathematics of Arrays is extremely general in its ability to represent 
multi-dimensional arrays.  Conceptually any array, independent of its 
dimensionality can be thought of as a one-dimensional array simply by forming
the vector containing all of the array's elements in some pre-chosen order.  
Since this process is done so frequently in MoA we define it as the operation 
{\em Ravel}. Thus: we define the operator $\Ravel$ to be a unary operator that
produces a vector $\vec v$ of the elements of the multi-dimensional array $A$
as:

\begin{equation}
{\vec v} = \Ravel A
\end{equation}
The vector $\vec v$ so produced is a linear array.  As such it is the
representation of the data with the {\em least} number of dimensions but 
the greatest number of elements in a given dimension.

In the opposite extreme, we can equally envision an array with the {\em most
number of dimensions}.  The {\em hyper-cube} is just such an array.  Each 
dimension has only two allowed values $0$ and $1$ and the number of dimensions
is equal to $\log_2(N)$ where $N$ is the total number of elements in the array.
The ordering of hyper-cube is chosen to correspond to {\em row-major} 
ordering\footnote{In general the choice of ordering is arbitrary 
(e.g. {\em row-major}, {\em column-major}, etc.) and is conveniently 
specified by the definition of the corresponding {\em gamma } function.}
so the array index of an element read from left to right corresponds to the
binary representation of the number of element in the array (assuming zero 
offset arrays, as is the case for the C/C++ language).

In between these two extremes, one can imagine a host of multi-dimensional 
arrays.  The various ways in which an array of length $N$ can be partitioned
is by given by the ways in which the number $N$ can be factored: each factor
corresponding to the length of a given dimension and the power of the given
factor corresponding to the number of dimensions having the given (factor) 
length.

\subsection{The Importance of the Array's Shape\label{shape}}

{\bf \em The power of the Conformal Computing approach lies in its ability to 
view the array using any multi-dimensional representation that 
is convenient}.  For
example, we often (1) pose the problem as a multi-dimensional array based on 
the structure of the underlying science or engineering problem, (2) increase the dimensionality of
the problem to represent a given partitioning of the data (block, cyclic, etc.)
over the hierarchy 
of the machine (cache, memory, paged memory, disk, network, processor, machine, grid-network, etc.). (3) It often is convenient to change the dimensionality of
the problem {\em again} by viewing it as a hyper-cube.  At this stage all of
the formal operations (matrix transformations, etc.) are carried out. Lastly
we change the dimensionality further by (4) projecting down to a linear
array for the final implementation on a computer.  The final form based on a 
linear array (i.e. the Operational Normal Form: ONF) is extremely efficient 
in that direct indexing of contiguous memory addresses is used.

The underlying view of MoA is that an array is specified by two vectors:
(1) the {\bf \em shape} vector, and (2) the one dimensional array of the 
array's elements, i.e. {\em layout}. For a given array $A$ the 
{\bf \em shape} vector is a vector
of length equal to the number of dimensions.  Each element of the shape 
gives the total number of elements in a given dimension.  As previously 
discussed (in Sec.~\ref{indshp}), the second vector is simply the {\em Ravel} of the array.

Just as we introduced an operator $\Ravel$ to obtain the {\em Ravel} of the 
array (written $\Ravel A$) we introduce the {\bf \em shape} operator 
$\rho$ to obtain the shape $\rho A$.  Thus the {\bf \em shape} vector $\vec s$
is obtained from the array $A$ by the assignment ${\vec s} = \rho A$.

The ability to change the dimensionality of an array is provided by the
{\bf \em reshape} operator written as $\hat \rho$. The {\bf \em reshape}
operator $\hat \rho$ is a binary operator that takes an array (the array
to be {\em reshaped}) as the right argument and a {\bf \em shape vector} 
(i.e. the {\em new shape}) as the left argument. This operation creates the 
new array by filling the elements of the new array sequentially from the 
{\em Ravel} of the old array.  Note that the shape gives us new ways to 
access the structure not the layout, which is still row or column major. 
That is, access changes without actually creating a new array.
The reshape-operator is 
completely general in that it can take as its left argument {\bf \em any
arbitrary shape} and is not constrained by shapes corresponding to the same number
of elements of the old array.  If the total number of elements of the new array
is larger than that of the old array, one simply starts at the beginning of 
the {\em Ravel} once one runs out of elements.  For reshaping into 
a smaller array one simply takes as many elements from the {\bf \em Ravel}
as will fit into the new array.

\subsection{MoA Operator Constructs and the $\psi$-Calculus}

The extreme flexibility to work in multi-dimensional arrays afforded by 
MoA results from the use of an advanced algebraic system that is similar to
standard Linear Algebra but is considerably generalized. The structure of this
algebra is summarized in Chap.~\ref{chap2}. Here we recall some of the most 
important notions of the theory and contrast it with similar (but limited) 
constructs in standard Linear Algebra.

So far we have introduced the operators {\bf \em Ravel} $\Ravel$, {\bf \em
shape} $\rho$, and {\bf \em reshape} $\hat \rho$.  Another very important 
operator for our present purposes is the {\bf \em transpose} operator 
$\transpose$.  The {\bf \em transpose} operator is a binary operator that
takes an array as its right argument and a permutation vector as its left 
argument and has the action of permuting the order of the elements of the
array by permuting the dimensions.\footnote{Note that the MoA/$\psi$-calculus
definition of $\transpose$ is part of the Fortran 95 standard and was 
introduced therein by Mullin.}  
All of the operators in the theory
are defined in terms of the effect on the indices of the array with shapes.  
The connection between an array's index and its elements is given by the {\bf \em psi} 
operator $\psi$.

We now demonstrate the use of the operator formalism to indicate the 
data restructuring discussed in the first paragraph of section~\ref{shape}.
In step (1) we pose the problem in terms of monolithic multi-dimensional 
arrays $A$, $B$, etc.  These we specify by their shapes and Ravel's:
\begin{equation}
{\vec v}_A = \Ravel A;
{\vec s}_A = \rho A.
\end{equation}
(2) Next we {\em reshape} the arrays to correspond to a given decomposition with 
\begin{equation}
A^\prime = {\vec s}\;^\prime {\hat \rho} A.
\end{equation}
(3) Next to carry out the operations of the theory (linear transformations) we 
transform to the hyper-cube representation,
\begin{equation}
A_H^\prime = <\!2\;2\;2\;\cdots\;2\!> {\hat \rho} A^\prime,
\end{equation}
(A series of numbers between angle brackets is used to denote a vector.) \\
(4) After carrying out all of the transformations required to reach a final 
form, that we denote by $B_H$ we transform back to a linear array:
\begin{equation}
{\vec v}_B = \Ravel B_H.
\end{equation}

\section{FFT in the Hyper-Cube}
\noindent
Suppose the input vector was
\begin{eqnarray}
\vec q \equiv < 0\;1\;2\;3\;4\;5\;6\;7\;8\;9\;10\;11\;12\;13\;14\;15>.
\end{eqnarray}
We put the index position into the contents of $\vec q$ so that we can see how
the indices move around during transformations.
Consequently, with an input length of $16$, there are $l = log_2 16 = 4$ cycles,
labeled by $j$, to the FFT:
\begin{equation}
\forall j; \;\;\; 0 \leq j < l.
\end{equation}
So, prior to the initial step, we restructure  $\vec q$:
\begin{eqnarray}
Z \equiv ( <l> \rshp 2) \rshp \vec q, 
\end{eqnarray}
(where ($<l> \rshp 2$) is a vector consisting of ``$l$" $2$'s; in this case
$l=4$).
\subsubsection{Step j=0}
The initial hyper-cube is given by:
\begin{eqnarray}
\left [ 
     \begin{array}{c}
        \left [ 
            \begin{array}{cc}
            \left [ 
            \begin{array}{ccc}
              & 0 \; & 1\; \\
              & 2 \; & 3 \; \\ 
            \end{array} 
        \right ]
&
        \left [ 
            \begin{array}{ccc}
              & 8  & 9  \\
              & 10 & 11 \\
            \end{array} 
        \right ] 
     \end{array} 
\right ] 
\\
\left [ 
     \begin{array}{cc}
       \left [ 
          \begin{array}{ccc}
            & 4\; & 5\; \\
            & 6\; & 7\;  \\ 
          \end{array} 
       \right ]
&
       \left [ 
          \begin{array}{cc}
            12 & 13   \\
            14 & 15 \\ 
          \end{array} 
       \right ]
      \end{array} 
\right ]
\end{array}
\right ].
\label{hyperinit}
\end{eqnarray}
\noindent
We now state the hyper-cube FFT in Conformal Computing notation and illustrate
its use.  In the following sections we present exhaustive detail as to how
it works by supplying the reader will all  of the steps of the derivation.

At each $j$ step we want to update all transposed pairs:
\begin{equation}
 {\tilde Z}_j \equiv (<0>   _{\psi} \Omega_{<1\;1>} Z_j )
_+ \Omega_{<0\;1>} (<1\;-1> 
_\times \Omega_{<1\;0>} ((<1>) _{\psi} \Omega_{<1\;1>}Z_j)), 
\label{ztilde}
\end{equation}
where,
\begin{eqnarray}
Z_j \equiv \vec t_j \transpose Z.
\label{zj}
\end{eqnarray}
with,
\begin{equation}
\vec t_j \equiv (((l-1)-j) \take \vec c\; ) \cat ((-1) \take \vec c\;)
\cat (j \take (((l-1)-j) \drop \vec c\;)).
\label{tvec}
\end{equation}
The vector ${\vec t}_j$ permutes the dimensions of the hyper-cube so that at each
step, neighboring elements are the ones that need to be combined for the FFT.
In Eq.~\ref{tvec}, the vector $\vec c$ is a zero-offset vector of integers
of length equal to the number of dimensions of the hyper-cube:
\begin{equation}
\vec c = \iota (l) 
\end{equation}
The vector ${\vec t}_j$ was found by observing the patterns 
that arise in the FFT. Examples of the hyper-cube permutations are given in the 
following three steps. The structure of $\vec t_j$ will be explored in some
detail shortly.

We know there are 4 cycles to the FFT since $l=log_2 16 = 4$. Recall, $ 0 \leq j < l$. In 
step 0, we want to index all pairs and  simultaneously update all components.
That is, all  pairs are updated by expressions in Eqs.~\ref{ztilde},~\ref{zj}, and~\ref{tvec}.

In step 1, we must permute the hyper-cube s.t. we can access the following permuted indices:
\begin{eqnarray}
\left [ \begin{array}{c}
\left [ 
\begin{array}{cc}
\left [ \begin{array}{ccc}
&0\; & 2\; \\
&1\; & 3\; \\ 
\end{array} 
\right ]
&
\left [ \begin{array}{ccc}
& 8\; & 10\;\\
& 9\; & 11\;\\
\end{array} 
\right ] 
\end{array} \right ] \\
\left [ \begin{array}{cc}
\left [ \begin{array}{ccc}
& 4\;& 6\; \\
& 5\;& 7\;\\
\end{array} 
\right ]
&
\left [ \begin{array}{cc}
12 & 14 \\
13 & 15 \\
\end{array} 
\right ]
\end{array} 
\right ]
\end{array}
\right ].
\end{eqnarray}
In step 2:
\begin{eqnarray}
\left [ \begin{array}{c}
\left [ 
\begin{array}{cc}
\left [ \begin{array}{ccc}
& 0\; & 4\;  \\
& 1\; & 5\; \\ 
\end{array} 
\right ]
&
\left [ \begin{array}{ccc}
& 8\; & 12 \; \\
& 9\; & 13 \; \\ 
\end{array} 
\right ] 
\end{array} \right ] \\
\left [ \begin{array}{cc}
\left [ \begin{array}{ccc}
& 2\; & 6\;  \\
& 5\; & 7\; \\ 
\end{array} 
\right ]
&
\left [ \begin{array}{cc}
10 & 14\\
11 & 15 
\end{array} 
\right ]
\end{array} 
\right ]
\end{array}
\right ].
\end{eqnarray}
\break
In Step 3:
\begin{eqnarray}
\left [ \begin{array}{c}
\left [ 
\begin{array}{cc}
\left [ \begin{array}{ccc}
& 0\; & 8 \;\\
& 1\; & 9 \; \\ 
\end{array} 
\right ]
&
\left [ \begin{array}{ccc}
& 4\; & 12\; \\
& 5\; & 13\; \\ 
\end{array} 
\right ] 
\end{array} \right ] \\
\left [ \begin{array}{cc}
\left [ \begin{array}{ccc}
&2\; & 10\; \\
&3\; & 11\; \\ 
\end{array} 
\right ]
&
\left [ \begin{array}{ccc}
& 6\; &14\;  \\
& 7\; & 15\; \\ 
\end{array} 
\right ]
\end{array} 
\right ]
\end{array}
\right ].
\end{eqnarray}

\section{Matrices, Arrays, Hyper-Cubes}
\noindent

\noindent
$\vec q$ is the input vector, we define:
\begin{eqnarray}
Z \equiv (<l> \rshp 2) \rshp \vec q, \\
\end{eqnarray}
and
$\forall \;\; 0 \leq j < l, $ let: 
\begin{equation}
 {\tilde Z}_j \equiv (<\!0\!>   _{\psi} \Omega_{<1\;1>} Z_j ) 
_+ \Omega_{<0\;1>} (<\!1\;-1\!> 
_\times\! \Omega_{<1\;0>} (<\!1\!> {}_{\psi} \Omega_{<1\;1>}Z_j)),
\label{ztildetwo}
\end{equation}
\noindent
where,
\begin{eqnarray}
Z_j \equiv \vec t_j \transpose Z,
\end{eqnarray}
with,
\begin{eqnarray}
\vec t_j \equiv (((l-1)-j) \take \vec c\; ) \cat (-1 \take \vec c\;)
\cat (j \take (((l-1)-j) \drop \vec c\;)). 
\end{eqnarray}

To summarize, $Z$ is the initial hyper-cube formed by the elements of the input 
vector $\vec q$ and $Z_j$ are the various permuted hyper-cubes.  The array 
$\vec c$ is 
a vector of integers that gets permuted to give the permutations of the 
hyper-cube.

In this chapter, we don't attempt to prove Eq.~\ref{ztildetwo} algebraically.  It 
summarizes the result of our experience with the FFT as illustrated in our 
two previous chapters.
Rather, we proceed with Eq.~\ref{ztildetwo}
as written and apply the machinery of Conformal Computing to derive the 
ONF.

It is helpful to consider the structure of $\vec t_j$ in detail.  Therefore we 
now \\ illustrate the construction of $\vec t_j$ for a complete example in 
Fig.~\ref{tvectab}. 
\break

\footnotesize
\begin{figure}[ht]
\[{\vec t_j}:\;\;\;\mbox{ suppose }l=8 \rightarrow \vec c=<0\;1\;2\;3\;4\;5\;6\;7>\]
\[0 \leq j < l \rightarrow j=0,1,2,3,4,5,6,7 \]
\vspace{.3in}

\begin{center}

\scriptsize

\begin{tabular}{|l|l|l|r|r|l|}\hline
$j$&$(((l-1)-j)\take \vec c)$&$(-1)\take \vec c$ & $(((l-1)-j)\drop \vec c)$ & $j \take (((l-1)-j)\drop \vec c)$ & $\vec t_j$ \\ \hline
$0$&$<0\; 1\; 2\; 3\; 4\; 5\; 6>$&$<7>$&$<7>$&$<>$&$<0\; 1\; 2\; 3\; 4\; 5\; 6\; 7>$ \\ \hline
$1$&$<0\; 1\; 2\; 3\; 4\; 5>$&$<7>$&$<6\;7>$&$<6>$&$<0\; 1\; 2\; 3\; 4\; 5\; 7\; 6>$ \\ \hline
$2$&$<0\; 1\; 2\; 3\; 4>$&$<7>$&$<5\;6\;7>$&$<5\;6>$&$<0\; 1\; 2\; 3\; 4\; 7\; 5\; 6>$ \\ \hline
$3$&$<0\; 1\; 2\; 3>$&$<7>$&$<4\;5\;6\;7>$&$<4\;5\;6>$&$<0\; 1\; 2\; 3\; 7\;4\; 5\; 6>$ \\ \hline
$4$&$<0\; 1\; 2>$&$<7>$&$<3\;4\;5\;6\;7>$&$<3\;4\;5\;6>$&$<0\; 1\; 2\; 7\;3\; 4\; 5\; 6>$ \\ \hline
$5$&$<0\; 1>$&$<7>$&$<2\;3\;4\;5\;6\;7>$&$<2\;3\;4\;5\;6>$&$<0\; 1\; 7\;2\; 3\; 4\; 5\; 6>$ \\ \hline
$6$&$<0>$&$<7>$&$<1\;2\;3\;4\;5\;6\;7>$&$<1\;2\;3\;4\;5\;6>$&$<0\;7\; 1\; 2\; 3\; 4\; 5\; 6>$ \\ \hline
$7$&$<>$&$<7>$&$<0\;1\;2\;3\;4\;5\;6\;7>$&$<0\;1\;2\;3\;4\;5\;6>$&$<7\;0\; 1\; 2\; 3\; 4\; 5\; 6>$ \\ \hline
\end{tabular}

\end{center}
\caption{Illustration of the construction of the index 
$\vec t_j$. \label{tvectab}}
\end{figure}
\normalsize

\section{Derivation}
We first reduce ${\tilde Z}_j$ then $\vec t_j \transpose Z$. The two derivations
are then merged.

We begin with Eq.~\ref{ztildetwo}, in which
$Z_j $ is entirely updated by taking the 0th component of all 
2 component vectors (pairs) and adding $<1 \;-1> \times$ the 1st component
of all 2 component vectors\footnote{We assume that weights
have been applied to $\vec q$.}. 
The easiest way to perform {\em $\psi$ reductions}, is to reduce an
expression's constituent pieces separately. Thus,
we'll rewrite Eq.~\ref{ztildetwo} as follows.

Let:
\begin{eqnarray}
{\tilde Z}_j & \equiv & A _+ \Omega_{<0\;1>} B, \nonumber \\
A & \equiv & ( <0> _{\psi} \Omega_{<1\;1>} Z_j ), \nonumber \\
B & \equiv & ( C _{\times} \Omega_{<1\;0>} D), \nonumber \\
C & \equiv & <1\;\;-1>, \nonumber \\
D & \equiv &  (<1>  _{\psi} \Omega_{<1\;1>} Z_j ), \nonumber \\
Z_j & \equiv & \vec t_j \transpose Z. \nonumber \\
\end{eqnarray}
\noindent
The quantity $Z$ is the restructured (i.e. hyper-cube) input vector. 

\subsection{Reduction of D}
Now, by applying the definition of $\Omega$ (see Sec.~\ref{omegadef}) we demonstrate
the reduction of $D$.
We have:
\begin{eqnarray}
D \equiv <1> _{\psi}  \Omega _{<1\;1>} Z_j.
\end{eqnarray}
That is, take the  {\em 1st} component of each vector in $Z_j$. 
If we look at the entire expression above we'll see that $A$
takes the  {\em 0th} component of each vector in $Z_j$.
Thus, except for a sign difference,  the derivation is the
same.  Consequently, the derivation for $A$ will be omitted.
\noindent
Applying the definition of Omega (see Sec.~\ref{omegadef}):
\begin{eqnarray}
 \xi_l = <1>,\; &  _g \Omega_{\vec d} = _{\psi} \Omega _{<1\;1>},\; & 
\xi_r = Z_j, \;\;\mbox{with}\;\; 
Z_j \equiv \vec t_j \transpose Z, \nonumber \\
\delta \xi_l = 1,\; & \vec d  \equiv <\sigma_l \; \sigma_r>, \; &
\delta \xi_r = l. 
\end{eqnarray}
That is $\xi_l$ is $<1>$, a vector, and is 1-dimensional. $\vec d$ is used
for partitioning information. Here $\sigma_l$ is 1, so we know we
want vectors from the left argument, $\xi_l$, and
vectors from $\xi_r$,  since $\sigma_r$ is 1. Continuing we have:
\[\begin{array}{cccc}
\rho \xi_l  =  <1>,  & g  = \psi, \;  & \rho \xi_r 
 \equiv  (<l>\rshp 2) \\
\sigma_l  =   1,  &\sigma_r  =  1. 
\end{array} \]

We now present all steps of the reduction of $D$ by simply applying the 
definition of the $\Omega$ operator as follows:
\begin{eqnarray}
m  & = & 0,\nonumber \\
\vec x  & = & <\;>, \nonumber \\
\vec u  & = &  0 \drop ((-1) \drop <1>)  \equiv  \Theta \nonumber, \\
\vec v & \equiv & 0 \drop ((-1) \drop \rho \xi_r)    
 \equiv ( <l-1>\rshp 2),\nonumber \\
\vec y & \equiv & (-1) \take \rho \xi_l  \equiv  <1>, \nonumber \\
\vec z  &\equiv&  (-1) \take \rho \xi_r    \equiv   <2>, \nonumber \\
\rho \xi_l & \equiv & \vec u \cat \vec x \cat \vec y  \equiv  <>\cat<>\cat<1> 
 \equiv  <1>, \nonumber \\
\rho \xi_r  & \equiv & \vec v \cat \vec x \cat \vec z  
\equiv  \vec v \cat  \vec z   \equiv  <\vec v\; z>  \equiv  ( <l> \rshp 2).
\end{eqnarray}
Thus, 
\begin{eqnarray}
\vec i & \equiv & <>, \nonumber \\ 
0  & \leq ^* & \vec j <^* \vec v, \nonumber \\
\vec k & \equiv &  <>, \nonumber \\
\vec w & \equiv &\rho ((<> \psi \xi_l) \psi (\vec j \psi \xi_r )), \nonumber \\
&\equiv & \rho (<1> \psi (\vec j \psi \xi_r)), \nonumber \\
& \equiv & (\tau <1>) \drop ((\tau \vec j ) \drop (\rho \xi_r))), \nonumber \\
& \equiv &  1 \drop ( <n-1> \drop (<n> \rshp 2)), \nonumber \\
& \equiv & 1 \drop <2>,\nonumber \\
\vec w & \equiv &  <>. 
\end{eqnarray}

Note, in the above, we made use of the identity $<\;> \psi \xi \equiv \xi$ that
holds (by definition) for any array $\xi$.
\subsection{Denotational Normal Form for D}
From above, the DNF for D is:
\noindent
\begin{equation}
D  \equiv   (<1>  _{\psi} \Omega_{<1\;1>} Z_j ).
\end{equation}
Thus, the shape of $D$ is defined by:
\begin{eqnarray}
\rho (\xi_l {_g} \Omega _{\vec d} \xi_r) & \equiv &( \vec u \cat \vec v \cat \vec x \cat \vec w ), \nonumber \\
\rho D & = & (<l-1> \rshp 2).
\end{eqnarray} 
Components are extracted using the index $\vec v$
with $0 \leq^* \vec v <^* (<l-1>\rshp 2$), as follows:
\begin{eqnarray}
\forall {\vec j}; \;\; 0 & \leq ^* & \vec j <^* \vec v, \nonumber \\
\vec j \psi D  & = & \vec j \psi (\xi_l ( _g \Omega _{\vec d} ) \xi_r), \nonumber \\
 &
= & (<> \psi \xi_l) \psi (\vec j \psi \xi_r), \nonumber \\
& = & <1> \psi (\vec j \psi Z_j), \nonumber \\
& = & (\vec j \cat 1) \psi Z_j.
\end{eqnarray}
The final result above is a scalar.

Note the simple heuristic way of thinking of $D$.  We can see that the 
array $D$ is simply the collection of all elements of $Z_j$ whose index vector
has the value $1$ in its right-most bit.  For example, if $Z_j$ has shape
$\rho Z_j = <\!2\;2\;2\;2\!>$, then the array $D$ contains the following 
elements:  $<\!0\;0\;0\;1\!> \psi Z_j$, $<\!0\;0\;1\;1\!> \psi Z_j$,
$<\!0\;1\;0\;1\!> \psi Z_j$, $<\!0\;1\;1\;1\!> \psi Z_j$, 
$<\!1\;0\;0\;1\!> \psi Z_j$, $<\!1\;0\;1\;1\!> \psi Z_j$,
$<\!1\;1\;0\;1\!> \psi Z_j$, and $<\!1\;1\;1\;1\!> \psi Z_j$. In like manner,
array $A$ has a $0$ in its rightmost bit.

This means that for step $j = 0$ we are working with the initial hyper-cube
of Eq.~\ref{hyperinit} the {\em ravel} of $D$ is given by:
\begin{equation}
\Ravel D = <\!1\;3\;9\;11\;5\;7\;13\;15\!>,
\end{equation}
while the {\em ravel} of $A$ is given by:
\begin{equation}
\Ravel A = <\!0\;2\;8\;10\;4\;6\;12\;14\!>.
\label{rava}
\end{equation}

\subsection{Reduction of  B}
Substituting the derivation for D we have:
\begin{eqnarray}
 \forall {\vec j}; \;\; 0 \leq^* \vec j <^* (<l-1>\rshp 2), 
\end{eqnarray}
\begin{eqnarray}
B & \equiv & C _{\times} \Omega _{<1\;0>} D \nonumber \\
 & = &  <1\;-1> {}_{\times}\! \Omega _{<1\;0>} 
(<1> _{\psi } \Omega _{<1\;1>} Z_j), \\
\vec j \psi B
& = & <1\;-1> {}_{\times}\! \Omega _{<1\;0>} (<1> {\psi} (\vec j \psi Z_j)), \nonumber \\ 
& = & <1\;-1> _{\times} \Omega _{<1\;0>} ((\vec j \cat 1) \psi Z_j).
\end{eqnarray}
Now we apply the definitions associated with $\Omega$:
\begin{eqnarray} 
\xi_l = <1\;-1>,\; & \xi_r = D, \nonumber \\
\delta \xi_l = 1,\; & _g \Omega_{\vec d} = \; _{\times }\! \Omega _{<1\;0>},\; 
& \delta \xi_r \equiv l-1, \nonumber \\
\rho \xi_l = <2>, & \vec d = < 1\;0 >,
& \rho \xi_r = <l-1>\rshp 2, \nonumber \\
\sigma_l = 1 & m = 0 & \sigma_r = 0.
\end{eqnarray}
Continuing:
\begin{eqnarray}
\vec x  = & 0 \take (-1) \drop <2> & = <>, \nonumber \\
\vec u  = & 0 \drop (-1) \drop <2>&  = <>, \nonumber \\
\vec v =  & 0 \drop 0 \drop \rho \xi_r & = (<l-1> \rshp 2), \nonumber \\
\vec y  = & (-1) \take <2> & = <2>, \nonumber \\
\vec z  = & 0 \take \rho \xi_r &  = <>.
\end{eqnarray}

\begin{eqnarray}
0 \leq^* & \vec i & <^* \vec u = <>,\;\;\;\;  \Rightarrow \;\;\;\; \vec i = <>, \nonumber \\
0 \leq^* & \vec j & <^* \vec v = (<l-1> \rshp 2), \nonumber \\
0 \leq^* & \vec k & <^*  \vec x = <>,   \Rightarrow  \vec k  = <>. 
\end{eqnarray}
\begin{eqnarray}
\vec w & \equiv & \rho (({\vec i}\cat {\vec j} \cat {\vec k})\psi 
(\xi_l(_g\!\Omega_{\vec d}\xi_r))), \nonumber \\ 
 \vec w = \rho 
 (<> \psi <1\;-1>) \times (  \vec j \psi D),
& = & \rho (<1\;-1> \times  ( \vec j \cat 1)\psi Z_j) )\nonumber \\
 \vec w & = & <2>.
\end{eqnarray}
\subsection{Denotational Normal Form for B}
From the above we conclude:
\begin{eqnarray}
B & \equiv & ( C _{\times} \Omega_{<1\;0>} D).
\end{eqnarray}
The shape of this expression is:
\begin{eqnarray}
\rho B & \equiv & \vec u \cat \vec v \cat \vec x \cat \vec w, \nonumber \\
& = & <l> \rshp 2.
\end{eqnarray}
Components of this expression are extracted with the vector $\vec j$ such 
that:
$\forall {\vec j},\;\;0 \leq^* \vec j <^* (<l-1>\rshp 2)$,

\begin{eqnarray}
(\vec i \cat \vec j \cat \vec k) \psi B & = & 
((\vec i \cat  \vec k ) \psi \xi_l  )
\times (( \vec j \cat \vec k ) \psi
\xi_r), \nonumber \\
\vec j \psi B & = & (<> \psi C) \times (\vec j \psi D), \nonumber \\
\vec j \psi B & = & <1\;-1> \times ( \vec j \cat <1>) \psi Z_j. 
\end{eqnarray}
This last result is a two-component vector $<\!d\;(-d)\!>$ and for each
value of $\vec j$ the variable $d$ takes on the components of the 
{\em ravel} of $D$ (i.e. $d = \{1,\;3,\;9,\;11,\;5,\;7,\;13,\;15\!\}$).

\subsection{Reduction of A}
We don't need to do this derivation since it is nearly {\em identical} to the
derivation for D. Thus it will be omitted. Therefore,
\begin{eqnarray}
A \equiv  <0> _{\psi } \Omega_{<1\;1>} Z_j, 
\end{eqnarray}
and,
$\forall \vec j \;\;\;0 \leq^* \vec j <^* (<l-1>\rshp 2)$,
\begin{eqnarray}
\vec j \psi  {\tilde Z}_j & = &  ((\vec j \cat 0) \psi Z_j) 
_+ \Omega_{<0\;1>} <1\;-1> \times  ( ( \vec j \cat 1) \psi Z_j ).
\end{eqnarray}

\subsection{Denotational Normal Form for A}
The DNF for A has the following shape:
\begin{eqnarray}
\rho A \equiv (<l-1> \rshp 2).
\end{eqnarray}
The components of A are obtained as:
\begin{eqnarray}
\vec j \psi A  = ( <0> \psi (\vec j \psi Z_j)) =  (\vec j \cat <0> ) \psi Z_j .
\end{eqnarray}
\noindent 
Note the similar result to that obtained for $D$ as expressed in Eq.~\ref{rava}.
\subsection{Final Reduction}
Consequently,
\noindent
$\forall {\vec j}; \;\; 0 \leq^* \vec j <^* (<l-1>\rshp 2)$,
\begin{eqnarray}
\vec j  \psi   {\tilde Z}_j & = & \vec j \psi (A _+ \Omega_{<0\;1>} B), \nonumber \\
& = & (  (\vec j \cat <0> ) \psi Z_j)) 
_+ \Omega_{<0\;1>} (<1\;-1> \times ( \vec j \cat <1>) \psi Z_j ).
\end{eqnarray}
Now, again we apply the definition of $\Omega$:
\begin{eqnarray}
\xi_l =  A,\;
& _g \Omega _{\vec d}  = _+\! \Omega _{<0 \;1>},\; 
     & \xi_r  = B, \nonumber \\
\delta \xi_l = l-1,\; & d = <\sigma_l\;\sigma_r>,\; 
     & \delta \xi_r = l,\nonumber \\
\rho \xi_l = (<l-1>\rshp 2 ),\;
     & \rho \xi_r = (<l> \rshp 2), \nonumber \\
\sigma_l = 0, \;
     & \sigma_r = 1. 
\end{eqnarray}
Thus,
\begin{eqnarray}
m & = &  l-1, \nonumber \\
\vec x & = & -(l-1) \take 0 \drop (<\!l-1\!> \rshp 2)  = (<l-1>\rshp 2), \nonumber \\
\vec u & = & -(l-1) \drop 0 \drop (<\!l-1\!> \rshp 2) = <\;>, \nonumber \\
\vec v & = & -(l-1) \drop 1 \drop (<\!l\!> \rshp 2)  = <\;>, \nonumber \\
\vec y & = & 0 \take (<\!l-1\!> \rshp 2)  = <>, \nonumber \\
\vec z & = & (-1) \take (<\!l\!> \rshp 2)= <2>. \nonumber  \\
\end{eqnarray}
Continuing,
\begin{eqnarray*}
0 \leq^* & \vec i & <^*  <>,\\
0 \leq^* & \vec j'&  <^*  <>,\\
0 \leq^* & \vec k  & <^* (<l-1>\rshp 2),
\end{eqnarray*}
\begin{eqnarray}
\vec w & = & \rho ( (\vec j \cat <0> ) \psi Z_j) 
+ (<1\;-1> \times ( (\vec j \cat <1> ) \psi Z_j)), \nonumber	\\
\vec w & = & <2>. 
\end{eqnarray}
This expression results from first taking components and then finding 
the shape of the result.  The components are obtained in the following 
(see Eq.~\ref{components}).

Note: $\vec Z_j$ is raveled (flattened) after the transpose, i.e.
\begin{equation}
\vec z_i \equiv \Ravel \vec t_i \transpose Z. 
\end{equation}
\subsection{Final Denotational Normal Form}
We now find the final DNF.
The shape is given by:
\begin{equation}
\rho \; (A _+ \Omega _{<0\;1>} B) 
   \equiv \vec u \cat \vec v \cat \vec x \cat \vec w =
<l> \rshp 2. 
\end{equation}
Components are obtained as:
\begin{eqnarray}
(\vec i \cat \vec j' \cat \vec k) \psi (A _+ \Omega_{<0\;1>} B) & \equiv &
((\vec i \cat \vec k) \psi A) +   (( \vec j' \cat \vec k) \psi B), \nonumber \\
 \vec k \psi (A _+ \Omega_{<0\;1>} B) & = &
 (\vec k \psi A) +              (\vec k \psi B ), \nonumber \\
 \vec j \psi A _+ \Omega_{<0\;1>} B & = & \vec j \psi A +              (\vec j \psi B ), \nonumber \\
& = & <0> \psi ( \vec j \psi Z _j ) + <1\;-1> \times  ( <1> \psi (\vec j \psi Z _j)), \nonumber \\
& = &  (( \vec j \cat 0) \psi Z _j ) + <1\;-1> \times  ( (\vec j \cat 1) \psi Z _j). \nonumber \label{components}\\
\label{finaldnf}
\end{eqnarray}
Note that both  $k$ and $j$ have the same limits: 
$0 \leq^* \vec k <^* (<l-1>\rshp 2$), and
$0 \leq^* \vec j <^* (<l-1>\rshp 2)$, respectively.
This is why we have changed from $\vec k$ to $\vec j$ in the third line above.
That is why we say $\vec j \cat 0$ or $\vec j \cat 1$ to index a scalar from $Z_j$.  

In simple terms, the right hand side of Eq.~\ref{finaldnf} can be written as:
\begin{equation}
a + <\!1\;-1\!>\times d = a + <\!d\;(-d)\!> = <\!(a + d)\;(a - d)\!>,
\label{simplest}
\end{equation}
which says, {\em take each element of $A$ (denoted by $a$) and add and 
subtract each corresponding element of $D$ (denoted by $d$)}.  This way of 
combining corresponding elements ({\bf assuming the FFT weights have already 
been included in $D$}) is called the {\em butterfly} operation (see 
Secs.~\ref{butter-order},~\ref{trans-weights} and Eqs.~\ref{xi1} and~\ref{xi2})
and the $n$ component
vector obtained by joining all of the vectors $<\!(a + d)\;(a - d)\!>$ gives
the vector that results (i.e. ${\tilde Z}_j$) after applying a single iteration 
of the FFT.

\section{Thinking in Hyper-Cubes}
What follows is a description of how each cache piece in the FFT can be viewed
as a hyper-cube with dimension  $\log_2 {n}$ with {\em n}  the cache length.
Unlike the discussion in Chaps.~\ref{part1} and~\ref{chap4} which realized each cache piece (i.e. physically moved data around) to bring locality
to the cache, we access the indices that are within the cache directly
without an intermediate realization between each transpose. 
Recall that the access patterns for the FFT predicted cache misses.
Consequently, with that anticipation, it becomes possible to determine how much to pre-fetch and how much to compute. Note that this process could easily  
be extended up the memory hierarchy which includes the network.

Now we are in each cache piece. Here we do {\bf not} want to realize each FFT transpose.
That is, we now want to determine the indices we need for $\log_2 {n}$ cycles
of the FFT since {\bf we now we have locality}. Our description of {\em the
butterfly}, is a hyper-cube reformulation using MoA algebra
and subsequently the $\psi$-calculus. We show how multiple transposes
on this hyper-cube can be expressed as an invariant number of loops (three
for this example),
starts, stops, and strides. We see through the formulation of the arithmetic needed for
the FFT  that we index pairs of components in the hyper-cube then assign the
same components their updated computed FFT portion. Pairs are assigned at once. Here we derive the transpose, i.e.
the final step. 
\subsection{Transpose Formulation of Butterfly}
Let $l \equiv \log_2{n}$, and $\forall j; \;\; 0 \leq j < l$,
let $\vec q$ denote the
input vector below, with $n \equiv 2^l$. Consequently, $l$, is the
dimension of the hyper-cube we'll define:
\begin{eqnarray}
 Z _j  \equiv \vec t_j \transpose ((< l> \rshp 2 )  \rshp {\vec q}\;).
\label{zjay}
\end{eqnarray}
That is, during each iteration, $j$, of the FFT on
the vector $\vec q$,  a new transpose vector
$\vec t_j$ is created and consequently a new transpose is performed, i.e.
$Z _j \equiv \vec t_j \transpose Z$. In this case, we 
DO NOT want to materialize the array.  We envision this algorithm to be
applied to a data vector $\vec q$ that fits in the cache.

Consider the vector of integers $\vec c \equiv \iota (l)$, where 
$l = log_2 n $ is the dimensionality of
the hyper-cube. Thus the original data vector $\vec q$ becomes
$(<l>  \rshp  2) \rshp \vec q$. Now we want to consider 
$\vec t_j \transpose (( <l> \rshp 2) \rshp \vec q\;)$. To do this we
consider indexing with $\vec i$ where $\vec i$ is a full index of the hyper-cube,
\begin{eqnarray}
\mbox{i.e. }  \forall \vec i; \;\;\;\; 0 \leq^* \vec i <^* (<l> \rshp 2).  
\end{eqnarray}
Thus, we wish to
$\psi $-reduce the following expression:
\begin{eqnarray} 
\vec i \psi (\vec t_j \transpose (( <l> \rshp 2) \rshp \vec q\;)),
\end{eqnarray}
where the transpose vector $\vec t_j$ is defined as:
\begin{equation}
 \vec t_j \equiv (((l-1)-j)\take ) \vec c\; ) \cat ((-1) \take \vec c\;)
\cat ( j \take ((( l-1) -j ) \drop \vec c\; )). 
\end{equation}

By the definition of transpose we have:
\begin{eqnarray}
\vec i \psi (\vec t_j \transpose (( <l> \rshp 2) \rshp \vec q\;))
= \vec i\;\! [ {\vec t}\; ] \psi (( <l> \rshp 2) \rshp \vec q\;).
\end{eqnarray}
Now define $Z \equiv  ( <l> \rshp 2) \rshp \vec q$.
Thus, 
\begin{eqnarray}
\vec i \psi (\vec t_j \transpose Z)
& = & \vec i\;\! [ {\vec t}\; ] \psi Z, \nonumber \\
& = &  ( \vec i\;\! [ (((l-1)-j) \take \vec c\;)]  
 \cat \vec i\;\! [(-1) \take \vec c\; ], \nonumber \\
&  & \cat 
\vec i\;\! [ j \take (((l-1) -j) \drop \vec c\;) ]) \psi Z. 
\end{eqnarray}
Now we take the index apart and apply the PCT (see Sec.~\ref{psicorrespond}). Thus, we have
\[
({\vec i}{\;}^{'}\!\! \cat \vec j \cat \vec k ) \psi  Z  \equiv  ( \vec k \psi ({\vec i}{\;}^{'}\!\! \cat \vec j  )\psi Z ) 
 \equiv  \vec k \psi (\vec j \psi ({\vec i}{\;}^{'} \psi Z )) \equiv 
{\vec k}\psi Y, 
\]
for any any ${\vec i}{\;}^{'}\!, \vec j, \vec k$. (in this case ${\vec i}{\;}^{'}$ is not the same as the above $\vec i\;$). 
Now we define: 
\begin{eqnarray}
\vec k \equiv \vec i\;\![j \take (((l-1)-j)\drop \vec c\;)], \\
Y \equiv (\vec i\;\![(((l-1)-j)\take \vec c\;)] \cat \vec i\;\![(-1) \take \vec c\; ]) \psi Z,
\end{eqnarray}
and we will now apply the PCT (see Sec.~\ref{psicorrespond}):
\begin{equation}
\Ravel (\vec k \psi Y) \equiv   
(\Ravel Y) [\gamma(\vec k;(\tau \vec k) \take (\rho Y)) 
\times \pi( (\tau \vec k)\drop (\rho Y)) + \iota (\pi ((\tau \vec k ) 
\drop (\rho Y)))]. 
\end{equation}
Now we determine the shape of $Y$,$(\rho Y)$ as:
\begin{eqnarray}
\rho Y = \rho ( \vec i\;\! [(((l-1)-j)\take \vec c\; )] \cat \vec i\;\![(-1) \take \vec c\; ]) \drop ( \rho  Z).
\end{eqnarray}
Thus,
\begin{eqnarray}
(\rho Z) & = & <l> \rshp 2 ,  \\
(\rho Y) & = &  (l-j)\drop (<l> \rshp 2), \nonumber \\
& = & <j> \rshp 2.
\end{eqnarray}
Next we need:
\begin{eqnarray}
\tau \vec k &=& j, \nonumber \\ 
(\tau \vec k) \drop (\rho Y) &=& j \drop (<j> \rshp 2) = <>.
\end{eqnarray}
Thus, $\pi ((\tau \vec k ) \drop (\rho Y)) = \pi (<>) = 1$, and $(\tau \vec k) \take (\rho Y) = 
j \take (<j> \rshp 2) \\ = <j> \rshp 2$. Using these results we apply the PCT 
to get:
\begin{eqnarray}
\Ravel (\vec i\;\![j &\take& (((l-1)-j) \drop \vec c\;)] \psi  Y, \nonumber \\ 
&=& ( \Ravel Y ) [\gamma(\vec i\;\![j \take (((l-1)-j) \drop \vec c\; )]; <j> \rshp 2) \times 1  + \iota (1)],  \nonumber \\
&=&( \Ravel Y ) [\gamma(\vec i\;\![j \take (((l-1)-j) \drop \vec c\; )]; <j> \rshp 2)]. 
\end{eqnarray}
Now we simply another step:
$j \take (((l-1)-j) \drop \vec c\; ) = ((l-1)-j)+\iota (j)$.
Thus,
\begin{eqnarray}
\Ravel (\vec i\;\![j & \take& (((l-1)-j) \drop \vec c\;)] \psi  Y  \nonumber \\
&=& ( \Ravel Y ) [\gamma(\vec i\;\![((l-1)-j) + \iota (j)]; (<j> \rshp 2)) ]. 
\label{ysub}
\end{eqnarray} 
Now we reduce Y further:
\begin{eqnarray}
 Y  & \equiv & ( \vec i\;\![(((l-1)-i) \take \vec c\; )] \cat
\vec i\;\! [ (-1) \take \vec c\; ]) \psi Z, \nonumber \\
& = & \vec i\;\![(-1) \take \vec c\;] \psi  ( \vec i\;\! [(((l-1)-j) \take \vec c\;)] \psi Z, \nonumber \\
& \equiv & \vec i\;\! [(-1) \take \vec c\; ] \psi W. 
\end{eqnarray} 
Where we have the definitions:
\begin{equation}
W \equiv  \vec i\;\![(((l-1)-j) \take \vec c\; )] \psi Z, 
\end{equation}
and,
\begin{equation}
{\vec k}{\;}^{'} \equiv \vec i\;\! [(-1)\take {\vec c}\;].
\end{equation}

Thus the PCT in this case is:
\begin{eqnarray}
(\Ravel ({\vec k}{\;}^{'} \psi W)) &  \equiv & ( \Ravel W) [ \gamma ( {\vec k}{\;}^{'}; (\tau {\vec k}{\;}^{'})\take (\rho W))   
\times \pi  ((\tau {\vec k}{\;}^{'}) \drop (\rho W)) \nonumber \\ 
&+& \iota
(\pi ((\tau {\vec k}{\;}^{'}) \drop ( \rho W)))].
\end{eqnarray}
Thus we need:
\begin{equation}
(\tau {\vec k}{\;}^{'})  = \tau (\vec i\;\![(-1)\take \vec c\;]) = \tau (\vec i\;\![l-1])=1, 
\end{equation}
and,
\begin{eqnarray}
\rho W & = &\rho (\vec i\;\![(((l-1)-j)\take \vec c\;)] \drop \rho Z), \nonumber \\ 
&=& (l-1-j) \drop (<l> \rshp 2),\nonumber \\ 
&=& <l-(l-1-j)> \rshp 2, \nonumber \\ 
 &=& <j+1> \rshp 2.
\end{eqnarray}
Thus:
\begin{eqnarray}
(\tau {\vec k}{\;}^{'}) \take \rho W &=& 1 \take (<j+1> \rshp 2) = <2>, \\ 
(\tau {\vec k}{\;}^{'} ) \drop \rho W &=& 1 \drop (<j+1> \rshp 2 ) = <j> \rshp 2.
\end{eqnarray}
Using these expressions we can now evaluate $(\Ravel Y)$ as:
\begin{eqnarray}
(\Ravel Y) & =& \Ravel ({\vec k}{\;}^{'} \psi W), \nonumber \\ 
&=& (\Ravel W)[\gamma(\vec i\;\![(-1)\take \vec c\;];<2>) \times \pi (<j> \rshp 2), \nonumber \\ 
&+& \iota (\pi(<j> \rshp 2))].
\end{eqnarray}
Now simplify $\vec i\;\![(-1) \take \vec c\;]=\vec i\;\![<\!l-1\!>]$. 
Now the $(\Ravel Y)$ expression is subscripted as:
\begin{eqnarray}
(\Ravel W)[\gamma(\vec i\;\! [<\!l-1\!>];& &<2>) \times \pi(<j> \rshp 2)+ \iota (\pi(<j> \rshp 2))]\nonumber \\ 
& & [\gamma(\vec i[((l-1)-j) +
\iota (j)];(<j> \rshp 2)].
\label{wsub}
\end{eqnarray} 
In other words, the second expression in brackets, in Eq.~\ref{wsub}, is a 
subscript (extracts an element) from the preceding expression (i.e. the first
part of Eq.~\ref{wsub} that has the form $(\Ravel W)[ \;\;]$ which is a 
sub-array since the expression in square brackets is a vector).
Next reduce $W \equiv {\vec k}{\;}^{''} \psi Z$. With ${\vec k}{\;}^{''} \equiv \vec i\;\![(((l-1)-j)\take \vec c\;)]$, $\rho Z = <l> \rshp 2$,
the PCT gives:
\begin{eqnarray}
\Ravel ({\vec k}{\;}^{''} \psi Z) & = & (\Ravel Z)[\gamma({\vec k}{\;}^{''};(\tau { \vec k}{\;}^{''})\take \rho Z)\times 
\pi((\tau {\vec k}{\;}^{''}) \drop \rho Z) \nonumber \\ 
&+& \iota (\pi((\tau {\vec k}{\;}^{''}) \drop \rho Z))].
\end{eqnarray}
Using:
\begin{eqnarray}
\tau {\vec k}{\;}^{''} & = & ((l-1)-j), \\ 
(\tau {\vec k}{\;}^{''}) \take \rho Z &=& <(l-1)-j) \take (<l> \rshp 2), \nonumber \\ 
&=& <(l-1)-j> \rshp 2. 
\end{eqnarray}
and,
\begin{equation}
(\tau {\vec k}{\;}^{''}  ) \drop \rho Z = <(l-(l-1-j))> \drop \rho Z  = <j+1> \rshp 2, 
\end{equation}
we obtain:
\begin{eqnarray}
(\Ravel W) &=& \Ravel ({\vec k}{\;}^{''} \psi Z), \nonumber \\ 
&=& (\Ravel Z)[\gamma(\vec i\;\![((l-1)-j) \take \vec c\; ] ; <\!((l-1)-j)\!> \rshp 2) \nonumber \\ 
& \times & \pi(<j+1> \rshp 2) + \iota(<j+1> \rshp 2)].
\end{eqnarray}
This is in a {\bf {\em transitional} ONF}. Why would this be true?
Notice,  when we go from $\psi$ indexing to $[\;\;]$ indexing, we
utilize $\gamma$, {\em index vectors}, and shapes. That is, we
start to use the {\em Psi Correspondence Theorem }. Thus, we
have created an ONF which does indeed calculate the {\em offset} from
the start of the array accessed {\em contiguously} in memory. However,
these indices i.e. offsets, are akin to {\em random accesses}. Ideally,
we want to be able to calculate {\em starts, stops}, and {\em strides}.
This representation can be {\em generally } fed to hardware, e.g. {\bf DMA}s,
{\bf FPGA}s, and ASICS,  etc.
We do this now.
The following {\em Generic Normal Form} illustrates the {\em idealized}
form discussed above. Here we have minimized the design to three deterministic
loops. That is, for any size problem represented in any dimension, we have
only three loops. We also know how to calculate {\em stop}, i.e. the
upper bound, and all {\em strides} as follows:
\[ \forall j; \;\;\;\;\; 0 \leq j < l, \]
\[ \forall m; \;\;\;\;\; 0 \leq m < \pi( <\!(l-1)-j\!> \rshp 2), \]
\[ \forall n; \;\;\;\;\;0 \leq n < \pi (<j> \rshp 2), \]
\[ \forall k; \;\;\;\;\; 0 \leq k < 2, \]
\begin{equation}
Z_j \equiv ( \Ravel Z) [(m \times (\pi <j+1>\rshp 2)) + (k\times \pi (<j>\rshp 2))
+n].
\label{final_loops}
\end{equation}
{\bf qed}
\footnote{ 
Notice that the $k$ loop cycles through the elements of $\iota \;(2)$. Also note that $\vec q$ (on which $Z$
is based) has the
weights applied prior to this step.}

Thus for each value of $j$ an array $Z_j$ is created according to 
Eq.~\ref{final_loops} by looping through the variables from fastest 
(innermost loop) to slowest (outermost loop) in the order $k$, $n$, and $m$. 
Now we verify that this is the correct ordering of the loops.  

For example, assume $l = 8$, $j = 4$, and define 
\begin{equation}
c \equiv \iota (l) = <\!0\;1\;2\;3\;4\;5\;6\;7\!>.  
\end{equation}
The ordering of the loops over variables $k$, $n$, and $m$ is related to the 
full index: 
\begin{equation}
\vec i \equiv <\!i_0\;i_1\;i_2\;i_3\;i_4\;i_5\;i_6\;i_7\!>,
\label{iind}
\end{equation}
used to select an element of the new array.  In this index, the variables
$i_p$ for $0 \le p < 8$ each take on the values $0 \le i_p < 2$.

In Eqs.~\ref{ysub} and~\ref{wsub}
we find the following pieces of $\vec i$ selected:
\begin{equation}
\vec i [<\!l-1\!>] = \vec i [<\!7\!>] \equiv i_7,
\label{ifirst}
\end{equation}
\begin{equation}
\vec i [((l-1) - j) + \iota (j)] = \vec i [<\!3\;4\;5\;6\!>] \equiv 
 <\!i_3\;i_4\;i_5\;i_6\!>,
\label{isecond}
\end{equation}
and,
\begin{equation}
\vec i [((l-1) - j) \take \vec c] = \vec i [<\!0\;1\;2\!>] \equiv
<\!i_0\;i_1\;i_2\!>.
\label{ithird}
\end{equation}
We see that as $\vec i$ cycles through all possible values, the piece given
by Eq.~\ref{ifirst}, cycles fastest followed by that in Eq.~\ref{isecond}
and the slowest cycling piece in Eq.~\ref{ithird}. Close examination of 
Eqs.~\ref{ysub} and~\ref{wsub} shows that these three pieces (Eqs.~\ref{ifirst},
~\ref{isecond}, and~\ref{ithird}) appear in $\gamma$ expressions involving the
following shapes: $<\!2\!>$, $(<\!j\!> \rshp <\!2\!>)$ and $(<\!(l-1) - j\!> \rshp 2)$
respectively.  As such, the corresponding $\gamma$ expressions take on
the bounds of the variables $k$, $n$, $m$ (from fastest to slowest).  This allows us to eliminate the $\gamma$ expressions in Eq.~\ref{ysub} and~\ref{wsub} 
to yield the final ONF of Eq.~\ref{final_loops}.

\begin{equation}
\end{equation}

{\section{Merging the two Derivations}
At each iteration, $j$, we deal with the quantity $Z_j$ defined  for 
$\;\; 0  \leq j <l$ in Eq.~\ref{zjay}.  Note also we deal with the vector 
index $\vec j$ (not
to be confused with the scalar index $j$).
The vector index takes on the values:
$\;\; 0 \leq^* \vec j <^* (<l-1> \rshp 2)$,
and when used to index $Z_j$ we obtain:
\begin{eqnarray}
\vec j \psi Z_j & \equiv & <\!(( \vec j \cat 0 ) \psi Z _j ) \;
 ((\vec j \cat 1 ) \psi Z _j )\!>.
\label{jofz}
\end{eqnarray}
But, $Z_j \equiv \vec t_j \transpose Z$, and $\forall {\vec i}; \;\;
0 \leq ^* \vec i <^* (<l> \rshp 2)$, we have:
\begin{equation}
\vec i \psi ( \vec t_j \transpose Z) \equiv \vec i\;\![{\vec t_j}\;] \psi Z. 
\end{equation}
The DNF previously given for the FFT in Eq.~\ref{finaldnf} thus becomes:
\[
((\vec j \cat 0)[{\vec t_j}\;] \psi Z) 
+ <1\;-1> \times (( \vec j \cat 1) [{\vec t_j}\;]) \psi Z , 
\]
and when reduced becomes:
\[ \forall j; \;\;\; 0 \leq j < l, \]
\[ \forall m; \;\;\; 0 \leq m < \pi (<\!(l-1)-j\!> \rshp 2), \]
\[ \forall n; 0 \leq n < \pi (<\!j\!> \rshp 2), \] 
\begin{eqnarray}
\lefteqn{
(\Ravel Z) [(m \times \pi (<\!j+1\!> \rshp 2)) + ((\iota (2)) \times
\pi (<\!j\!> \rshp 2)) + n] }\nonumber \\
 &  \equiv &
(\Ravel Z) [(m \times \pi (<\!j+1\!> \rshp 2)) + (0 \times
\pi (<\!j\!> \rshp 2)) + n] \nonumber \\
&+& <1\;-1>  \times 
(\Ravel Z)[m \times  \pi (<\!j+1\!> \rshp 2) \nonumber \\
\;& +& ( 1 \times \pi (<\!j\!> \rshp 2 )) + n]. 
\end{eqnarray}
Notice, that computationally we'd really be calculating the strides
in the registers, not the control structures as in the classical design of 
the FFT. 
Finally, let $ f \equiv \pi ( <\!j\!> \rshp 2)$, and let $g \equiv
 \pi (<\!j+1\!> \rshp 2)$ then,
\begin{eqnarray}
\lefteqn{ (\Ravel Z) [(m \times g) + n + (<0\;1> \times f)]} \nonumber \\
& \equiv &
(\Ravel Z)[(m \times g)+n] \; + <1\;-1> \times
(\Ravel Z)[(m \times g)+(f+n)]. 
\label{final}
\end{eqnarray}
And we finally make contact with Eq.~\ref{simplest} by writing the the right 
hand side of Eq.~\ref{final} as:
\begin{equation}
<\!(a + d)\;(a - d)\!>,
\label{next_simplest}
\end{equation}
if we identify $a$ and $d$ with:
\begin{equation}
a \equiv (\Ravel Z)[(m \times g)+n], 
\end{equation}
and,
\begin{equation}
d \equiv (\Ravel Z)[(m \times g)+(f+n)]. 
\end{equation}

Thus the entire array of Eq.~\ref{final} at step $j$ (scalar $j$) is an array 
of shape $(<\!l\!> \rshp 2)$ consisting of the collection of two component
vectors given in Eq.~\ref{next_simplest}, each of which being indexed by the 
$l - 1$ component index vector $\vec j$ of shape $(<\!l - 1\!> \rshp 2)$.

This is the final result which is relatively simple considering 
the lengthy derivation required to produce it. 

\section{Implementation and Performance}

This section recaps the implementation and resulting performance details that 
were presented in the first two chapters of this series. We emphasize that 
the present chapter is a formal derivation of a practical algorithm that 
was implemented and tested in Chaps.~\ref{part1} and~\ref{chap4}.  

The innermost part of the FFT algorithm is shown as implemented in Fortran
90 in Fig.~\ref{code}.  This algorithm is essentially the non-cache optimized
FFT that was taken from Ref.~\cite{cpc}.

\begin{figure}
\begin{center}
\includegraphics{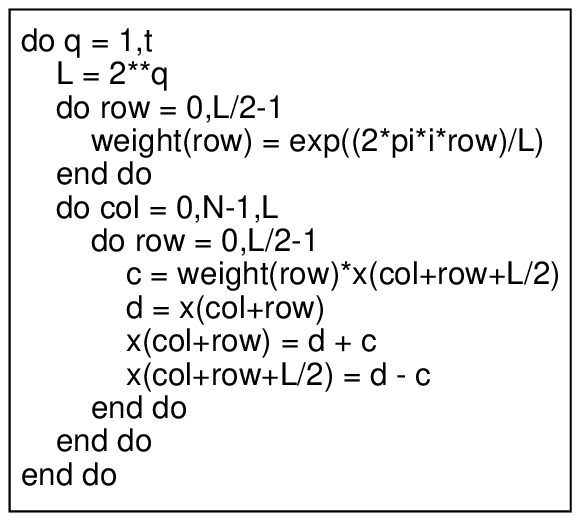}
\end{center}
\caption{\label{code} Fragment for the most important piece of the CC code
(radix 2, in-place, butterfly). In this fragment $t=log_2(N)$ is the power
of $2$ corresponding to the total number of array elements, $N$, and
$x$ is the array being transformed.}
\end{figure}

The ONF of Eq.~\ref{final} is equivalent to the cache optimized FFT that was
implemented and tested in the first two chapters in this series.  In practice, 
however, Eq.~\ref{final} implies a simpler control structure that increments
by unity rather than a power of two as illustrated in the code fragment
of Fig.~\ref{code}.  We expect the hyper-cube formulation of Eq.~\ref{final}
to be somewhat faster due to this simpler control structure.  This conjecture
is currently being tested.

Two other important
code fragments implement the reshape-transpose operation and 
the final transpose that carry out data rearrangements for the cache optimized
FFT.  The reshape-transpose operation is given
below as implemented in C++ in Fig.~\ref{trans_rshp_code} and the 
final transpose is illustrated in Fig.~\ref{final_trans_code}.

In Figs.~\ref{code},~\ref{trans_rshp_code}, and~\ref{final_trans_code}
we intentionally present implementations in Fortran 90 and C++ to emphasize
that our derivations serve as a prescription for implementation in any language.

\renewcommand{\baselinestretch}{1}
\begin{center}
\begin{figure}[ht]
\scriptsize
\begin{quotation}
\begin{tt}
\noindent

\begin{verbatim}
void trans_rshp(complex *datvec,complex *temp,int nmax,
                int csize)
{
        int iind,jind,kind,kmax,jmax,imax,index,c2size;
        int rows,arg;
        int max(int a,int b);

//      The routine carries out the transpose-reshape operation
        index=0;

        rows = nmax/csize;
        imax = rows-1;

        c2size = int(pow(csize,2.0));
        jmax = max(0,nmax/c2size-1);


        for(jind=0;jind<=jmax;jind++)
        {
            for(iind=0;iind<=imax;iind++)
            {

                temp[index] = datvec[jind+csize*iind];
                index += 1;
            }
        }

        for(iind=0;iind<=nmax-1;iind++)
        {
            datvec[iind] = temp[iind];
        }
}
\end{verbatim}

\end{tt}
\end{quotation}
\caption{\label{trans_rshp_code}
C++ code fragment implementing the reshape-transpose in terms of
index manipulations.
}
\end{figure}
\end{center}

\renewcommand{\baselinestretch}{1}
\begin{figure}[ht]
\scriptsize
\begin{quotation}
\begin{tt}
\noindent

\begin{verbatim}
void final_trans(complex *datvec,complex *temp,int nmax,
                 int lcap,int csize)
{
/*      dvar = # of 2's in the hyper-cube
 *      lcap = # of transpose-reshapes (T-rho) that have to be undone
 */
        logc = log(float(csize))/log(2.0);
        sig = lcap*logc;
        smax = pow(float(2),sig);

        dvar = log(float(nmax))/log(2.0);
        tmax = pow(2.0,dvar-sig);

        index = 0;
        for(tind=0;tind<=tmax-1;tind++)
        {
            for(sind=0;sind<=smax-1;sind++)
            {
                temp[index] = datvec[sind*tmax + tind];
                index += 1;
            }
        }

        for(index=0;index<=nmax-1;index++)
        {
            datvec[index] = temp[index];
        }
}

\end{verbatim}

\end{tt}
\end{quotation}

\caption{\label{final_trans_code}
C++ code fragment implementing the final transpose bringing the data back
into the correct order.
}
\end{figure}

\begin{figure}[t]
\small
\begin{centering}
\begin{center}
\includegraphics{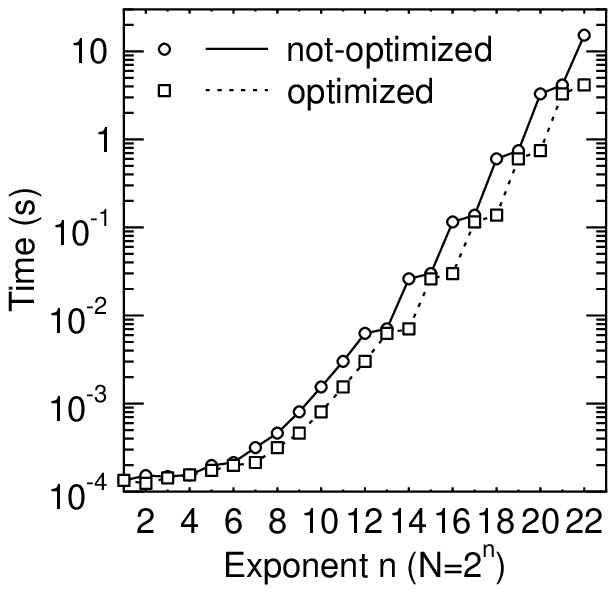}
\end{center}
\end{centering}
\caption{\label{time_vs_size} Comparison of the cache-optimized FFT with our
previous $\psi$-designed FFT~\cite{mullin.small:} indicating reproducible
enhanced performance. The data in this figure
represent the raw timing data obtained by running the experiments in a
single-processor, dedicated, non-shared environment on the Maui SP2 machine
``squall" (one of $2$, 375Mhz Nighthawk-2, IBM SP2 nodes).
Reproducibility was demonstrated by comparison of the results of
five separate runs which produced nearly identical results (not shown). The
slope of the curve reveals the speed of various levels of the memory hierarchy
as amplified in Fig.~\ref{enhance}.}
\end{figure}

\begin{figure}[t]
\small
\begin{centering}
\begin{center}
\includegraphics{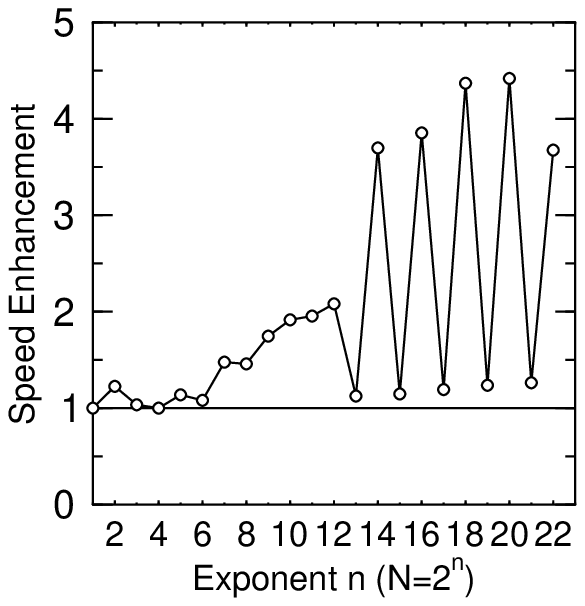}
\end{center}
\end{centering}
\caption{\label{enhance} Performance enhancement of the cache-optimized FFT
as compared with our previous $\psi$-designed FFT showing a factor of four
enhancement for some of the largest sizes. The data plotted in this
figure is the ratio of the time for the non-optimized routine divided by that for
the cache-optimized routine. The changing slope of the optimization curve 
reveals
various levels of the memory hierarchy.  For roughly $2^1 \le N \le 2^6$
the speed is most likely dominated by the speed of the registers.  For
$2^6 \le N \le 2^8$ the speed is dominated by L1 and L2 cache and for
$2^9 \le N \le 2^{12}$ the speed corresponds to main memory.  For $N \ge 2^{13}$
paged memory (of size $4kB$) dominates.  The jumps in performance correspond
to the presence or absence of page faults.
}
\end{figure}

The performance results for our cache-optimized FFT algorithm 
(presented in the two previous chapters) 
are presented in
Figs.~\ref{time_vs_size} and~\ref{enhance}.  These experiments were
run in a single-processor, dedicated, non-shared environment on the
IBM SP2 machine ``squall" at the Maui High-Performance Supercomputer
center~\cite{maui:}.
Specifications for the machine are quoted in the caption to
Fig.~\ref{time_vs_size}.

In the first figure (Fig.~\ref{time_vs_size}) we plot the time vs. input
data length.  There are two curves, one for our new cache-optimized FFT
and one for a similar run with no cache optimization.  Direct comparisons
are possible since both curves are produced by the same code.  For the
non-cache-optimized run, we simply chose the blocking size $c$ (specified as
a parameter at run time) to be greater than or equal to the length of the data
vector.
We see that the curves have essentially the same shape but the cache-optimized
one is shifted to the right by one power of two compared to the non-optimized
one.  Thus for a {\it given run time}, we can legitimately claim a factor of two
speed-up.

The results presented in Fig.~\ref{enhance} emphasize the improved performance
for a {\it fixed data size} by taking the ratio of the run time for the
non-optimized run to that for the cache-optimized one.  We see that for some of 
the largest sizes considered, a factor on the order of $4$ speedup is achieved.

Figure~\ref{enhance} is also enlightening in that it highlights the various
levels of the memory hierarchy.  A change in slope of run-time vs. size
indicates the crossing of a boundary between one level of the memory hierarchy
and another.  For example from the results of Fig.~\ref{enhance} we can make
the following estimates.  For roughly $2^1 \le N \le 2^6$ the speed is most
likely dominated by the speed of the registers.  For
$2^6 \le N \le 2^8$ the speed is dominated by L1 and L2 cache and for
$2^9 \le N \le 2^{12}$ the speed corresponds to main memory.  For $N \ge 2^{13}$
paged memory (of size $4kB$) dominates.  The large jumps in performance (i.e.
factors of $4$ for some of the largest sizes) correspond to the presence or absence of
page faults.

In general, the performance of the algorithm is a tradeoff
between the increased speed obtained by having more data in the cache (with
increasing $c$) and the cost of actually moving the data around.  One might
naively guess that the best performance would be obtained by choosing $c$ to
be equal to the cache size.  However, we find, the best performance by choosing
blocking sizes $c$ given by small powers of $2$.  In other words, it is
more economical to move data around many times within the cache than it is
to move large blocks into and out of cache due to the extreme speed of the
cache.  In a sense, the operating system is able to overlap computation and IO
using small blocking sizes $c$.  Direct comparisons of our {\em non-cache 
optimized} routine to library routines showing comparable, and in most cases,
superior performance were presented in 
Refs.~\cite{lenss3,lenss4,mullin.small:,mul03} and Chap.~\ref{intro}.

\section{Conclusion}

We have presented a derivation of the FFT using the techniques of Conformal
Computing in the framework of a hyper-cube data structure.  The final result is
given in Eq.~\ref{final} and is extremely simple given the lengthy derivation
that led to it.  The structure is very simple (one can clearly see the
three loops over the variables $k$, $n$ and $m$ for each step $j$ in the 
FFT) and is independent of the length
of the input data vector.  The addresses associated with the loop variables 
$k$, $n$ and $m$ will be evaluated in registers.
As such, the present implementation is expected to be faster than the result
of the previous two 
chapters.
This is because the outermost
loop variable $j$ is successively incremented by unity in contrast to the traditional approach in which it is incremented by a power of $2$.  Effort is currently underway to
test this conjecture.

This chapter also serves as a continuing {\em in depth} tutorial introduction
to the methods of Conformal Computing applied to a non-trivial example.
Every step of every calculation has been presented in full detail and is
based on the introduction of the Conformal Computing machinery in 
earlier chapters.
These 
techniques are extremely powerful in that they allow one to eliminate temporary
arrays through the use of direct indexing (note the role played by the loop
variables $k$, $n$, $m$, in Eq.~\ref{final}).  Another important aspect of this
approach is that once the analysis is carried out, the resulting ONF 
(see Eq.~\ref{final}) is a {\bf prescription for building an efficient computer
program in any convenient language for implementation in software or hardware.  
Also, since the same formalism
is used to describe the machine as is used to describe the science and/or engineering problem, one is empowered to reason mathematically about the correctness
and efficiency of the implementation.} 


%
%
%

\chapter{Density Matrix Operations for a Quantum Computer}
\label{chap6}

\section{Chapter Summary}

This chapter is concerned with the efficient manipulation of sparse matrix
operations that arise in the simulation of a quantum computer. In
particular, matrix multiplication traditionally employed to effect
the {\em gating} operation of a particular qubit or collection of qubits is
replaced by an equivalent operation involving direct indexing of the
matrix elements.  As such, the efficiency of a quantum simulator will
be greatly enhanced due to the elimination of the need for temporary arrays.
The algorithm is completely general and applies to the {\em gating} operation
of arbitrary collections of qubits.  The algorithm we present allows one
to do a number of generalized matrix operations in a {\em single step} thus
eliminating the need for large temporary arrays.

\section{Quantum Computing: Motivation for a Matrix Problem with  Arbitrary 
Array Access Patterns}

We now give a brief overview of the motivation for the present problem.
The dream of Quantum Computing is the realization of a {\em quantum computer} in
which data is represented by the states of a physical system such as the spin
of an electron or proton.  The physical picture one should imagine (to the 
extent that quantum processes can be imagined) is that of a spin or collection 
of spins interacting with electromagnetic fields.  One possible embodiment of 
a quantum computer would be to utilize an apparatus closely resembling that used
in Magnetic Resonance Imagining (MRI).  Individual spins are manipulated 
through application of pulses of electromagnetic fields.  

The primary interest
in Quantum Computing is the promise of greatly increased computing capability
due to the inherently parallel nature of data storage and computation resulting
from the superposition principle of quantum mechanics.  For example, it has
been theoretically proven that certain algorithms requiring exponential time
on a classical computer can be solved in polynomial time on a quantum 
computer~\cite{shor}.
We shall say no more about Quantum Computing in this chapter and we turn to now 
the specific matrix problem to be solved.  Further background material can 
be found in Refs.~\cite{cirac,ekert,lomonaco,acm,qinfo} and references 
therein.

The algorithm we present is based on the techniques of {\em Conformal 
Computing}: a rigorous mathematical approach to  the construction of 
computer programs based on an algebra of abstract data types ({\em 
A Mathematics of Arrays}) and an indexing calculus ({the \em $\psi$-Calculus}).
One of the most important aspects of this approach is the ability to compose 
a sequence of algebraic manipulations in terms of array shapes and direct 
indexing.  The net result is the elimination of temporary arrays, which leads 
to significant performance improvements.  Another important point of this 
approach is that the mathematics used to describe the problem is the same as 
that used to describe the details of the hardware. Thus at the end of a 
derivation the resulting final expression can simply be translated into 
portable, efficient code for implementation in hardware and/or software.  
Another important attribute of the Conformal Computing approach is the 
ability to mathematically 
prove that the resulting implementation is maximally efficient given a set of 
metrics (e.g. speeds of memory levels, processors, networks, etc).
The details of this approach have been presented in detail elsewhere in
Refs.~\cite{mul03,cpc} and in Chap.~\ref{chap2}

The reader should not be misled by the name {\em Conformal Computing}.  
Conformal in this sense is not related to {\em Conformal Mapping} or similar 
constructs from mathematics although it was inspired by these concepts in which
certain properties are preserved under transformations.  In particular,
by {\em Conformal Computing} we mean a mathematical system that {\it conforms}
as closely as possible to the underlying structure of the hardware.

\section{Quantum Evolution: the Density Matrix}

Linear Algebra is the natural context in which to describe operations
in a quantum computer and the central quantity is the density matrix.  The
density matrix provides a complete description of the time evolution of 
the quantum system.  Changes to the system under the application of 
{\em gating} operations are represented as unitary transformations of the 
density matrix.  In practice these transformations are carried out by 
multiplying the density matrix on the right and left by a unitary matrix and
its inverse respectively.  In general, for any given operation, we require 
only a sparse collection of elements from the density matrix to be 
rearranged.  The specific arrangement of the required elements in the matrix 
depends on which spin or collection of spins are being manipulated.

We propose herein a sparse matrix algorithm that eliminates the need 
to store large sparse unitary matrices corresponding to specific gating 
operations.  In effect we use direct indexing to effectively move
the required density matrix elements onto the diagonal to achieve 
block-diagonal form.  Then the gating operations are applied to these 
elements in a simplified form.  Note, we are not {\bf \em actually}
moving elements of the density matrix around but rather we are carrying out
such operations {\bf \em virtually} through direct indexing.  The net result
is an algorithm requiring fewer floating point multiplies and less storage.

We focus on the following problem.  Given the density matrix, for an arbitrary
quantum operation on an arbitrary number of states (qubits) we wish (for
computational convenience) to rearrange the data so as to place the
required elements on the diagonal in block-diagonal form.  Using the 
techniques of Conformal Computing we have
found a way to do this {\em in one step}. 

In the following we consider $16\times 16$ matrices corresponding to 
a situation in which we have four qubits and we are considering only 
operations involving two qubits at a time.  We choose this situation 
merely for the purpose of illustration.  The algorithm that we present 
in this chapter is applicable to arbitrary numbers of qubits (i.e. $2^n$ for
some non-negative integer $n$) and arbitrary collections of qubits to 
be gated at any one time.

Consider the following possible arrangements for a $16\times 16$ density 
matrix for which we wish to manipulate two spins (qubits).
\begin{figure}[h]
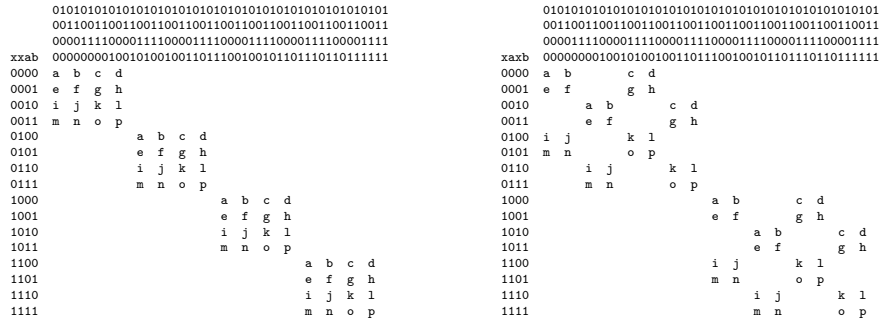

\tiny
\begin{minipage}{2.5in}
\input{xxab}
\end{minipage} \ \
\begin{minipage}{2.5in}
\input{xaxb}
\end{minipage}
\caption{\label{qubits}Accessing qubits.  We wish to rearrange a pattern
such as that on the right into one such as that on the left {\bf \em virtually}
using direct indexing.}
\end{figure}
\normalsize

We wish to rearrange a pattern such as the one on the right into a
block-diagonal form such as that on the left. The transformation is effected
by viewing the $2^n\times 2^n$ density matrix as a $2^{2n}$ dimensional 
hyper-cube and carrying out a certain rearrangement of the indices of the 
hyper-cube.  Note that the labeling of the entries in the matrix 
corresponds to the indices of the hyper-cube.  These indices -- the hyper-cube
coordinates -- are simply the integers in the base-2 representation.
 
For a $2^n\times 2^n$ density matrix $D$, we say that the {\em shape} of the 
array (a vector giving the lengths of its dimensions) is $\rho D = 
<\!2^n\;2^n\!>$ (see Chap.~\ref{chap2} for elements of the 
theory).  Now we 
{\em reshape} the array into a hyper-cube $D_h$.  
Now the {\em shape} of the hyper-cube is a vector of $2^n$ $2$'s, that is,
$<\!2\;2\;\cdots 2\!>$.

The rearrangement we seek is a certain permutation of the indices of the 
hyper-cube.  We write the block-diagonal matrix $D_b$ as
$ D_b = (\rho D) \rshp ({\vec t} \transpose D_h)$, 
where the vector $\vec t$ is a permutation
vector and the operator $\transpose$ corresponds to transposing the indices
of the hyper-cube as specified by the permutation vector.  For the specific 
example above, we have ${\vec t} = <\!0\;2\; 1\; 3\; 4\; 6\; 5\; 7\!>$.
In the following, we present an algorithm for determining the general 
permutation vector.

\section{The Algorithm}

We want to view the {\em qubits} as {\em coordinates} in 
a {\em hyper-cube}, that is formed by reshaping (restructuring) the density 
matrix.  

Based  on which qubits we wish to gate, i.e. move to the diagonal, 
we desire to create a permutation vector that permutes the indices of
the hyper-cube. This is accomplished by applying MoA's
binary transpose (see Chap.~\ref{chap2}). Consequently, all gated bits are moved to the 
diagonal
in a block fashion. The blocks are square, i.e. $2^q \times 2^q$. with
$q$ denoting the number of qubits. {\bf \em Note that the design scales to  
multiple density hyper-cubes, and processing of each gate could be
performed in parallel.} We'll first look at an explicit example, i.e.
a $16 \times 16$ density matrix where we gate various combinations of
2 qubits. Thus we start with a $2^4\times 2^4$ density matrix which
gets restructured to a $2^8$ hyper-cube. That is, we structure the density 
matrix, denoted by $D$ to be a hyper-cube, $D_h$, 
defined as follows:
\begin{eqnarray}
D_h = (<8> \rshp 2  ) \rshp D,
\end{eqnarray}
where the {\em reshape} operator $\rshp$ changes the shape of the array.
Consequently, a 2-dimensional  structure has been transformed into
an 8-dimensional structure. We want to now develop an algorithm that
transposes $D_h$ and by doing so eliminates the need for
numerous permutation matrices, i.e. matrices that must be multiplied
with the density matrix, $D$, to have the effect of moving the desired
qubits to the diagonal.

Let's begin with 4, $4 \times 4$ diagonal block matrices. That means that 
bits 0 and 2 are gated (counting from right to left), i.e. the pattern in 
Fig.~\ref{xaxb}.
Consequently, we will use {\em ab} to denote which bits to gate in the density 
matrix.   Note that here we talk about  {\em ab} or $2$ bits. In general, 
we'll say $\tau {\vec a}$ bits. This means that whenever we want
to reference a patterned sparse array of gate indices, we must permute
them such that they are on the diagonal, blocked as $4 \times 4$ sub-matrices.
Figures,~\ref{xaxb}, and~\ref{xabx}, demonstrate some
of the ways two bits may be gated. Notice also what we have done is 
transform row, column 
indices of $D$ into their base 2 representation. We will now show
how to combine the later to form $D_h$ coordinates.





\begin{figure}
\small
\begin{center}
\includegraphics[width=6cm]{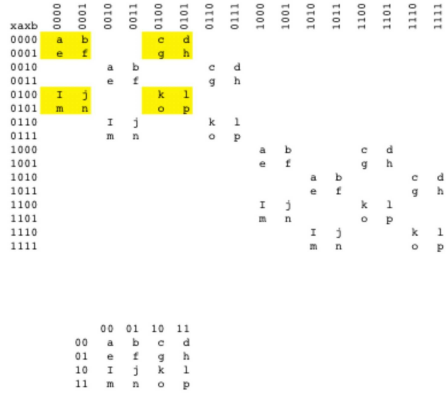}
\end{center}
\caption{\label{xaxb} The xaxb permutation.}
\normalsize
\end{figure}

\begin{figure}
\small
\begin{center}
\includegraphics[width=6cm]{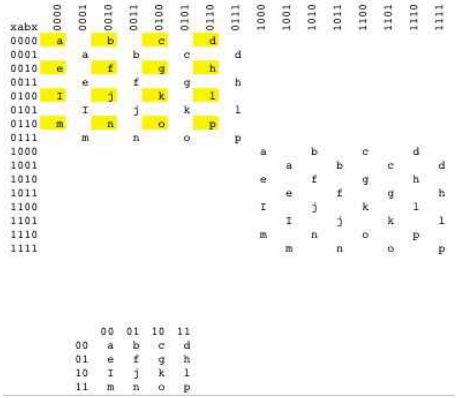}
\end{center}
\caption{\label{xabx} The xabx permutation.}
\normalsize
\end{figure}





When an algorithm is defined using MoA and reduced using it's 
$\psi$-calculus, a normal form is 
revealed. When processor/memory/ ... / hierarchies
are added, i.e. {\em increasing the dimension of the algorithm}, the iteration
space and data flow over each level is also normalized.
Consequently, we can describe the physics using a data structure indicative
of its quantum nature. The same algebra can partition the problem into
blocks indicative of the processor/memory/.../ hierarchy used for
execution.

\subsection{Assumptions in the Example:}
\begin{description} 
\item[Gated Bits:] Let a and b denote which bits to gate.  In our examples we 
look at the following bit patterns: 

\begin{itemize} 
\item xaxb: bits 0 and 2, (Fig.~\ref{xaxb})
\item xabx: bits 1 and 2, (Fig.~\ref{xabx})
\end{itemize}
There are others, but for the purpose of illustration we will only consider 
the above patterns. 

\item[Bit Ordering:]
Bits are numbered from right to left. That is, 
 $1\; 1\; 1\; 0\;$ is  used to evaluate its decimal equivalent:
\[	( 1 * 2^3 )  + ( 1 * 2^2  )  +  ( 1 * 2^1   ) + ( 0 *  2^0   ). \]
\item[Vector Ordering:] 
Indexing is numbered from left to right
That is,  a vector, $<\; 1 \;1 \;1\; 0 \;> $ when indexed would yield:
\[< 1 1 1 0 >[0]  = 1, \]
when the 0th index is accessed and,
\[< 1 1 1 0 >[3]  = 0, \] 
when accessing the third.

\item[ Example(cont.): From qubits to permutation vector:] 
\end{description}

\begin{tabular} {lllllll}
bits 0,2: & xaxb  $\rightarrow$ & $3\;2\;1\;0$ & $\;\;\;$ & $3\;2\;1\;0$ & $\;\;\;\;\;$  &  bit ordering \\
          &                     & $0\;1\;2\;3$ & $\;\;\;$ & $0\;1\;2\;3$ & $\;\;\;\;\;$  &  index ordering \\
          &                     & $0\;2\;1\;3$ & $\;\;\;$ & $0\;2\;1\;3$ & $\;\;\;\;\;$  &  swap bits 1 and 2 \\
          & &  $<\!0\;2\;1\;3\;4\;6\;5\;7\!>$ & & & &  is the transpose vector   \\
& & & & & & \\
bits 1,2: & xabx  $\rightarrow$ &$3\;2\;1\;0$ & $\;\;\;$ & $3\;2\;1\;0$ & $\;\;\;\;\;$  & bit ordering \\
          &                     &$0\;1\;2\;3$ & $\;\;\;$ & $0\;1\;2\;3$ & $\;\;\;\;\;$  & index ordering \\
          &                     &$0\;1\;3\;2$ & $\;\;\;$ & $0\;1\;3\;2$ & $\;\;\;\;\;$  & swap bits 2 and 3 \\
          &                     &$0\;3\;1\;2$ & $\;\;\;$ & $0\;3\;1\;2$ & $\;\;\;\;\;$  & swap bits 1 and 2 \\
         & &  $<\!0\;3\;1\;2\;4\;7\;5\;6\!>$ & & & & is the transpose vector \\
& & & & & & \\
\end{tabular}

Let $\vec t$ denote a transpose vector. Thus, given an arbitrary index
vector $\vec i$ s.t.
$0 \leq^* \vec i <^* 2^{2n}$  of the $D_h$,\footnote{The notation 
$\sigma f^*\vec v$ and $\vec v f^* \sigma $ denotes a binary operation 
$f$ s.t. $f$ is a
boolean operation, e.g. $\leq$ or $<$, and $\sigma$ is applied point-wise
to each component of $\vec v$, e.g. $\sigma \leq \vec v[i]\; \; \forall 
\; \; 0 \leq i <  (\tau n)$, denotes
comparison of the elements of the vector ${\vec i}$.} 
composed as above, 
we permute all indices by $\vec t$, i.e. the permuted index:
\begin{equation}
 \vec i\; [ \vec t\;] \equiv \vec t \transpose \vec i
\end{equation}
acting on  the original array is equivalent to the index $\vec i$ acting
on the permuted array which, consequently, never needs to be materialized.
The permutation vector $\vec t$ effectively
moves all gated indices to the 
diagonal.

Indices are calculated and addressed directly from the
original array stored in memory thus, eliminating intermediate
arrays or permutation matrices. We have shown that we can
algebraically represent the physics, algebraically describe an 
all-at-once operation that is algebraically decomposable to present and 
future architectural platforms (even quantum).
The algebra remains the same throughout
the problem, the decomposition over processor/memory/FPGA, the mapping, 
and the architectural abstraction, with verifiable designs.

\subsection{Final Expression and Normal Form}
Given a density matrix $D$  such that the shape is given by:
$\rho D  =  < 2n\;  2n >$, we restructure to a hyper-cube, $D_h$.
Let $\vec s$  denote the shape of $D_h$ s.t. \\
$\vec s =   < (<\!2n\!> \rshp 2) >$
(i.e. the shape is a vector consisting of $2n$, 2's).
Then 
\[D_h = \vec s  \rshp D \]
Using $\vec t$, the transpose vector previously defined,
perform a binary transpose:     
\[			\vec t   \transpose  D_h  \]
Now, all matrices defined by bits chosen are on the diagonal.
Note: Restructuring back to $D$ and indexing creates no new
          arrays because of $\psi$-reduction.

Now $\psi$-reduce to normal form $\rightarrow$ Generic Design.
\begin{eqnarray}
      \forall\;  \vec i\;\  \mbox{s.t.}\;  0 \le{}^* \vec i  <{}^* <\! (<\!2n\!> \rshp 2)\!>,  \\     
\vec i \psi (\vec t \transpose D) \equiv  @D + \gamma( \vec i\; [ \vec t\; ]  ; \vec s )
\label{generic}
\end{eqnarray}
This is the Generic Normal Form.  Equation~\ref{generic} denotes the address of
 an element of the density matrix $D$ in terms of the address of the first
element $@D$ plus the offset $\gamma( \vec i [ \vec t ]  ; \vec s )$.  The
quantity $\gamma( \vec i [ \vec t ]  ; \vec s )$ is the polynomial that 
generates an address from the index vector and the shape.
                                 
\section{Conclusion}
We have presented a general algorithm for increasing the efficiency of 
certain key operations that arise in the solution of the evolution equations
for a quantum system in the density matrix formalism.  In particular, standard
matrix multiplication operations that arise in the description of the quantum
{\em gating} operation have been eliminated.  The equivalent operation is 
now represented in terms of a direct indexing of the appropriate matrix 
elements.  The present approach will increase the efficiency of simulations
of quantum computers through the elimination of temporary arrays and parallel
processing. 

In effect we use direct indexing to effectively move
the required density matrix elements onto the diagonal to achieve 
block-diagonal form.  Then the gating operations are applied to these 
elements in a simplified form.  Note, we are not {\bf \em actually}
moving elements of the density matrix around but rather we are carrying out
such operations {\bf \em virtually} through direct indexing.  The net result
is an algorithm requiring fewer floating point multiplies and less storage.

\section{ Acknowledgments}
We wish to acknowledge many interesting discussions with R. M. Mattheyses 
of GE Global Research, Niskayuna, NY who introduced us to the density 
matrix problem as it relates to quantum simulators such as Quantum Express
(https://www.research.ge.com/quantum/index.jsp). 


%
%
%

\chapter{Conclusions and Grand Challenges}

\section{Conclusions}

We have presented a self-contained introduction to the methods of Conformal 
Computing and have illustrated their use.  The introduction presented 
a survey of the history of these techniques that have been developed over the
past three decades.  We also presented a review of our published work in 
which we found considerable performance gains of important algorithms such
as the FFT in comparison with well-tuned library routines.  Considerable 
reference material was also provided illustrating a variety of applications
to algorithms of interest to science and engineering.

In Chap.~\ref{chap2} a survey of the Mathematics of Arrays (MoA) and 
$\psi$-calculus was given.  These techniques are the two cornerstones of 
the Conformal Computing approach.  The MoA is similar to the algebra of 
the APL programming language.  Indeed, MoA was inspired by APL and was 
an outgrowth of research built on Sylvester's {\em Universal Algebra}.
The MoA represents a substantial improvement over APL's algebra, 
however, in that
a number of mathematical anomalies  have been corrected in the 
MoA's theoretical foundations.  
We now wish to emphasize an important point: {\em \bf
Conformal Computing is NOT APL and only has algebraic similarities in 
common with APL. In particular there is no concept of an indexing calculus
in APL AND MoA is NOT a programming language. It is a Mathematical Theory.} 
Conformal Computing's indexing calculus (i.e. the $\psi$-calculus)
is a crucial aspect of our approach that facilitates the construction of 
efficient computer programs through the derivation of the {\em Operational
Normal Form} (ONF) from the {\em Denotational Normal Form} (DNF).  This 
process, called $\psi$-reduction is unique to the Conformal Computing 
approach and has been successfully exploited in a number of important contexts.
Neither APL nor any other programming language contain the $\psi$-calculus. 
Other languages,
including APL, e.g. Fortran 95, ZPL, etc. contain indexing rules NOT
theories and none can be reduced to a {\em normal form} that allows
one to prove the equivalence of programs.

The present monograph represents the
most complete and self-contained account of the Conformal Computing approach
to date.
The bulk of this monograph is contained in Chaps.~\ref{part1} 
through~\ref{chap6} in which previously unpublished work is presented.  This
work illustrates the techniques in extensive detail and serves as an in-depth 
look at the workings of the Conformal Computing approach.  Chapter~\ref{part1}
presents a new {\em cache-optimized} FFT primarily in the language of 
traditional linear algebra.  As such it bridges the gap between traditional
mathematics of science and engineering and the MoA and $\psi$-calculus.
In addition, Chap.~\ref{part1} demonstrates the importance of the new  
algorithm in which speedups on the order of factors of $2$ to $4$ are achieved.
We emphasize that these are improvements over our previously published 
work that was shown to be competitive or superior to well-tuned library
routines.~\cite{cpc,mullin.small:} 

The full machinery of the Conformal Computing approach, as applied to our 
{\em cache-optimized} routine, is presented in Chaps.~\ref{chap4} 
and~\ref{chap5}.  Chapter~\ref{chap4} presents details of the $\psi$-reduction
process for two key steps in the algorithm: (1) the {\em reshape-transpose} 
operation and the (2) {\em final transpose} operation.

We are finding that it is often convenient to work with multi-dimensional
arrays that have been restructured ({\em reshaped}) as {\em hyper-cubes}.
In this approach, an algorithm is developed and expressed as certain
operations on the index vector of the hyper-cube.  The hyper-cube approach
to the {\em cache-optimized} FFT is presented in detail in Chap.~\ref{chap5}.

The power of the hyper-cube approach is illustrated in another example in 
Chap.~\ref{chap6} in the context of a quantum simulator.  In this example,
an algorithm on the hyper-cube index vector was developed that allows one
to efficiently select arbitrary sparse collections of matrix elements and
{\em \bf virtually} move them to achieve a block-diagonal form for further
processing.

The ability to eliminate {\em unnecessary}
temporary arrays and to compose algorithm steps leads to efficient code. The
ONF is a mathematical prescription for how the code is to be built in any 
language of choice for software and hardware applications. Our approach
uses a common formalism to represent both the computational problem as well
as the underlying hardware.  In this way it becomes possible to mathematically
reason about optimal performance of a given algorithm in a given computational
environment given a set of performance metrics (e.g. speed of the various
levels of the memory/processor/network hierarchy).  The {\em cache-optimized}
FFT is an example of this in which details of the hardware (e.g. the cache 
size) are specified parametrically at run time.  Natural extensions of this
approach to include processors, networks, etc. are possible and are the subject of 
future research.  As architectures get smaller, eventually quantum, issues
such as heat, power, etc. will become essential parameters that
Conformal Computing can address.
The authors have attempted to present {\em Conformal 
Computing} in as much detail as possible and in a tutorial style so as to 
enable others to apply these powerful techniques to their own research.  

\section{Grand Challenges and Future Directions}

The Conformal Computing approach is poised to provide solutions to a number
of important problems facing the field of large-scale and embedded computing.
However, {\em Grand Challenges} in Computational Science  are a necessary
prerequisite.  Such {\em Challenges} include:
\begin{itemize}
\item The Ability to Prove Two Executables are Equivalent.
\item The Ability to Provide {\em Intentional Information} Regarding
Loops, Variables, etc. to a {\em Translator  (or Compiler)} Such that
Various Instatiations are Built, e.g. OpenMP, MPI, forks, threads, etc.
\item Tools to Theorize About Performance Prior to Building Code.
\item Identifying Which Array (Matrix, Tensor) Based Algorithms are Common
Across Domains: LU, QR, etc.
\item Develop Theories Similar to Conformal Computing.
\item Determine the Theoretical Foundations Behind {\em Expression Templates}
Such that an Interface or Tool can be Built to Interface {\em Normal Forms}
to {\em How to Build}  Instantiations.
\item Determine how Languages Such as Matlab and Fortran 90 Can Provide
MoA and Psi Calculus to Users.
\end{itemize}



\backmatter

\bibliographystyle{unsrt}
\bibliography{hpf_combined,paper1,paper3,paper4,cpc,siam}

\begin{thebibliography}{10}
\bibitem{maui:}
We would like the thank the maui high performance computing center for access
  to their {IBM SP}.
\end{thebibliography}


\end{document}